\documentclass[12pt]{article}
\usepackage{amssymb}
\usepackage{a4}
\usepackage{authblk}
\usepackage{amsmath}
\usepackage{graphicx}
\usepackage{subcaption}

\usepackage{comment}
\usepackage{hyperref}
\allowdisplaybreaks[4]

\evensidemargin \oddsidemargin
\def\1{{\bf 1}}
\def\id{\mbox{id\,}}
\def\ot{\!\otimes\!}
\def\ots{\otimes_{\star}}
\def\F{{\cal F}}
\def\bF{\mbox{$\overline{\cal F}$}}

\def\f{{\scriptscriptstyle {\cal F}}}

\def\cross{\mbox{$\rule{0.7pt}{1.3ex}\!\times $}}
\def\cocross{\mbox{$\times \!\rule{0.3pt}{1.1ex}\,$}}
\def\ra{\rangle}
\def\la{\langle}

\def \A{{\cal A}}
\def \B {{\cal B}}
\def \BM {{\cal B}_{\scriptscriptstyle M_c}}
\def \BMst {{\cal B}_{{\scriptscriptstyle M_c}\star}}\def \C {{\cal C}}
\def \C {{\cal C}}
\def \D {{\cal D}}
\def \E {{\sf E}}

\def \I {{\cal I}}
\def \J {{\cal J}}

\def \Ms {{\scriptscriptstyle M}}

\def \IM {{\cal I}_{\scriptscriptstyle M_c}}
\def \IMst {{\cal I}_{{\scriptscriptstyle M_c}\star}}

\def \M {{\cal M}}

\def \Q {{\cal Q}}
\def \QM {{\cal Q}_{\scriptscriptstyle M_c}}
\def \QMst {{\cal Q}_{{\scriptscriptstyle M_c}\star}}
\def \X {{\cal X} }
\def \Z {{\cal Z} }

\def\R{\mbox{$\cal R$}}
\def\r{{\scriptscriptstyle {\cal R}}}
\def\bR{\mbox{$\overline{\cal R}$}}

\newcommand{\trc}{\triangleright}

\newcommand{\ltlc}{\stackrel{\scriptscriptstyle \triangleleft}{}}
\def\g{\mathfrak{g}}
\def\k{\mathfrak{k}}

\def\gm{{\bf g}}
\newcommand{\XiM}{\Xi_{\scriptscriptstyle M}}

\newcommand{\XM}{{\cal X}^{\scriptscriptstyle M}}
\newcommand{\Xip}{\Xi_{\scriptscriptstyle \perp}}
\newcommand{\Xips}{\Xi_{{\scriptscriptstyle \perp}\star}}
\newcommand{\Up}{U_{\scriptscriptstyle \perp}}

\newcommand{\Vp}{V_{\scriptscriptstyle \perp}}

\newcommand{\Np}{N_{\scriptscriptstyle \perp}}

\newcommand{\Omps}{\Omega_{\scriptscriptstyle \perp\star}}
\newcommand{\Omp}{\Omega_{\scriptscriptstyle \perp}}
\newcommand{\Pp}{\mathrm{pr}_{\scriptscriptstyle \perp}}
\newcommand{\Pps}{\mathrm{pr}_{{\scriptscriptstyle{\perp}}\star}}
\newcommand{\Pt}{\mathrm{pr}_{ t}}
\newcommand{\Pts}{\mathrm{pr}_{t\star}}

\DeclareMathOperator{\Rad}{Rad}
\newcommand{\ic}{\ensuremath{\mathcal{I}}}
\newcommand{\Pd}{\mathbb{P}_{\mathbb{C}}^2}
\newcommand{\Pn}{\mathbb{P}_{\mathbb{C}}^n}

\def\TT{{\mathcal T}}
\def\Xis{{\Xi_\star }}

\def\rR{\mathsf{R}}
\def\ric{\mathsf{Ric}}
\def\tT{\mathsf{T}}

\newcommand{\CC}{\mathbb{C}}
\newcommand{\RR}{\mathbb{R}}

\newcommand{\NN}{\mathbb{N}}
\newcommand{\KK}{\mathbb{K}}

\def\nn{\nonumber \\}

\newcommand{\be}{\begin{equation}}
\newcommand{\ee}{\end{equation}}
\newcommand{\bea}{\begin{eqnarray}}
\newcommand{\eea}{\end{eqnarray}}
\newcommand{\ba}{\begin{array}}
\newcommand{\ea}{\end{array}}
\newtheorem{prop}{Proposition}
\newtheorem{lemma}[prop]{Lemma}
\newtheorem{theorem}[prop]{Theorem}

\newtheorem{example}[prop]{Example}
\newtheorem{remark}[prop]{Remark}
\def\sq{\mbox{\rlap{$\sqcap$}$\sqcup$}}
\newenvironment{proof}[1]{\vspace{5pt}\noindent{\bf Proof #1}\hspace{6pt}}%
{\hfill\sq}
\newcommand{\bp}{\begin{proof}}
\newcommand{\ep}{\end{proof}\par\vspace{10pt}\noindent}
\begin{document}

\title{Twisted Quadrics and Algebraic Submanifolds in $\RR^n$}

%\author{Gaetano Fiore\footnote{gaetano.fiore@unina.it}, Thomas Weber\footnote{thomas.weber@unina.it} \\\\
%Dip. di Matematica e Applicazioni, Universit\'a ``Federico II''\\and         I.N.F.N., Sez. di Napoli,         Complesso MSA, V. Cintia, 80126 Napoli, Italy}

%\author[1,3]{Francesco D'Andrea?\footnote{francesco.dandrea@unina.it}}
\author[1,2]{Gaetano Fiore\footnote{gaetano.fiore@unina.it, \ gaetano.fiore@na.infn.it}}
\author[1]{Davide Franco\footnote{davide.franco@unina.it}}
\author[2,3]{Thomas Weber\footnote{thomas.weber@unina.it}}
\affil[1]{Dipartimento di Matematica e Applicazioni “Renato Caccioppoli”,\protect\\
Universit\`a degli Studi di Napoli "Federico II",\protect\\
Via Cintia, Monte S. Angelo, 80126, Napoli, Italia
\vspace{0.3cm}}

\affil[2]{I.N.F.N., Sezione di Napoli,\protect\\
Complesso universitario di Monte S. Angelo ed. 6,
%via Cintia,
80126, Napoli, Italia
\vspace{0.3cm}}

\affil[3]{Dipartimento di Scienze e Innovazione Tecnologica,\protect\\ 
Universit\`a degli Studi del Piemonte Orientale “Amedeo Avogadro”,\protect\\
Viale Teresa Michel 11, 15121 Alessandria, Italia}

\date{October 15, 2020}

\maketitle

\abstract{We propose a general procedure to construct noncommutative deformations of an algebraic submanifold $M$ of $\mathbb{R}^n$, specializing the procedure [G. Fiore, T. Weber, Twisted submanifolds of $\mathbb{R}^n$, arXiv:2003.03854] valid for smooth submanifolds. We use the framework of twisted differential geometry of Aschieri et al. (Class. Quantum Grav. 23, 1883–1911, 2006), whereby the commutative pointwise product is replaced by the $\star$-product determined by a Drinfel’d twist. We actually simultaneously construct noncommutative deformations of \textit{all} the algebraic submanifolds $M_c$ that are level sets of the $f^a(x)$, where $f^a(x) = 0$ are the polynomial equations solved by the points of $M$, employing twists based on the Lie algebra $\Xi_t$ of vector fields that are tangent to all the $M_c$. The twisted Cartan calculus is automatically equivariant under twisted $\Xi_t$ . If we endow $\mathbb{R}^n$ with a metric, then twisting and projecting to normal or tangent components commute, projecting the Levi-Civita connection to the twisted $M$ is consistent, and in particular a twisted Gauss theorem holds, provided the twist is based on Killing vector fields. Twisted algebraic quadrics can be characterized in terms of generators and $\star$-polynomial relations. We explicitly work out deformations based on abelian or Jordanian twists of all quadrics in $\mathbb{R}^3$ except ellipsoids, in particular twisted cylinders embedded in twisted Euclidean $\mathbb{R}^3$ and twisted hyperboloids embedded in twisted Minkowski $\mathbb{R}^3$ [the latter are twisted (anti-) de Sitter spaces $dS_2$, $AdS_2$].
}

\tableofcontents

\section{Introduction}

%We could mention e.g. 
%\cite{Aschieri2006,AschieriCastellani2009,BayFlaFroLicSte,Connes,Drinfeld1983,DVM1996,DubMadMasMou96,FioreWeber2020,Kon97,Woronowicz1989} and others...

The concept of a  submanifold $N$ of a manifold $M$ plays a fundamental role in mathematics and physics.
A metric, connection, ..., on $M$ uniquely induces a metric, connection, ..., on $N$.
%  (see e.g. \cite{Kobayashi1996}). 
Algebraic submanifolds of affine spaces such as 
$\RR^n$ or $\CC^n$ are paramount for their simplicity and their special properties.
In the last few decades  the program of generalizing differential geometry into  so-called Noncommutative Geometry (NCG)  has made a remarkable progress \cite{Connes,Lan97,Madore99,Majid2000,GraFigVar00};
 NCG might provide a suitable framework for a theory of quantum spacetime allowing the
quantization of gravity (see e.g. \cite{DopFreRob95,Aschieri2006}) or for unifying fundamental interactions (see e.g. \cite{ConLot91,ChaConvan}).
Surprisingly, the question whether, and to what extent, a notion of a submanifold is possible 
 in  NCG   has received little systematic attention  
(rather isolated exceptions are e.g. Ref.  \cite{Masson1995,
TWeber2019,Dan19,NguSch20}).
On several noncommutative (NC) spaces one can make sense of special classes of NC submanifolds,
but some aspects of the latter may depart  from their commutative counterparts. For instance,
from the $SO_q(n)$-equivariant noncommutative algebra ``of functions on the  quantum Euclidean
space $\RR^n_q$", which is generated by $n$ non-commuting coordinates $x^i$, one can obtain
the one $\A$ on the quantum Euclidean sphere $S^{n-1}_q$  by imposing that the [central and
$SO_q(n)$-invariant] ``square distance 
from  the origin'' $r^2=x^ix_i$ be 1. But the $SO_q(n)$-equivariant  differential calculus on $\A$ (i.e. the corresponding $\A$-bimodule $\Omega$ of 1-forms) remains of dimension  $n$ instead of  $n-1$;  the 1-form $dr^2$ cannot be set to zero, and actually the graded commutator \
$\left[\frac 1{q^2-1}r^{-2} dr^2,\,\cdot\,\right]$ \ acts as  the exterior derivative  \cite{Fio06JPCS,Ste96JMP,FioMad00,CerFioMad01}.

In \cite{FioreWeber2020} the above question is systematically addressed   
within the framework of deformation quantization \cite{BayFlaFroLicSte}, in the particular approach based on  Drinfel'd twisting \cite{Drinfeld1983} of Hopf algebras; a general procedure to  construct noncommutative generalizations of  smooth submanifolds  $M\subset \RR^n$, of the Cartan calculus, and of
(pseudo)Riemannian geometry on $M$ is proposed.
In the present work we proceed studying more in detail algebraic submanifolds $M\subset \RR^n$, 
in particular quadrics, using tools of algebraic geometry.
Considering $\CC^n$ %(or another affine space) 
instead of  $\RR^n$ seems viable, too.

\medskip
Assume that the  algebraic submanifold $ M\subset \RR^n$ consists of solutions $x$ of the %system of
 equations  
\be
f^a(x)=0,\qquad a=1,2,...,k<n,                  \label{DefIdeal}
\ee
where $f\equiv(f^1,...,f^k):\RR^n\mapsto \RR^k$ are polynomial
functions fulfilling the irreducibility conditions listed in Theorem~\ref{DecoCpol};
in particular, the Jacobian matrix $J=\partial f/\partial x$ is of rank $k$  on  some non-empty
open subset ${\cal D}_f\subset\RR^n$, and $M$ more precisely
consists  of the points of  ${\cal D}_f$  fulfilling (\ref{DefIdeal}). One easily shows that 
${\cal E}_f:=\RR^n\setminus{\cal D}_f$ is
empty or of zero measure\footnote{Let $J_\alpha$ be the 
$k\!\times\! k$ submatrices of $J$,  $j_\alpha $  their determinants, ${\cal E}_\alpha:=\{x\in\RR^n \: |\: j_\alpha(x)=0\}$,  $\alpha=1,2,...,\binom{n}{k}$. %The algebraic set 
\ ${\cal E}_f=\bigcap_\alpha {\cal E}_\alpha$. 
%If $ j_\alpha\equiv 0$, then ${\cal E}_\alpha=\RR^n$ can be omitted.
At least one 
polynomial function
$j_\alpha(x)$ is not identically zero; hence ${\cal E}_\alpha$ has codimension 1  and zero measure, and so has ${\cal E}_f$. 
}.
By replacing in (\ref{DefIdeal}) \ $f^a(x)\mapsto f^a_c(x):=f^a(x)\!-\!c^a$, 
\ with  $c \equiv(c^1,...,c^k)\in f\left({\cal D}_f\right)$, \ we
 obtain a $k$-parameter family of embedded manifolds $M_c$ ($M_0=M$) of dimension $n\!-\!k$  that are level sets of $f$. 
Embedded %(level set) 
algebraic submanifolds $N\subset M$ can be obtained by
adding more polynomial equations of the same type to (\ref{DefIdeal}).
Let  $\X$ be the $*$-algebra (over  $\CC$) of polynomial
 functions $P:\RR^n\to\CC$, restricted to ${\cal D}_f$.
The $*$-algebra $\XM$ of  complex-valued polynomial functions on $M$
can be expressed as the quotient of  $\X$ over the ideal $\C\subset\X$ of  polynomial functions 
vanishing on $M$:
\be
\XM:=\X/\C\equiv\big\{\, [\alpha]:=\alpha+\C \:\: |\:\: \alpha\in\X\big\};            \label{quotient}
%\ni a+\C=:.
\ee
In appendix \ref{RealNullstellensatz}, after recalling some basic notions and notation in algebraic geometry,  we prove 
%the following 

\begin{theorem} Assume  that $J$ is of rank $k$ on a non-empty
open subset \ ${\cal D}_f\subset\RR^n$, \ so that 
the system (\ref{DefIdeal}) defines an algebraic submanifold
 \ $M\subset {\cal D}_f$ \ of dimension $n-k$. \
In addition, assume
that \ $M$ is irreducible in $\CC^n$; \ this is the case e.g. if there exists a $k$-dimensional affine subspace 
\ $\pi \subset \RR^n$ \ meeting $M$ in \ $s:=\prod%\limits
_{a=1}^k \deg f^a$ \ points. 
Then \   %$\C=%\bigoplus_{a=1}^k \X f^a\X= 
             %\bigoplus\limits_{a=1}^k \X f^a$
$\C$ is the complexification of the ideal %$(f^1,...,f^k)$ 
%l'ho levato per non confonderlo con f della riga dopo (1)
generated by the $f^a$ in $\RR[x^1,...,x^n]$, 
\ i.e. for all $h\in\C$ there exist  $h^a\in\X$ such that
\be
h(x)=\sum_{a=1}^k h^a(x)f^a(x)=\sum_{a=1}^k f^a(x)h^a(x).        \label{DecoCeq}
\ee
\label{DecoCpol}
\end{theorem}
%In other words, the proposition states that under such assumptions the  $\X$-bimodule $\C$ is spanned by the left-hand sides (lhs) of (\ref{DefIdeal}). 
(In the smooth context, i.e. with $f^a,h,h^a\!\in\!  C^\infty({\cal D}_f)$,  (\ref{DecoCeq}) holds if $J$ is of rank $k$ on ${\cal D}_f$ \cite{FioreWeber2020}.) \
$\X^{\scriptscriptstyle N}$ is the quotient  of  $\XM$ over 
the ideal generated by further equations of type (\ref{DefIdeal}), 
or equivalently of $\X$  over the  ideal generated by all such equations. Identifying vector fields with derivations (first order differential operators), we denote as \
$\Xi:=\{X=X^i\partial_i \:\: |\:\:  X^i\in \X\}$ \ the Lie
algebra of polynomial  vector fields $X$ on ${\cal D}_f$ (here and below we abbreviate $\partial_i\equiv \partial /\partial x^i$) and
\begin{equation}
\begin{split}
    \Xi_\mathcal{C}&=\{X\in\Xi~|~X(f^a)\in\mathcal{C}
    \text{ for all }a\in\{1,\ldots,k\}\},\\
    \Xi_{\mathcal{C}\mathcal{C}}&=\{X\in\Xi~|~X(h)\in\mathcal{C}
    \text{ for all }h\in\X\}\subset\Xi_\C.
\end{split}
\end{equation}
The former is a Lie $*$-subalgebra of $\Xi$, while the latter is a Lie $*$-ideal; both
are $\X$-$*$-subbimodules.
By Theorem~\ref{DecoCpol} the latter decomposes as $\Xi_{\C\C}=\bigoplus_{a=1}^k f^a\Xi$.
We identify  the Lie algebra $\XiM$ of vector fields tangent to $M$  with
that of derivations of $\XM$, namely with % the quotient
\be
\XiM:=\Xi_\C/\Xi_{\C\C}\equiv\big\{\, [X]:=X+\Xi_{\C\C} \:\: |\:\: X\in\Xi_\C\big\}.                                     \label{quotient'}
\ee 

A general framework  for deforming $\X$ into a family - depending on a formal parameter  $\nu$ - of noncommutative  algebras $\X_\star$ over  $\CC[[\nu]]$
(the ring of formal power series in    $\nu$ with coefficients in $\CC$) is Deformation Quantization  \cite{BayFlaFroLicSte,Kon97}: as a module over  
$\CC[[\nu]]$
$\X_\star$  coincides with $\X[[\nu]]$, but the commutative pointwise product $\alpha \beta$ of $\alpha,\beta\in\X$ ($\CC[[\nu]]$-bilinearly extended to $\X[[\nu]]$)
%$m(a\otimes b)=ab$
is deformed into a possibly noncommutative (but still associative) product,
\be
\alpha\star \beta=\alpha \beta+\sum\nolimits_{l=1}^\infty \nu^l B_l(\alpha,\beta),
\ee
where $B_l$ are suitable bidifferential operators of degree $l$ at most.
%, in particular $B(1,\alpha)=0=B(\alpha,1)$. 
We wish to deform $\XM$ into a noncommutative algebra $\XM_\star$  in the form of  a quotient
\be
\XM_\star:=\X_\star/\C_\star\equiv\big\{\, [\alpha]:=\alpha+\C_\star \:\: |\:\: \alpha\in\X_\star\big\}, \label{quotientstar}
%\ni \alpha+\C=:.
\ee
with 
%the same $\star$-product  and 
$\C_\star$ a two-sided ideal of  $\X_\star$,
% such that $\C_\star= \C[[\nu]]$ holds at least as an equality of vector spaces.  %over $\CC[[\nu]]$.
and fulfilling itself  $\XM_\star=\XM[[\nu]]$ as an equality of 
$\CC[[\nu]]$-modules. To this end we  require that $\C_\star= \C[[\nu]]$, 
% actually holds as an algebra equality, 
%i.e. that $\C[[\nu]]$ be also a two-sided ideal of  $\X_\star$, 
i.e. that \ $c\star \alpha,\, \alpha\star c\in \C[[\nu]]$  for all $\alpha\in\X$, $c\in\C$, \  so that \ $(\alpha+c)\star(\alpha'+c')-\alpha\star \alpha'\in \C[[\nu]]$  for all $\alpha,\alpha'\in \X[[\nu]]$ and $c,c'\in\C[[\nu]]$. \ 
As a result, taking the quotient would commute  with deforming the product: \ $(\X/\C)_\star=\X_\star/\C_\star$. \
As argued in \cite{FioreWeber2020}, these conditions are fulfilled if\footnote{In fact, for
all $c\equiv \sum_{a=1}^kf^a c^a\in \C$ ($c^a\in\X$) (\ref{cond1}) implies 
$c=\sum_{a=1}^kf^a \star c^a$  and, for all $\alpha\in\X$, by the associativity of $\star$,
$ c\star \alpha=(\sum_{a=1}^kf^a \star c^a)\star \alpha=\sum_{a=1}^kf^a \star (c^a\star \alpha)
=\sum_{a=1}^kf^a (c^a\star \alpha)\in \C[[\nu]]$; and similarly for $\alpha\star c$. \\
It is not sufficient to require that \ $\alpha\star f^a\!-\!\alpha f^a$, $f^a\star \alpha\!-\!f^a\alpha$ 
 \ belong to $\C[[\nu]]$ to obtain the same results.}, for all $\alpha\in\X$, $a=1,..,k$,
\be
\alpha\star f^a=\alpha f^a=f^a\star \alpha\qquad\Leftrightarrow\qquad B_l(\alpha,f^a)=0=B_l(f^a,\alpha)\qquad \forall l\in\mathbb{N}
\label{cond1}
\ee 
(this implies that the $f^a$ are central in $\X_\star$,  again).
The quotient (\ref{quotientstar}) also appears in the context of deformation
quantization of Marsden-Weinstein reduction \cite{Bordemann,GuttWaldmann}.
A more algebraic approach to deformation quantization of reduced spaces
is given in the recent article \cite{Dippell}.

In  \cite{Drinfeld1983} Drinfel'd   introduced  a general deformation quantization procedure of 
universal enveloping algebras $U\g$ (seen as Hopf algebras) of Lie groups  $G$
and of their module algebras, based on {\it twisting}; a {\it twist} is a suitable
element (a 2-cocycle, see section \ref{TwistAlgStruc})
\be
\F=\1\ot\1+\sum_{l=1}^\infty \nu^l \sum_{I_l} \F^{I_l}_{1} \otimes\F^{I_l}_{2} \in (U\g\otimes U\g) [[\nu]]                                   \label{twist}
\ee
(here $\otimes=\otimes_{\CC[[\nu]]}$, and
tensor products are meant completed in the $\nu$-adic topology); 
$\F$ acts on the tensor product of any two $U\g$-modules or module algebras, 
in particular algebras of functions on any smooth manifolds $G$ acts on,
%$G$-manifold (i.e. smooth manifolds $G$ acts on)
including some symplectic manifolds\footnote{However this quantization procedure does 
not apply to every Poisson manifold:
there are several symplectic manifolds, e.g. the symplectic $2$-sphere and the
symplectic Riemann surfaces of genus $g>1$, which do not admit a $\star$-product induced by a Drinfel'd twist (c.f. \cite{Thomas2016,FrancescoThomas2017}). Nevertheless, if one
is not taking into account the Poisson structure, every $G$-manifold can be quantized
via the above approach.}  \cite{Aschieri2008}.
Given a generic smooth manifold $M$,   the authors of 
\cite{Aschieri2006} pick up  $\g\equiv\XiM$, the Lie algebra of 
smooth vector fields on $M$ (and of the infinite-dimensional Lie group of 
diffeomorphisms of $M$), and  the $U\XiM$-module algebra 
$\XM=C^\infty(M)$; $ \F_{1}^{I_l},\F_{2}^{I_l}$ seen as differential operators acting on $\XM$ have order  $l$ at most and no zero-order term. 
The corresponding deformed product reads
\be
\alpha\star \beta:=\alpha \beta+  \sum\nolimits_{l=1}^\infty \nu^l \sum\nolimits_{I_l}\bF_{1}^{I_l}   (\alpha) \:\:\bF_{2}^{I_l} (\beta)\,,
 \label{starprod}
\ee
where  \ $\bF\equiv\F^{-1}=\1\ot\1+\sum_{l=1}^\infty \nu^l \sum_{I_l} \bF_{1}^{I_l}\otimes\bF_{2}^{I_l}$ \ is the inverse of the twist. In the sequel we will use Sweedler notation with suppressed summation 
symbols and abbreviate \ $\F=\F_1 \otimes\F_2$, \ $\bF=\bF_1 \otimes\bF_2$; \
in the presence of several copies of $\F$ we distinguish the summations
by writing \ $\F_1\ot\F_2$, \ $\F'_1\ot\F'_2$, \ etc.
Actually Ref. \cite{Aschieri2006} twists not only $U\XiM, \XM$ 
into new Hopf algebra $U\XiM^\f$
and   $U\XiM^\f$-equivariant module algebra $\XM_\star$,  but also the
 $U\XiM$-equivariant $\XM$-bimodule of differential forms on $M$, their tensor powers,
the Lie derivative, and the geometry on $M$ (metric, connection, curvature, torsion,...) - if present -, into 
deformed counterparts.

Here and in \cite{FioreWeber2020}, as  in  \cite{Masson1995}, we  take the algebraic characterization (\ref{quotient}), (\ref{quotient'}) as the starting point  for defining submanifolds in NCG, but use a twist-deformed differential calculus on it.
%\footnote{The derivation-based approach to differential calculi of Dubois-Violette and Michor 
%\cite{DVM1996}, which was used in \cite{Masson1995}, does not encompass several differential calculi (e.g. quantum group covariant ones), or requires algebra extensions to succed (see e.g. \cite{CerFioMad01}). 
%The approaches to the differential calculus  \`a la  Connes \cite{Connes}
%and Woronowicz \cite{Woronowicz1989} 
%(which include  the one we are considering here) are  more general:
%the bimodule of noncommutative differential 1-forms is the primary object  from which the whole calculus  can be derived by imposing the Leibniz rule and nilpotency of the exterior derivative. As a result, the dual module consists of noncommutative vector fields which are no longer derivations.}.
Our twist is based on the Lie subalgebra (and  $\X$-bimodule) $\g\equiv\Xi_t\subset\Xi$ defined by
\be
\Xi_t:=\{X\in\Xi \:\:\: |\:\:\: X (f^1)= 0,\;...,\:X (f^k) =0  \}\subset\Xi_\C,          \label{defeqXis}
\ee 
which consists of vector fields  tangent to {\it all} submanifolds $M_c$ (because they fulfill $X(f^a_c)=0$ for all $c\in\RR^k$) at all points. 
%In the sequel we will often simplify the notation dropping the symbol of Lie derivative 
%and identifying a Lie derivatives $\L_X$ with the vector field $X$ itself. 
As  in \cite{FioreWeber2020}, we note that, 
applying this deformation procedure to the previously defined \ $\X$ \ with a 
twist \ $\F\in U\Xi_t\otimes U\Xi_t [[\nu]]$, \ we 
satisfy (\ref{cond1}) and therefore obtain   a deformation $\X_\star$ of $\X$ such that 
for all $c\in f({\cal D}_f)$ \
$\X^{\scriptscriptstyle M_c}_\star\!=\!\X^{\scriptscriptstyle M_c}[[\nu]]=\X_\star/\C_{\star}^c$; \ moreover, $\Xi_{\scriptscriptstyle M_c\star}\!=\!\Xi_{\scriptscriptstyle M_c}[[\nu]]=\Xi_{\C^c\star}/\Xi_{\C\C^c\star}$,  \ see section \ref{TwistSmoothSubman}. 
In other words, we obtain a noncommutative deformation, in the sense of deformation quantization
 and in the form of quotients as in (\ref{quotient}), (\ref{quotient'}), of the $k$-parameter family of embedded algebraic manifolds $M_c\subset\RR^n$. 
For every $X\in\Xi_\C$ there is an element in the equivalence
class $[X]$ that belongs to $\Xi_t$, namely its tangent projection $X_t$; hence we can work with the latter.
$\X_\star,\Xis,...$ are $U\Xi^\f$-equivariant,
while $\X^{\scriptscriptstyle M_c}_\star,\XiM{}_{\star},\Xi_{t\star},...$ are $U\Xi_t^\f$-equivariant.
If $\F$ is unitary or real, then $U\Xi^\f$ and $\X_\star,\Xis,...$  admit $*$-structures (involutions) making them a Hopf $*$-algebra and $U\Xi^\f$-equivariant (Lie) $*$-algebras
respectively; thereby $U\Xi_t^\f$ is a Hopf $*$-subalgebra and $\X^{\scriptscriptstyle M_c}_\star,\Xi_{t\star},...$ are $U\Xi_t^\f$-equivariant (Lie) $*$-subalgebras.

In passing, we recall that sometimes,
if a Poisson manifold $M$  is symmetric under  a solvable Lie group $G$ like $\RR^d$, the Heisenberg  or the $"ax+b"$ group, one can construct  even a 
{\it strict} (i.e. non-formal)  deformation quantization \cite{Rieffel}  of $C^\infty(M)$  such that the $\star$-product remains invariant under $G$ itself (or a cocommutative Hopf algebra), see e.g. \cite{Rieffel,BieBonMae2007}.

The plan of the paper will be as follows. 

Section \ref{Preli} reviews: Hopf algebras, their module algebras and twisting\cite{Chari1995,ES2010,Ka95,Majid2000,Mo93,Drinfeld1983,Aschieri2014,GiZh98}  (section \ref{TwistAlgStruc}); their application \cite{Aschieri2006,AschieriCastellani2009} to the differential geometry on a generic manifold (section \ref{TwistedNCGonM});  twisting of smooth submanifolds of $\RR^n$ as developed in \cite{FioreWeber2020} (section \ref{TwistSmoothSubman}).

In section \ref{TwistDiffGeomAlgSubman} we apply this procedure to algebraic submanifolds $M\subset \RR^n$. For simplicity we stick to  $M$  of codimension 1,
and we assume
 that there is a Lie subalgebra $\g$ (of dimension at least 2) of both $\Xi_t$ and 
 the Lie algebra ${\sf aff}(n)$ of the affine group ${\sf Aff}(\RR^n)=\RR^n\cocross GL(n)$ of $\RR^n$;  the level sets of  $f(x)$ 
of degree 1 (hyperplanes) or 2 (quadrics) are of this type.
Choosing a twist
$\F\in U\g\otimes U\g[[\nu]]$ we find that the algebra  $\X%=\mathrm{Pol}^\bullet(\RR^n)
$
of polynomial functions (with complex coefficients) in the set of Cartesian coordinates $x^1,...,x^n$ is deformed so that 
every $\star$-polynomial of degree $k$ in $x$ equals an ordinary polynomial of the same degree in $x$,
and vice versa. This implies in particular that the polynomial relations $x^ix^j-x^jx^i=0$ 
(whence the commutativity of $\X$), 
as well as the ones   (\ref{DefIdeal}) defining the ideal 
$\C$,  can be expressed as $\star$-polynomial relations {\it of the same degree}, so that 
$\X_\star$, $\XM_\star=\X_\star/\C_\star$ can be defined globally 
in terms of generators and polynomial relations,
 and moreover the subspaces $\widetilde{\X}^q$, $\widetilde{\XM}^q$ of 
$\X$, $\XM=\X/\C$ consisting of polynomials of any degree $q$
in $x^i$ coincide as $\CC[[\nu]]$-modules with their deformed counterparts
$\widetilde{\X}_{\star}^q$, $\widetilde{\XM}_{\star}^q$;
 in particular  their dimensions   (hence the Hilbert-Poincar\'e series  of both $\X$ and $\XM$) remain the same under deformation - an important (and often overlooked) property that guarantees the smoothness of the deformation. 
The same occurs with the $\X_\star$-bimodules  
and algebras $\Omega^\bullet_\star$ of differential forms, that of differential operators, etc. We convey all these informations into  what we name the {\it differential calculus algebras} $\Q^\bullet,\QM^\bullet$ on $\RR^n,M$ respectively (generated by the Cartesian coordinates, their
differentials, and a basis of vector fields, subject to appropriate relations; they are graded
by the form degree and filtered by both the degrees in the $x^i$ and in the  vector fields), and their deformations $\Q^\bullet_\star,\QMst^\bullet$ (see sections  %\ref{SectionDCA}, @Gaetano: we erased this section
\ref{Qstar}, \ref{QMstar}).
 
In section \ref{quadricsR^3} we discuss in detail deformations, induced by unitary 
twists%\footnote{Also real twists would be possible} 
of  abelian \cite{Reshetikhin1990} or Jordanian \cite{Ogievetsky1992}   type, of all families
of quadric surfaces embedded in $\RR^3$, except ellipsoids.  
The deformation of each element of every class is interesting by itself, as 
a novel example  of a NC manifold. 
Endowing $\RR^3$ with the Euclidean (resp. Minkowski) 
metric gives the circular cylinders (resp. hyperboloids and cone)
a  Lie algebra 
$\mathfrak{k}\subset\Xi_t$ 
%%
%of the group $G$ 
of isometries of dimension at least 2; choosing a twist $\F\in U\k\otimes U\k [[\nu]]$
we thus find twisted (pseudo)Riemannian $M_c$ (with the metric given by the
twisted first fundamental form) that are symmetric  under 
the Hopf algebra $U\mathfrak{k}^{\f}$ (the ``quantum group of isometries");
the twisted  Levi-Civita connection on $\RR^3$ (the exterior derivative)
projects to the twisted Levi-Civita connection on  $M_c$,
while the twisted curvature can be expressed 
in terms of the twisted second fundamental form through a twisted Gauss theorem.
Actually, the metric, Levi-Civita connection, intrinsic and extrinsic curvatures of 
any circular
cylinder or hyperboloid, as elements in the appropriate tensor spaces, remain undeformed; the twist enters only their action on twisted tensor products of vector fields.
The twisted hyperboloids  can be seen as twisted (anti-)de Sitter spaces $dS_2,AdS_2$.

In  appendices  \ref{RealNullstellensatz},\ref{OtherProofs} 
we recall basic notions %and notation 
in algebraic geometry and prove most   theorems.

We recall that (anti-)de Sitter spaces, which can be represented as solutions of  \
$2f_c(x)\equiv (x^1)^2\!+\!...\!+\!(x^{n-1})^2\!-\!(x^n)^2\!-\!2c\!=\!0$ in Minkowski $\RR^n$,  are  maximally symmetric cosmological solutions to the Einstein equations of general relativity with a nonzero cosmological constant $\Lambda$  in spacetime dimension $n\!-\!1$, and play a prominent role in present cosmology and theoretical physics  (see e.g. \cite{Dod03,Mal98}).
Interpreting $x$ in Minkowski $\RR^n$ as relativistic $n$-momentum, rather than
position in spacetime, then the same equation  represents the dispersion relation
of a relativistic particle of square mass $2c$. In either case it would be interesting to study the physical consequences of twist deformations.
On the mathematical side, directions for further investigations include: 
submanifolds of $\CC^n$ (rather than $\RR^n$),  just 
dropping  $*$-structures and the related constraints on the twist; 
 twist deformations of the (zero-measure) algebraic set ${\cal E}_f$.

\smallskip
Finally, we mention that 
in \cite{FioPis18,FioPis19,Pis20} an alternative approach to introduce
NC (more precisely, fuzzy)  submanifolds $S\subset\RR^n$
has been proposed and applied to  spheres, projecting the algebra of observables of a quantum particle
in $\RR^n$, subject to a confining potential with a very sharp
minimum on $S$, to the Hilbert subspace with energy below a certain cutoff.

\smallskip
Everywhere we consider vector spaces $V$ over the field $\KK\in\{\RR,\CC\}$;
we denote by $V[[\nu]]$ the $\KK[[\nu]]$-module of formal power series in $\nu$ with coefficients in $\KK$. We shall denote by the same symbol a $\KK$-linear map
$\phi\colon V\to W$ and its $\KK[[\nu]]$-linear
extension $\phi\colon V[[\nu]]\to W[[\nu]]$.

%If two twists lead to the same triangular structure $\R$, then they lead to the same commutation relations among $x^i, dx^i,\partial_i,...$, but in general to different deformed defining relations $\hat f^a(x\star)=0$.

\section{Preliminaries}
\label{Preli}

\subsection{Hopf algebras and their representations}
\label{TwistAlgStruc}

{\bf Hopf algebras.} \ \
We recall that a  
\textit{Hopf algebra} $(H,\mu,\eta,\Delta,\epsilon,S)$ over $\KK$  
 is an associative unital algebra $(H,\mu,\eta)$ over $\KK$ 
[$\mu\colon H\ot H\to H$ is the product: $\mu(a\ot b)\equiv a\cdot b$ for   $a,b\in H$, 
$\eta\colon\KK\to H$ with $\eta(1)=:\1$ is the unit] endowed with a coproduct,
counit, antipode $\Delta,\epsilon,S$. While   $\Delta,\epsilon$ are algebra maps, 
$S$ is an anti-algebra map; they have to fulfill a number of properties (see e.g. \cite{Chari1995,Majid2000,ES2010}), namely
$(\Delta\otimes\mathrm{id})\circ\Delta=(\mathrm{id}\otimes\Delta)\circ\Delta=:\Delta^{(2)}$
(coassociativity),
$(\epsilon\otimes\mathrm{id})\circ\Delta=\mathrm{id}
=(\mathrm{id}\otimes\epsilon)\circ\Delta$
(counitality), 
$\mu\circ(S\otimes\mathrm{id})\circ\Delta
=\eta\circ\epsilon
=\mu\circ(\mathrm{id}\otimes S)\circ\Delta$
(antipode property). We shall use Sweedler's notation with suppressed summation symbols
for the coproduct $\Delta$ and its $(n\!-\!1)$-fold iteration
\bea
\Delta^{(n)}: H \to (H)^{\ot n},\qquad\qquad \Delta^{(n)}(a)=
%\sum_I  g^I_{(1)} \otimes g^I_{(2)} \otimes ...\otimes g^I_{(n)}
a_{(1)} \otimes a_{(2)} \otimes ...\otimes a_{(n)}.
\eea
A $*$-involution on a $\KK$-algebra $\A$ is an involutive, anti-algebra map $*\colon \A\to \A$ 
such that $(\lambda a+\rho b)^*=\overline{\lambda}a^*+\overline{\rho}b^*$
for all $a,b\in \A$ and $\lambda,\rho\in\KK$ (here $\overline{\lambda}$ denotes the complex conjugation of $\lambda$). A  \textit{Hopf $*$-algebra} $(H,\mu,\eta,\Delta,\epsilon,S,*)$ over $\KK$  is a Hopf algebra endowed with a $*$-involution such that, 
for all $a,b\in H$,
\begin{equation}\label{eq12}
%    (a\cdot b)^*=b^*\cdot a^*,\quad
    \1^*=\1,\quad
    \Delta(a)^{*\ot *}=\Delta(a^*),\quad
    \epsilon(a^*)
    =\overline{\epsilon(a)}\quad
    \text{ and }~
    S[S(a^*)^*]    =a.
\end{equation}
The universal enveloping algebra (UEA) $U\g$
of a $\KK$-Lie algebra $(\g,[\cdot,\cdot])$ is a Hopf algebra; $\Delta,\epsilon,S$ are determined by their actions on $\1$ and on {\it primitive} elements, i.e. $g\in\g$:
\begin{equation}
    \Delta(g)
    =g\ot 1+1\ot g,\quad
    \epsilon(g)
    =0\quad
    \text{ and }\:\:
    S(g)
    =-g.
\end{equation}
It is cocommutative, i.e. $\tau\circ\Delta=\Delta$, where $\tau$ is the flip,
$\tau(a\otimes b)=b\otimes a$.
If there is a $*$-involution $*\colon\g\to\g$ on $\g$ such that $[g,h]^*=[h^*,g^*]$ for all
$g,h\in\g$, the UEA $U\g$ becomes a Hopf $*$-algebra with respect to the extension
$*\colon U\g\to U\g$.

Replacing everywhere in the above definition $\KK$ by the commutative ring $\KK[[\nu]]$
one obtains the definition of a Hopf ($*$-)algebra over $\KK[[\nu]]$.
For any Hopf  ($*$-)algebra over $\KK$ the $\KK[[\nu]]$-linear extension
 (with completed tensor product
in the $\nu$-adic topology) is trivially a Hopf ($*$-)algebra over $\KK[[\nu]]$.
Other ones can be obtained by twisting (see below).

\medskip
{\bf Hopf algebra modules and module algebras.} \ \
Given an associative unital algebra $\A$ over $\KK$,
a $\KK$-vector space $\M$ is said to be a \textit{left $\A$-module} if it is endowed with a
$\KK$-linear map $\trc\colon\A\ot\M\to\M$
such that $a\trc(b\trc s)=(a\cdot b)\trc s$ and $\1\trc s=s$ for all $a,b\in\A$ and $s\in\M$.
Similarly right $\A$-modules are defined. An $\A$-bimodule is a left and a right $\A$-module
with commuting module actions. A $\KK$-linear map $\phi\colon\M\to\M'$ between left $\A$-modules
is said to be \textit{$\A$-equivariant} if $\phi$ intertwines the $\A$-module actions, i.e. if
$\phi(a\trc s)=a\trc\phi(s)$ for all $a\in\A$ and $s\in\M$.
For a Hopf $*$-algebra $H$, a left $H$-module $\M$ is said to be a \textit{left $H$-$*$-module}
if there is a $*$-involution $*\colon\M\to\M$ on $\M$ such that
\begin{equation}
    (a\trc s)^*
    =S(a)^*\trc s^* \:
    \text{ for all }a\in H \:
    \text{ and } \: s\in\M.
\end{equation}
Similarly, right $\A$-$*$-modules and $\A$-$*$-bimodules are defined. An element
$s\in\M$ of a left $H$-module is said to be \textit{$H$-invariant} if $a\trc s=\epsilon(a)s$
for all $a\in H$.
An associative unital ($*$-)algebra $\A$ is said to be a \textit{left $H$-($*$-)module algebra}
if $\A$ is a left $H$-($*$-)module such that
\begin{equation}
    \xi\trc(a\cdot b)
    =(\xi_{(1)}\trc a)\cdot(\xi_{(2)}\trc b)\quad
    \text{ and }\quad
    \xi\rhd 1
    =\epsilon(\xi)1                \label{Leibniz}
\end{equation}
for all $\xi\in H$ and $a,b\in\A$. More generally, an $\A$-($*$-)bimodule $\M$ for a left
$H$-module ($*$-)algebra $\A$ is said to be an \textit{$H$-equivariant $\A$-($*$-)bimodule} if $\M$ is a left $H$-module such that
\begin{equation}
    \xi\trc(a\cdot s\cdot b)
    =(\xi_{(1)}\trc a)\cdot(\xi_{(2)}\trc s)\cdot(\xi_{(3)}\trc b)
   \quad \text{ [and } \:
    (a\cdot s\cdot b)^*
    =b^*\cdot s^*\cdot a^*]
\end{equation}
hold for all $\xi\in H$, $a,b\in\A$ and $s\in\M$, 
where we denoted the $\A$-($*$-)module actions by $\cdot$.

Similarly one defines module ($*$-)algebras and (equivariant) (bi-)($*$-)modules over
$\KK[[\nu]]$, and trivially obtains istances of them from their  $\KK$-counterparts by  $\KK[[\nu]]$-linear extension.

\medskip
{\bf Drinfel'd twist deformation.} \ \ 
Fix a Hopf algebra $H$ over $\KK$. A \textit{(Drinfel'd) twist} on $H$ 
is an element $\F=\1\ot\1+\mathcal{O}(\nu)\in(H\ot H)[[\nu]]$ 
of the form (\ref{twist})
satisfying the $2$-cocycle property 
\begin{equation} \label{cocycle}
    (\F\ot\1)(\Delta\ot\mathrm{id})(\F)
    =(\1\ot\F)(\mathrm{id}\ot\Delta)(\F)
\end{equation}
and the normalization property  \ $(\epsilon\ot\mathrm{id})(\F)
    =\1    =(\mathrm{id}\ot\epsilon)(\F)$. \
Every twist is invertible as a formal power series. We denote the inverse twist
by $\bF$ and suppress summation symbols, employing the \textit{leg notation}: $\F=\F_1\ot\F_2$, $\bF=\bF_1\ot\bF_2$,
and $\F_1\ot\F_2\ot\F_3$ for the expression at both sides of (\ref{cocycle}).
In the presence of several copies of $\F$ we write $\F=\F'_1\ot\F'_2$ for the second copy
etc. to distinguish the summations.
To every twist we assign an element $\beta:=\F_1\cdot S(\F_2)\in H[[\nu]]$. It is invertible
with inverse given by $\beta^{-1}=S(\bF_1)\cdot\bF_2\in H[[\nu]]$.

Let $\F$ be a Drinfel'd twist on $H$. Then
$H^\f=(H[[\nu]],\mu,\eta,\Delta_\f,\epsilon,S_\f)$ is a Hopf algebra over 
$\KK[[\nu]]$, where the twisted coproduct and antipode are defined by
\begin{equation} \label{inter-2}
    \Delta_\f(\xi)
    =\F\Delta(\xi)\bF\quad 
    \text{ and }\quad
    S_\f(\xi)
    =\beta S(\xi)\beta^{-1}
\end{equation}
for all $\xi\in H$\footnote{Here one could replace $\beta^{-1}$ by
$S(\beta)$, as $S(\beta)\beta\in\mbox{Centre}(H)[[\lambda]]$.}. 
Again, we shall use Sweedler's notation with suppressed summation symbols
for the coproduct $\Delta_\f$ and its $(n\!-\!1)$-fold iteration
\bea
\Delta_\f^{(n)}: H \to (H)^{\ot n},\qquad\qquad \Delta_\f^{(n)}(a)=
%\sum_I  g^I_{(\hat 1)} \otimes g^I_{(\hat 2)} \otimes ...\otimes g^I_{(\hat n)}
a_{\widehat{(1)}} \otimes a_{\widehat{(2)}} \otimes ...\otimes a_{\widehat{(n)}}.
\eea
If $\A$ is a left $H$-module algebra then $\A_\star
=(\A[[\nu]],\star,1)$ is a left $H^\f$-module algebra with respect to the
product 
%$a\star b     =(\bF_1\trc a)\cdot(\bF_2\trc b)$ \  for   $a,b\in\A[[\nu]]$;
(\ref{starprod}), [now abbreviated as
\ $a\star b     =(\bF_1\trc a)\cdot(\bF_2\trc b)]$ \  for   $a,b\in\A[[\nu]]$;
this implies the twisted Leibniz rule
\be
\ba{l} g\trc (a \star b)=
%\sum_I\! \left[g^I_{(\hat1)}\!\!\trc  a\right] \!\star\! \left[g^I _{(\hat 2)}\!\!\trca'\right]
\big(g_{\widehat{(1)}} \trc  a\big)  \star \big(g_{\widehat{(2)}} \trc b\big),\text{ for all }g\in H^\f.
\ea \label{TwistedLeibniz}
\ee
More generally, if $\A$ is a left $H$-module algebra and $\M$ an $H$-equivariant
$\A$-bimodule, then $M_\star=\M[[\nu]]$ becomes  (cf. \cite{Aschieri2014}~Theorem~3.5) an $H^\f$-equivariant $\A_\star$-bimodule,
with respect to the undeformed Hopf algebra action and the twisted module actions
\begin{equation} \label{starprod'}
    a\star s
    =(\bF_1\trc a)\cdot(\bF_2\trc s)\quad 
    \text{ and }\quad 
    s\star a
    =(\bF_1\trc s)\cdot(\bF_2\trc a)\:
    \text{ for all }\: a\in\A\text{ and }s\in\M
\end{equation}
on $\M_\star$.
 If $H$ is cocommutative then in general
$H^\f$ is not, but it is {\it  quasi-cocommutative}, i.e.
\begin{equation}
    \xi_{\widehat{(2)}}\ot\xi_{\widehat{(1)}}
    =\R\cdot\Delta_\f(\xi)\cdot\bR\quad 
    \text{ for all }\:
    \xi\in H^\f,
\end{equation}
where $\R:=\F_{21}\bF\in(H\ot H)[[\nu]]$ is the \textit{triangular structure}
or \textit{universal $\R$-matrix}. 
$\R$  has inverse 
$\bR=\F\bF_{21}=\R_{21}\in(H\ot H)[[\nu]]$ and
further satisfies the so-called \textit{hexagon relations}
\begin{equation}\label{hexagon}
    (\Delta_\f\ot\mathrm{id})(\R)
    =\R_{13}\R_{23}\quad 
    \text{ and }\quad 
    (\mathrm{id}\ot\Delta_\f)(\R)
    =\R_{13}\R_{12}.
\end{equation}
As the representation theory of a Hopf algebra $H$ is monoidal, the $\KK$-tensor product
$\M\ot\M'$ of two left $H$-modules is also a left $H$-module, via the $H$ action
$\xi\trc(s\ot s')=\xi_{(1)}\trc s\ot\xi_{(2)}\trc s'$. The $\star$-tensor product
\begin{equation}\label{startensor}
    s\ot_\star s'
    :=\bF_1\trc s\ot\bF_2\trc s',\quad
    s\in\M,~s'\in\M'
\end{equation}
is the corresponding monoidal structure on the representation theory of $H^\f$, since
\begin{equation}
    \xi\trc(s\ot_\star s')
    =\xi_{\widehat{(1)}}\trc s\ot_\star\xi_{\widehat{(2)}}\trc s'
\end{equation}
for all $\xi\in H^\f$, i.e. $(\M\ot\M')_\star=\M_\star\ot_\star\M'_\star$.
Consider \cite{Aschieri2014} for more information.

The algebra $(H[[\nu]],\star)$ itself is a  $H^\f$-module algebra, and one can build a triangular Hopf algebra $H_\star=(H[[\nu]],\star,\eta,\Delta_\star,\epsilon,S_\star,\R_\star)$  isomorphic  to $H^\f=(H[[\nu]],\mu,\eta,\Delta_\f,\epsilon,S_\f,\R)$, with isomorphism $D:H_\star\to H^\f$ given by 
$D(\xi):=(\bF_1\trc \xi)\bF_2=\F_1\,\xi\, S\!\left(\F_2\right)\,\beta^{-1}$
and inverse by $D^{-1}(\phi)=\bF_1 \,\phi\,\beta\, S\!\left(\bF_2\right)$
\cite{GurMajid1994,Aschieri2006} (cf. also \cite{Fio98JMP,Fiore2010}). In other words, \
$D(\xi\star \xi')=D(\xi)D(\xi')$, \ and \
$\Delta_\star,S_\star,\R_\star$ \ are related to \ $\Delta_\f,S_\f,\R$ \ by the relations
\be \label{HF-HstarREL}
\Delta_\star=(D^{-1}\otimes D^{-1})\circ  \Delta_\f \circ D,
\qquad S_\star=  D^{-1}\circ S_\f \circ D,
\qquad \R_\star= (D^{-1}\otimes D^{-1})(\R).
\ee
One can think of $D$ also as a
change of generators within $H[[\nu]]$.

If $H$ is a Hopf $*$-algebra, and the twist is either \ \textit{real} \ [namely, if $\F^{*\ot *}=(S\ot S)(\F_{21})$] \ or \textit{unitary} \ (namely, if
$\F^{*\ot *}=\bF$), \ then one can make both $H^\f$  and  $H_\star$ into Hopf $*$-algebras in such a way that  twisting transforms the $H$ $*$-modules and module 
$*$-algebras into $H^\f$  and  $H_\star$ $*$-modules and module 
$*$-algebras, respectively.  
In fact, if $\F$ is real  then also $\beta^*=\beta$, while  
$\R^{*\ot *}=(\beta\otimes\beta)^{-1}\bR(\beta\otimes\beta)=(\beta\otimes\beta)\bR(\beta\otimes\beta)^{-1}$, and
 $H^\f$ endowed with the $*$-involution
\begin{equation}\label{star"}
    \xi^{*_\f}:=\beta\xi^*\beta^{-1},\quad 
    \text{ for }\:\xi\in H^\f,
\end{equation}
is a triangular Hopf $*$-algebra (in fact, $\R^{*_\f\ot *_\f}=\bR$); moreover, $\A_\star$, $\M_\star$ are a left $H^\f$-module $*$-algebra and  a $H^\f$-equivariant $\A_\star$-$*$-bimodule when
endowed with the undeformed $*$-involutions  (cf. \cite{Majid2000}~Proposition~2.3.7). In particular
$(H[[\nu]],\star,*)$ is a left $H^\f$-module $*$-algebra.
Actually $D$ is an isomorphism of the triangular
Hopf $*$-algebra $(H_\star,*)$ onto the one $(H^\f,*_\f)$, see
\cite{Aschieri2006,Majid2000} for more information.
If $\F$ is a unitary twist,  
then also $\R$ is, $\beta^*=S\!\left(\beta^{-1}\right)$,
and $H^\f$ endowed with the undeformed $*$-involution is a Hopf $*$-algebra; moreover, 
$\A_\star, \M_\star$ are respectively a left $H^\f$-module
$*$-algebra and  an $H^\f$-equivariant $\A_\star$-$*$-bimodule when
endowed with the twisted $*$-involutions 
\begin{equation}\label{eq03}
    a^{*_\star}
    =S(\beta)\trc a^*,~~~~~~~~
    s^{*_\star}
    =S(\beta)\trc s^*,
\end{equation}
where $a\in\A[[\nu]]$ and $s\in\M[[\nu]]$ (cf. \cite{Fiore2010}). In particular
$(H[[\nu]],\star,*_\star)$ is a left $H^\f$-module $*$-algebra. Actually,
one finds that $(H_\star,*_\star)$ is a triangular Hopf $*$-algebra,
in particular $\Delta_\star\circ *_\star=(*_\star\otimes *_\star)\circ\Delta_\star $, 
$S_\star\circ *_\star\circ S_\star\circ *_\star=\id$,
and   $D:(H_\star,*_\star)\to(H^\f,*)$ is an isomorphism of  triangular
Hopf $*$-algebras, see Proposition 18 in \cite{FioreWeber2020}.

%\begin{prop}
%If $\F$ is unitary, then $D$ is an isomorphism of the triangular
%Hopf $*$-algebra $(H_\star,*_\star)$ onto the one $(H^\f,*)$.
%In particular, $\Delta_\star\circ *_\star=(*_\star\otimes *_\star)\circ\Delta_\star $, 
%$S_\star\circ *_\star\circ S_\star\circ *_\star=\id$.
% \label{IsomorHopf} \end{prop}

For their simplicity, here we shall only use \textit{abelian} or the following \textit{Jordanian} Drinfel'd twists on UEAs:
\begin{enumerate}
\item[i.)]
For a finite number $n\in\mathbb{N}$ of pairwise commuting elements $e_1,\ldots,e_n,f_1,\ldots,f_n\in\mathfrak{g}$ we set \
$P:=\sum_{i=1}^ne_i\otimes f_i\in\mathfrak{g}\otimes\mathfrak{g}$, \
$P'=\frac 12\sum_{i=1}^n(e_i\otimes f_i-f_i\otimes e_i)$. \
Then
\be
\mathcal{F}=\exp(i\nu P)\in( U\g \otimes U\g )[[\nu]]  \label{abeliantwist}
\ee
is a Drinfel'd twist on $ U\g $ (\cite{Reshetikhin1990}); it is said of
{\it abelian} (or {\it Reshetikhin}) type. 
It is unitary if $P^{^*\otimes^*}=P$; 
this is e.g. the case if the $e_i,f_i$ are anti-Hermitian or Hermitian.
The twist $\F'=\exp(i\nu P')$  is both unitary and real, 
leads to the same $\R$ and makes $\beta=\1$,
whence $S_\f=S$, and the $*$-structure remains undeformed also for 
$H$-$*$-modules and module algebras, see (\ref{eq03}).

\item[ii.)]
Let $H,E\in\mathfrak{g}$ be elements of a Lie algebra such that
$
[H,E]=2E.
$
Then 
\be
\mathcal{F}
=\exp\left[\frac{1}{2}H\otimes\log(\1+i\nu E)\right]
\in( U\g \otimes U\g )[[\nu]]  \label{Jordaniantwist}
\ee
defines a {\it Jordanian} Drinfel'd twist \cite{Ogievetsky1992}.
If $H$ and $E$ are anti-Hermitian, $\mathcal{F}$ is unitary.
%More sophisticated twists can be obtained using this 
%as a prototype \cite{Borowiec2005,Pachol2017,BorMelMelPac18}.
\end{enumerate}

\subsection{Twisted differential geometry}
\label{TwistedNCGonM}

Here we recall some results obtained in \cite{Aschieri2006,AschieriCastellani2009}.
We apply the notions overviewed in the previous section choosing as Hopf $*$-algebra $H=U\Xi$, where $\Xi:=\Gamma^\infty(TM)$ denotes the
Lie $*$-algebra of smooth vector fields on a smooth manifold $M$, as a 
left $H$-module $*$-algebra the $*$-algebra $\X=\mathcal{C}^\infty(M)$ of smooth $\KK$-valued functions on $M$, as $H$-equivariant symmetric $\X$-$*$-bimodules  $\Xi$ itself, the space
$\Omega=\Gamma^\infty(T^*M)$  of differential $1$-forms on $M$,
as well as their tensor (or wedge) powers.
The Hopf $*$-algebra action on $\X$, $\Xi$ and $\Omega$ is given by the
extension of the Lie derivative: for $X,Y\in\Xi$, $f\in\X$ and $\omega\in\Omega$ we have 
\begin{equation}
    \mathcal{L}_Xf
    =:X(f),~~~~~
    \mathcal{L}_XY
    =[X,Y],~~~~~
    \mathcal{L}_X\omega
    =(\mathrm{i}_X\mathrm{d}+\mathrm{d}\mathrm{i}_X)\omega
\end{equation}
and we set $\mathcal{L}_{XY}=\mathcal{L}_X\mathcal{L}_Y$, $\mathcal{L}_\1=\mathrm{id}$.
Henceforth we denote such an extension by $\trc$.

\subsubsection{Twisted tensor fields}

The \textit{tensor algebra} $\TT:=\bigoplus_{p,r\in\NN_0}\TT^{p,r}$ on $M$ is defined as 
the direct sum of the $\KK$-modules
\begin{equation}
    \TT^{p,r}
    :=\underbrace{\Omega\ot\ldots\ot\Omega}_{p\text{-times}}\ot
    \underbrace{\Xi\ot\ldots\ot\Xi}_{r\text{-times}}
\end{equation}
for $p,r\geq 0$, $p+r>0$, where we set $\TT^{0,0}:=\X$. 
Here and below $\otimes$  stands for $\otimes_{\X}$ (rather than $\otimes_{\KK}$), 
namely $T\otimes f T'= T f\otimes T'$ for all $f\in\X$. 
Every $\TT^{p,r}$ is an
$H$-equivariant $\X$-$*$-bimodule with respect to the module actions
\begin{equation*}
    \xi\trc(\omega_1\ot\ldots\ot\omega_p\ot X_1\ot\ldots\ot X_r)
    =[\xi_{(1)}\trc\omega_1]\ot\ldots\ot[\xi_{(p)}\trc\omega_p]\ot
    [\xi_{(p+1)}\trc X_1]\ot\ldots\ot[\xi_{(p+r)}\trc X_r],
\end{equation*}
\begin{equation*}
    h\cdot(\omega_1\ot\ldots\ot\omega_p\ot X_1\ot\ldots\ot X_r)\cdot k
    =(h\cdot\omega_1)\ot\ldots\ot\omega_p\ot X_1\ot\ldots\ot(X_r\cdot k)
\end{equation*}
for all $\xi\in H$ and $h,k\in\X$.
This induces the structure of an $H$-equivariant $\X$-$*$-bimodule on $\TT$. 
In particular, for all $T,T'\in\TT$, $\xi\in H$ and $h,k\in\X$ the relations
\begin{equation}
\begin{split}
    \xi\trc(T\ot T')
    =&\xi_{(1)}\trc T\ot\xi_{(2)}\trc T',\\
    h\cdot(T\ot T')\cdot k
    =&(h\cdot T)\ot(T'\cdot k),\\
    (T\cdot h)\ot T'
    =&T\ot(h\cdot T')
\end{split}
\end{equation}
hold. Let $T\in\TT^{p,r}$. On a local chart $(U,x)$ of $M$ there are unique functions
$T^{\lambda_1,\ldots,\lambda_r}_{\mu_1,\ldots,\mu_p}\in\mathcal{C}^\infty(U)$ such that
$T=T^{\lambda_1,\ldots,\lambda_r}_{\mu_1,\ldots,\mu_p}
\mathrm{d}x^{\mu_1}\ot\ldots\ot\mathrm{d}x^{\mu_p}
\ot\partial_{\lambda_1}\ot\ldots\ot\partial_{\lambda_r}$, where $\{\partial_i\}$ is the
dual frame of vector fields on $U$ corresponding to $\{x^i\}$, i.e.
$\langle\partial_i,\mathrm{d}x^j\rangle=\delta_i^j$ and we sum over repeated indices.

Consider a (in particular, unitary or real) Drinfel'd twist $\F$ on $H$.
Applying the results of Section~\ref{TwistAlgStruc}
to $H$, $\X$, $\Xi$, $\Omega$ and $\TT$ we obtain the
following: $H^\f=U\Xi^\f$ is a Hopf ($*$-)algebra, $\X_\star$ is a left
$H^\f$-module ($*$-)algebra, while $\Xi_\star,\Omega_\star,\TT_\star$ are $H^\f$-equivariant
$\X_\star$-($*$-)bimodules. The $H^\f$-actions are given by the \textit{$\star$-Lie derivative}
$\mathcal{L}^\star_\xi T:=\mathcal{L}_{\overline{\f}_1\trc\xi}(\bF_2\trc T)$
for all $\xi\in H^\f$ and $T\in\TT_\star$.
On $\star$-vector fields $X,Y\in\Xi_\star$, the $\star$-Lie derivative
\begin{equation}
    \mathcal{L}^\star_XY
    =[\bF_1\trc X,\bF_2\trc Y]
    =X\star Y-(\bR_1\trc Y)\star(\bR_2\trc X)
    =:[X,Y]_\star
\end{equation}
structures $\Xi_\star$ as a \textit{$\star$-Lie algebra}. This means that $[\cdot,\cdot]_\star$
is twisted skew-symmetric, i.e. $[Y,X]_\star=-[\bR_1\trc X,\bR_2\trc Y]_\star$ and satisfies the twisted Jacobi identity
$[X,[Y,Z]_\star]_\star=[[X,Y]_\star,Z]_\star +[\bR_1\trc Y,[\bR_2\trc X,Z]_\star]_\star
$
for all $X,Y,Z\in\Xi_\star$. Furthermore, $[\cdot,\cdot]_\star$ is $H^\f$-equivariant, i.e.
$\xi\trc[X,Y]_\star=[\xi_{\widehat{(1)}}\trc X,\xi_{\widehat{(2)}}\trc Y]_\star$ and
$\star$-vector fields act on $\X_\star$ as \textit{twisted derivations}, i.e.
\begin{equation}
    \mathcal{L}^\star_X(f\star f')
    =\mathcal{L}^\star_X(f)\star f'
    +(\bR_1\trc f)\star\mathcal{L}^\star_{\overline{\r}_2\trc X}f'
\end{equation}
for all $X\in\Xi_\star$ and $f,f'\in\X_\star$.
By setting $\A=\TT$  we can apply the results of section \ref{TwistAlgStruc}, 
in particular define a
deformed tensor algebra $\TT_\star$ with associative $\star$-tensor product defined by eq. (\ref{startensor}). This can be decomposed as
$\TT_\star=\bigoplus_{p,r\in\NN_0}\TT^{p,r}_\star$, where $\TT^{0,0}_\star:=\X_\star$
and for  $p+r>0$
\begin{equation}
    \TT^{p,r}_\star
    :=\underbrace{\Omega_\star\ot_\star\ldots\ot_\star\Omega_\star}_{p\text{-times}}
    \ot_\star\underbrace{\Xi_\star\ot_\star\ldots\ot_\star\Xi_\star}_{r\text{-times}}.
\end{equation}
In particular, for all $T,T'\in\TT_\star$, $h,k\in\X_\star$ and $\xi\in H^\f$
\begin{equation}
\begin{split}
    \xi\trc(T\ot_\star T')
    =&\xi_{\widehat{(1)}}\trc T\ot_\star\xi_{\widehat{(2)}}\trc T',\\
    h\star(T\ot_\star T')\star k
    =&(h\star T)\ot_\star(T'\star k),\\
    (T\star h)\ot_\star T'
    =&T\ot_\star(h\star T').
\end{split}
\end{equation}
The third formula shows that $\otimes_\star$ is actually    $\otimes_{\X_\star}$, the tensor product over $\X_\star$. 
Let $T\in\TT^{p,r}_\star$.
On any local chart $(U,x)$ of $M$ there unique functions 
$T^{\lambda_1,\ldots,\lambda_r}_{\star\mu_1,\ldots,\mu_p}\in\mathcal{C}^\infty(U)[[\nu]]$
such that
\begin{equation}
    T
    =T^{\lambda_1,\ldots,\lambda_r}_{\star\mu_1,\ldots,\mu_p}\star
    \mathrm{d}x^{\mu_1}\ot_\star\ldots\ot_\star\mathrm{d}x^{\mu_p}\ot_\star
    \partial_{\lambda_1}\ot_\star\ldots\ot_\star\partial_{\lambda_r}.
\end{equation}
Higher order differential forms are defined by the twisted skew-symmetrization of 
$\ot_\star$
\begin{equation}
    \omega\wedge_\star\omega'
    :=(\bF_1\trc\omega)\wedge(\bF_2\trc\omega')
    =\omega\ot_\star\omega'
    -\bR_1\trc\omega'\ot_\star\bR_2\trc\omega
\end{equation}
({\it $\star$-wedge product}, an associative unital product),
and we define $\Omega^\bullet_\star:=(\Lambda^\bullet\Omega_\star,\wedge_\star)$
to be the twisted exterior algebra of $\Omega$ (see \cite{TWeber2019} for more information).

\medskip
The dual pairing $\la~,~\ra$ between vector fields and 1-forms can be equivalently considered
as $\X$-bilinear maps $\Xi\ot\Omega\to\X$ or $\Omega\ot\Xi\to\X$; for all arguments
$X\in\Xi$, $\omega\in\Omega$ these maps have the same images, which we respectively 
denote by the lhs and right-hand side (rhs) of the identity \ $\la X,\omega\ra=\la \omega,X\ra$. 
They have %two 
distinct twist deformations (\textit{$\star$-pairings}) defined by  
\bea\label{starpairing}
(T,T')~&\mapsto &\la T,T'\ra_\star
:=\left\la\bF_1\trc T,\bF_2\trc T'\right\ra~,
\eea
with \ $(T,T')=( X,\omega)$ and $(T,T')=(\omega,X)$ \ respectively.
They satisfy
\begin{equation}\label{equiv-linearity}
\begin{split}
    \langle T,T'\rangle_\star 
    =&\langle\bR_1\trc T',\bR_2\trc T\rangle_\star,\\[4pt]
     \xi\trc\langle T,T'\rangle_\star
    =&\langle\xi_{\widehat{(1)}}\trc X,\xi_{\widehat{(2)}}\trc\omega\rangle_\star,\\[4pt]
   \langle h_1\star T\star h_2,T'\star h_3\rangle_\star
    =&h_1\star\langle T,h_2\star T'\rangle_\star\star h_3
\end{split}
\end{equation}
for all $\xi\in H^\f$, $X\in\Xi_\star$, $\omega\in\Omega_\star$,
$(T,T')=( X,\omega)$ or $(T,T')=(\omega,X)$, 
and $h,h_1,h_2,h_3\in\X_\star$. Moreover, $\langle X,\mathrm{d}h\rangle_\star
    =\mathcal{L}^\star_Xh$. 
As one can extend the ordinary %commutative 
pairing to higher tensor powers setting
\be
\la T_p\otimes...\otimes T_1,T'_1\otimes...\otimes T'_p\otimes \tau\ra
:=\la T_p\la\ldots\la T_1,T'_1\ra,\ldots\ra,
T'_p\ra
\: \tau~ , \label{ExtPairing}
\ee
for all $\tau\in\TT^{p,r}$ (the image will belong again to $\TT^{p,r}$)
provided $(T_i,T'_i)\in\Xi\ot\Omega$ or  $(T_i,T'_i)\in\Omega\ot\Xi$ for all $i$, 
so can one extend \ $\la ~,~\ra_\star$ \ to the corresponding twisted tensor powers
using the same formula (\ref{starpairing}). 
Due to the `onion structure' of   (\ref{ExtPairing}) (i.e. the order of the $T_i$ and of the $T'_i$ are opposite of each other), properties (\ref{equiv-linearity}) are preserved,
namely the $\star$-paring is $H^\f$-equivariant, as well as left, right and middle $\X_\star$-linear (if we chose a different order in (\ref{ExtPairing})  the deformed definition would
need copies of $\R$ acting on the $T_i,T'_i$).

\subsubsection{Twisted covariant derivatives and metrics}

A \textit{twisted covariant derivative} (or \textit{connection}) is a $\KK[[\nu]]$-linear map
$\nabla^\f\colon\Xi_\star\ot_{\KK[[\nu]]}\TT_\star\to\TT_\star$ fulfilling,
for all $X,Y\in\Xi_\star$, $h\in\X_\star$, $T,T'\in\TT_\star$ and $\omega\in\Omega_\star$,
\begin{equation}
    \nabla^\f_Xh
    =\mathcal{L}^\star_Xh,
\end{equation}
\begin{equation}
    \nabla^\f_{h\star X}T
    =h\star(\nabla^\f_XT),
\end{equation}
\begin{equation}\label{eq04}
    \nabla^\f_X(T\ot_\star T')
    =[\bR_1\trc\nabla^\f_{\overline{\r}'_2\trc X}(\bR''_2\trc T)]
    \ot_\star[(\bR_2\bR'_1\bR''_1)\trc T']
    +(\bR_1\trc T)\ot_\star(\nabla^\f_{\overline{\r}_2\trc X}T),
\end{equation}
\begin{equation}\label{eq05}
    \nabla^\f_X\langle Y,\omega\rangle_\star
    =\langle\bR_1\trc[\nabla^\f_{\overline{\r}'_2\trc X}(\bR''_2\trc Y)],
    (\bR_2\bR'_1\bR''_2)\trc\omega\rangle_\star
    +\langle\bR_1\trc Y,\nabla^\f_{\overline{\r}_2\trc X}\omega\rangle_\star.
\end{equation}
Its \textit{curvature} $ \rR^\f_\star$ and \textit{torsion} $\tT^\f_\star$ maps respectively 
act on  all $X,Y,Z\in\Xi_\star$ through
\begin{equation}\label{TANDR}
\begin{split}
    \tT^\f_\star(X,Y)
    :=&\nabla^\f_XY-\nabla^\f_{\overline{\r}_1\trc Y}(\bR_2\trc X)-[X,Y]_\star,\\[6pt]
    \rR^\f_\star(X,Y,Z)
    :=&\nabla^\f_X\nabla^\f_YZ
    -\nabla^\f_{\overline{\r}_1\trc Y}\nabla^\f_{\overline{\r}_2\trc X}Z
    -\nabla^\f_{[X,Y]_\star}Z
\end{split}
\end{equation}
and are left  $\X_\star$-linear maps
$\tT^\f_\star\colon\Xi_\star\ot_\star\Xi_\star\to\Xi_\star$ and 
$\rR^\f_\star\colon\Xi_\star\ot_\star\Xi_\star\ot_\star\Xi_\star\to\Xi_\star$ fulfilling
\begin{equation}
%\begin{split}
    \tT^\f_\star(Y,X)    =-\tT^\f_\star(\bR_1\trc X,\bR_2\trc Y),\qquad
    \rR^\f_\star(Y,X,Z) =-\rR^\f_\star(\bR_1\trc X,\bR_2\trc Y,Z).
%\end{split}
\end{equation}
They are in one-to-one
correspondence with elements
$\tT^\f\in\Omega^2_\star\ot_\star\Xi_\star$, 
$\rR^\f\in\Omega_\star\ots\Omega^2_\star\ot_\star\Xi_\star$ such that
\be
 \tT^\f_\star(X,Y)=\la\, X \ots Y, \tT^{\f}\,\ra_\star,\qquad 
 \rR^\f_\star(X,Y,Z)=\la\, X \ots Y\ots Z, \rR^{\f}\,\ra_\star.
\ee
Setting $\F=1\ot 1$ it follows that $\R=1\ot 1$ and the definitions of twisted connection, torsion, curvature give the algebraic notion of connection, torsion,
curvature of differential geometry. 
Consider a (classical) connection $\nabla\colon\Xi\ot\TT\to\TT$ on $M$ and
its \textit{equivariance Lie algebra} $\mathfrak{e}\subseteq\Xi$
(cf. \cite{FioreWeber2020}).
The latter is a Lie subalgebra of the Lie algebra of vector fields defined by
\begin{equation}
    \mathfrak{e}
    =\{\xi\in\Xi~|~
    \xi\trc(\nabla_XT)
    =\nabla_{\xi\trc X}T
    +\nabla_{X}(\xi\trc T)
    \text{ for all }X\in\Xi,~T\in\TT\}.
\end{equation}
It follows that $\nabla$ is $U\mathfrak{e}$-equivariant, i.e.
$\xi\trc(\nabla_XT)=\nabla_{\xi_{(1)}\trc X}[\xi_{(2)}\trc T]$ for all
$\xi\in U\mathfrak{e}$, $X\in\Xi$ and $T\in\TT$. If $\F\in(U\mathfrak{e}\ot U\mathfrak{e})
[[\nu]]$ is a Drinfel'd twist, then
\begin{equation}  \label{twistedNabla}
    \nabla^\f_XT
    :=\nabla_{\overline{\f}_1\trc X}(\bF_2\trc T)
\end{equation}
defines an $U\mathfrak{e}^\f$-equivariant twisted connection
$\nabla^\f\colon\Xi_\star\ot_{\mathbb{K}[[\nu]]}\TT_\star\to\TT_\star$; 
then eqs.(\ref{eq04}-\ref{eq05}) reduce to 
\begin{equation}
\begin{split}
    \nabla^\f_X(T\ot_\star T')
    =&(\nabla^\f_XT)\ot_\star T'
    +(\bR_1\trc T)\ot_\star(\nabla^\f_{\overline{\r}_2\trc X}T'),\\[6pt]
    \nabla^\f_X\langle Y,\omega\rangle_\star
    =&\langle\nabla^\f_XY,\omega\rangle_\star
    +\langle\bR_1\trc Y,\nabla^\f_{\overline{\r}_2\trc X}\omega\rangle_\star
\end{split}
\end{equation}
for all $X,Y\in\Xi_\star$, $T,T'\in\TT_\star$ and $\omega\in\Omega_\star$
(cf. \cite{FioreWeber2020}~Proposition~2).

A \textit{metric} on $M$  is a non-degenerate element
$\gm=\gm^\alpha\ot\gm_\alpha\in(\Omega\ot\Omega)[[\nu]]$ such that
$\gm=\gm_\alpha\ot\gm^\alpha$. We can view $\gm$ as an element
$\gm=\gm^A\ot_\star\gm_A\in\Omega_\star\ot_\star\Omega_\star$ with
$\gm^A\ot\gm_A=\F_1\trc\gm^\alpha\ot\F_2\trc\gm_\alpha$.
A twisted connection $\nabla^\f$ such that $\tT^\f=0$ and $\nabla^\f\gm=0$ is said to be
a   {\it Levi-Civita} (LC) connection for $\gm$. The associated Ricci tensor map and Ricci scalar  of  $\nabla^\f$
are respectively defined by 
\be
\ric^{\f}_\star\colon\Xi_\star\ots\Xi_\star\to\Xi_\star, \quad
\ric^{\f}_\star(X,Y):=\la \theta^i, \rR^{\f}_\star(e_i,X,Y)\ra_\star, \qquad
%=\la \theta^i \ots e_i \ots X \ots Y,\rR^{\f}\ra_\star
\mathfrak{R}^{\f}:=\ric^{\f}\left(\gm^{-1A},\gm^{-1}{}_{A}\right)
\ee
(sum over $\alpha,A,i$), where
$\{ e_i\}$, $\{ \theta^i\}$ are $\star$-dual bases of $\Xi_\star,\Omega_\star$, in the sense
$\la  e_i,\theta^j\ra_\star=\delta^j_i$. 
%The tensor $e_i\ots\theta^i$ is $H^\f$-invariant.  % (?)
One easily finds \ $\ric^{\f}(X,Y)=\la \theta^i \ots e_i \ots X \ots Y,\rR^{\f}\ra_\star$. \ 

For a (pseudo-)Riemannian manifold $(M,\gm)$ we define the Lie subalgebra
\begin{equation} \label{Killing}
    \k:=\{\xi\in\Xi~|~
    \xi\trc\gm(X,Y)
    =\gm(\xi\trc X,Y)
    +\gm(X,\xi\trc Y)
    \text{ for all }X,Y\in\Xi\}\subseteq\Xi
\end{equation}
of \textit{Killing vector fields}. If $\nabla\colon\Xi\ot\TT\to\TT$ is the
\textit{Levi-Civita} (LC) covariant derivative on $(M,\gm)$ 
[i.e. $\tT=0$ and 
$\mathcal{L}_X\gm(Y,Z)=\gm(\nabla_XY,Z)+\gm(Y,\nabla_XZ)$
for all $X,Y,Z\in\Xi$] and $\mathfrak{e}$ the corresponding equivariance Lie algebra,
we obtain $\k\subseteq\mathfrak{e}$ by the Koszul formula. 

 The following results are taken from \cite{AschieriCastellani2009,FioreWeber2020}. 
If $\F\in(U\k\ot U\k)[[\nu]]$ is a twist ``based on Killing vector fields", then (\ref{twistedNabla}) defines a twisted LC connection $\nabla^\f\colon\Xi_\star\ot_{\mathbb{K}[[\nu]]}\TT_\star\to\TT_\star$, and moreover
\begin{equation} \label{twistedmetric2}
    \gm_\star(X,Y)
    :=\left\langle X,\left\langle Y,\gm^A\right\rangle_\star\gm_A\right\rangle_\star
    =\gm\left(\bF_1\trc X,\bF_2\trc Y\right)=\la \, X\ots Y\,,\,\gm\,\ra
\end{equation}
for all $X,Y\in\Xi_\star$. \ 
$\nabla^\f$ is the unique LC connection with respect to $\gm_\star$; equivalently
\begin{equation}
    \mathcal{L}^\star_X[\gm_\star(Y,Z)]
    =\gm_\star(\nabla^\f_XY,Z)
    +\gm_\star(\bR_1\trc Y,\nabla^\f_{\overline{\r}_2\trc X}Z)
\end{equation}
for all $X,Y,Z\in\Xi_\star$. This \textit{twisted metric map} \ $\gm_\star\colon\Xi_\star\ot_\star\Xi_\star\to\X_\star$ \ as well as the twisted curvature
and Ricci tensor maps, are left $\X_\star$-linear  in the first argument and right $\X_\star$-linear  in the last argument.
Also the twisted Ricci tensor map is  in one-to-one
correspondence with an element \
$\ric^\f\in\Omega_\star\ot_\star\Omega_\star$ \ such that
\ $\ric^{\f}_\star(X,Y)=\la X\ots Y,\ric^\f\ra_\star$, \ by the non-degeneracy of the 
$\star$-pairing.
 The twisted  curvature, Ricci tensor and Ricci scalar are $U\k^\f$-invariant 
and coincide with their undeformed counterparts as elements
\be\label{TRspaces}
\rR^\f=\rR\in(\Omega\ot\Omega^2\ot\Xi)[[\nu]], \qquad
\ric^\f=\ric\in(\Omega \ot\Omega)[[\nu]], \qquad \mathfrak{R}^{\f}=\mathfrak{R}\in\X.
\ee

\subsection{Twisted smooth submanifolds of \texorpdfstring{$\RR^n$}{Rn} of codimension 1}
\label{TwistSmoothSubman}

Here we collect the main results of \cite{FioreWeber2020} regarding  a smooth submanifold $M\subset\mathcal{D}_f\subseteq\RR^n$ whose points $x$ solve the single equation $f(x)=0$.
More generally, the solutions $x\in\mathcal{D}_f$ of 
\begin{equation}
    f _c(x):=f (x)-c =0, \qquad\qquad c\in f(\mathcal{D}_f)\subseteq\RR,
\end{equation}
define a smooth manifold  $M_c$; varying $c$ 
we obtain a whole 1-parameter family of embedded submanifolds $M_c\subseteq\RR^n$
%(the level sets of $f$) 
of dimension $n\!-\!1$. %It follows that $(f^a_c)^*=f^a_c$. 
In \cite{FioreWeber2020} $\X$ stands for
the $*$-algebra of smooth functions on $\mathcal{D}_f$, and also
$\X^M=\mathcal{C}^\infty(M),\Xi_{\Ms},\Xi_t,...$ are understood in the smooth context.

\medskip
{\bf Twist deformation of tangent and normal vector fields.} \ \
%Here we  recall the results \cite{FioreWeber2020} on twisted tangent and normal vector fields
%for a unitary or real Drinfel'd twist $\F\in(U\Xi_t\ot U\Xi_t)[[\nu]]$.
%The classical analogues of definitions and results follow by setting $\F=1\ot 1$.
According to Section~\ref{TwistAlgStruc} $\Xi_\star$ is a $\X_\star$-bimodule with
$\X_\star$-subbimodules $\Xi_{\mathcal{C}\star},\Xi_{\mathcal{C}\mathcal{C}\star},
\Xi_{t\star}$. We further define the $\X_\star$-bimodule
\begin{equation}
    \Omega_{\scriptscriptstyle{\perp}\star}
    :=\{\omega\in\Omega_\star~|~\langle\Xi_{t\star},
    \omega\rangle_\star=0\}.
\end{equation}
\noindent
By {\bf Proposition 9 in \cite{FioreWeber2020}}, 
the $\X_\star$-bimodules $\Xi_{\mathcal{C}\star},\Xi_{t\star}$ and \
$\Xi_{M\star}    =:\Xi_{\mathcal{C}\star}/\Xi_{\mathcal{C}\mathcal{C}\star}$ \
are $\star$-Lie subalgebras of $\Xi_\star$ while $\Xi_{\mathcal{C}\mathcal{C}\star}$ is a
$\star$-Lie ideal. Furthermore, we obtain the decomposition \
$    \Omega_{\scriptscriptstyle{\perp}\star}
    =\X_\star\star\mathrm{d}f 
    =\mathrm{d}f\star\X_\star$, \
and the twisted exterior algebras
$\Xi^\bullet_\star,\Xi^\bullet_{t\star},\Xi^\bullet_{\mathcal{C}\star},
\Xi^\bullet_{\mathcal{C}\mathcal{C}\star},\Xi^\bullet_{M\star},
\Omega^\bullet_{\scriptscriptstyle{\perp}\star}$
are $U\Xi^\f_t$-equivariant $\X_\star$-bimodules.
$\Xi_{t\star},\Xi_{\C\star}, \Xi_{\C\C\star},\XiM{}_{\star},
\Omega_{\scriptscriptstyle{\perp}\star}$
resp. coincide as  $\CC[[\nu]]$-modules with  $\Xi_t[[\nu]],\Xi_\C[[\nu]],
\Xi_{\C\C}[[\nu]],\XiM[[\nu]],\Omega_{\scriptscriptstyle{\perp}}[[\nu]]$. 

Let $\gm=\gm^\alpha\ot\gm_\alpha\in\Omega\ot\Omega$ be a (non-degenerate) metric
on $\mathcal{D}_f$ with inverse $\gm^{-1}=\gm^{-1\alpha}\ot\gm^{-1}_\alpha$. 
\begin{equation}\label{eq30}
    \Xi_{\scriptscriptstyle{\perp}}
    :=\{X\in\Xi~|~\gm(X,\Xi_t)=0\}, \qquad
     \Omega_t
    :=\{\omega\in\Omega~|~\gm^{-1}(\omega,\Omega_{\scriptscriptstyle{\perp}})=0\}
\end{equation}
are the $\X$-bimodules
of normal vector fields and tangent differential forms. The open subset where the
restriction \ $    \gm^{-1}_{\scriptscriptstyle{\perp}}  :=
\gm^{-1}|_{\Omega_{\scriptscriptstyle{\perp}}  \otimes\Omega_{\scriptscriptstyle{\perp}}}     \colon\Omega_{\scriptscriptstyle{\perp}}\ot\Omega_{\scriptscriptstyle{\perp}}     \to\X $
is non-degenerate is denoted by $\mathcal{D}'_f\subset\mathcal{D}_f$. If $\gm$
is Riemannian $\mathcal{D}'_f=\mathcal{D}_f$. From now on we denote the restrictions
of $\Xi,\Xi_t,\Xi_{\scriptscriptstyle{\perp}},\Omega,\Omega_t,
\Omega_{\scriptscriptstyle{\perp}}$ to $\mathcal{D}'_f$ by the same symbols and by
$\mathfrak{k}\subseteq\Xi_t$ the Lie subalgebra of Killing vector fields with respect
to $\gm$ which are also tangent to $M_c\subseteq\mathcal{D}'_f$. The deformed analogues of (\ref{eq30}) 
\begin{equation}
    \Xi_{\scriptscriptstyle{\perp}\star}
    :=\{X\in\Xi_\star~|~\gm_\star(X,\Xi_{t,\star})=0\}, \qquad
    \Omega_{t\star}
    :=\{\omega\in\Omega_\star~|~
    \gm_\star^{-1}(\omega,\Omega_{\scriptscriptstyle{\perp}\star})=0\}
\end{equation}
can be defined for any twist $\F\in(U\Xi_t\ot U\Xi_t)[[\nu]]$.
Henceforth in this section $\F\in(U\k\ot U\k)[[\nu]]$. 

\medskip
\noindent
If $\F\in(U\k\ot U\k)[[\nu]]$ then by {\bf Proposition 10 in \cite{FioreWeber2020}},  
%where $\k$ is the Killing Lie algebra of $\gm$,
there are direct sum decompositions
\begin{equation}\label{eq14}
    \Xi_\star
    =\Xi_{t\star}\oplus\Xi_{\scriptscriptstyle{\perp}\star},\qquad
    \Omega_\star
    =\Omega_{t\star}\oplus\Omega_{\scriptscriptstyle{\perp}\star}
\end{equation}
into orthogonal $\X_\star$-bimodules, with respect to 
$\gm_\star$ and  $\gm_\star^{-1}$ respectively.
$\Xi_{t\star}$ is a $\star$-Lie subalgebra of $\Xi_\star$, $\Omega_{t\star}$ and
$\Xi_{\scriptscriptstyle{\perp}\star}$ are orthogonal with respect to the
$\star$-pairing and actually $\Omega_{t\star}=\{\omega\in\Omega_\star~|~
\langle\Xi_{\scriptscriptstyle{\perp}\star},\omega\rangle_\star=0\}$.
Furthermore, the restrictions
\begin{equation}
\begin{split}
    \gm_{\scriptscriptstyle{\perp}\star}
    :=&\gm|_{\Xi_{\scriptscriptstyle{\perp}\star}\ot_\star
    \Xi_{\scriptscriptstyle{\perp}\star}}\colon
    \Xi_{\scriptscriptstyle{\perp}\star}\ot_\star
    \Xi_{\scriptscriptstyle{\perp}\star}\to\X_\star,~~~~~~~~
    \gm_{t\star}
    :=\gm|_{\Xi_{t\star}\ot_\star\Xi_{t\star}}\colon
    \Xi_{t\star}\ot_\star\Xi_{t\star}\to\X_\star,\\
    \gm^{-1}_{\scriptscriptstyle{\perp}\star}
    :=&\gm^{-1}|_{\Omega_{\scriptscriptstyle{\perp}\star}\ot_\star
    \Omega_{\scriptscriptstyle{\perp}\star}}\colon
    \Omega_{\scriptscriptstyle{\perp}\star}\ot_\star
    \Omega_{\scriptscriptstyle{\perp}\star}\to\X_\star,~~~~
    \gm^{-1}_{t\star}
    :=\gm^{-1}|_{\Omega_{t\star}\ot_\star\Omega_{t\star}}\colon
    \Omega_{t\star}\ot_\star\Omega_{t\star}\to\X_\star,
\end{split}
\end{equation}
are non-degenerate.
$\Xi_{t\star},\Omega_{{\scriptscriptstyle{\perp}}\star},\Xi_{{\scriptscriptstyle{\perp}}\star},\Omega_{t\star}$ \
resp. coincide  with \ $\Xi_t[[\nu]],\Omp[[\nu]],\Xip[[\nu]],\Omega_t[[\nu]]$ as  $\CC[[\nu]]$-modules; 
and similarly for their $\star$-tensor (and -wedge) powers.
The orthogonal projections
$\mathrm{pr}_{t\star}\colon\Xi_\star\to\Xi_{t\star}$,
$\mathrm{pr}_{\scriptscriptstyle{\perp}\star}\colon\Xi_\star
\to\Xi_{\scriptscriptstyle{\perp}\star}$,
$\mathrm{pr}_{t\star}\colon\Omega_\star\to\Omega_{t\star}$ and
$\mathrm{pr}_{\scriptscriptstyle{\perp}\star}\colon\Omega_\star
\to\Omega_{\scriptscriptstyle{\perp}\star}$ and their (unique) extensions to
multivector fields and higher rank forms
are the $\CC[[\nu]]$-linear extensions of their classical counterparts. They, as well as $\Xi^\bullet_\star,\Xi^\bullet_{t\star},
\Xi^\bullet_{\scriptscriptstyle{\perp}\star},\Omega^\bullet_\star,\Omega^\bullet_{t\star},
\Omega^\bullet_{\scriptscriptstyle{\perp}\star}$,  are
$U\k^\f$-equivariant. 

The induced metric ({\it first fundamental form})  for the family of submanifolds $M_c\subseteq\mathcal{D}'_f$, where $c\in f(\mathcal{D}'_f)$, stays undeformed: \ 
$\gm_t^\f:=(\Pts\ot\Pts)(\gm)=(\Pt\ot\Pt)(\gm)=:\gm_t$.

\medskip
Defining $\Omega_{\mathcal{C}\star}\!:=\!\{\omega\in\Omega_\star~|~
\langle\Xi_{\scriptscriptstyle{\perp}\star},\omega\rangle_\star\subseteq\mathcal{C}[[\nu]]\}$
and
$\Omega_{\mathcal{C}\mathcal{C}\star}\!:=\!\Omega_\star\!\star\! f 
=f \!\star\!\Omega_\star$, we further obtain
\begin{equation}
    \Omega_{\Ms\star}
    =\Omega_{\mathcal{C}\star}/\Omega_{\mathcal{C}\mathcal{C}\star}
    =\{[\omega]=\omega+\Omega_{\mathcal{C}\mathcal{C}\star}~|~
    \omega\in\Omega_{\mathcal{C}\star}\}.
\end{equation}
The following proposition assures that every element of $\Xi_{\Ms\star}$ can be
represented by an element in $\Xi_{t\star}$ and
every element of $\Omega_{\Ms\star}$ can be
represented by an element in $\Omega_{t\star}$.

\medskip
\noindent
{\bf Proposition 11 in \cite{FioreWeber2020}}. \
{\it
For \ $X\in\Xi_{\mathcal{C}\star}$, \ $\omega\in\Omega_{\mathcal{C}\star}$ \ the tangent projections \ $X_{t\star}:=\mathrm{pr}_{t\star}(X)\in\Xi_{t\star}$, \
$\omega_{t\star}:=\mathrm{pr}_{t\star}(\omega)\in\Omega_{t\star}$ \
respectively belong to \ $[X]\in\Xi_{\Ms\star}$ \ and \ $[\omega]\in\Omega_{\Ms\star}$.}

\medskip
Let $\nabla$ be the LC connection corresponding to $(\mathcal{D}_f,\gm)$ and
 $\nabla^\f$ be the twisted LC connection
corresponding to $\gm_\star$. The induced twisted   {\it second fundamental form} and 
{\it LC connection}  on the family of submanifolds $M_c$ are \
$II^\f_\star:= \mathrm{pr}_{\scriptscriptstyle{\perp}\star}\!\circ\!
\nabla^\f|_{\Xi_{t\star}\otimes_\star\Xi_{t\star}}
\colon\Xi_{t\star}\ot_\star\Xi_{t\star}\to\Xi_{\scriptscriptstyle{\perp}\star}$
\ and
\ $\nabla\!_t^{\,\f}:=\Pts\circ\nabla^\f|_{\Xi_{t\star}\otimes_{\mathbb{K}[[\nu]]}\Xi_{t\star}}
\colon\Xi_{t\star}\ot_{\mathbb{K}[[\nu]]}\Xi_{t\star}\to\Xi_{t\star}$
\ respectively; the latter  yields the curvature $\rR^\f_{t\star}$ via (\ref{TANDR}).
We now summarize results of {\bf Propositions 3, 12 and 13 in \cite{FioreWeber2020}}. \
As %the  twisted first fundamental form, 
$\gm_t^\f=\gm_t$,
also the twisted second fundamental form,  curvature,
 Ricci tensor and Ricci scalar on $M$
are $U\k^\f$-invariant and coincide with the undeformed ones as elements
\be\label{tangentgIIR}
\ba{ll}
%\gm_t^\f=\gm_t\in(\Omega_t \otimes\Omega_t)[[\nu]],\quad 
II^\f=II\in(\Omega_t \otimes\Omega_t\ot\Xip)[[\nu]],\qquad 
&\rR^\f_{t}=\rR_{t}\in(\Omega_t\ot\Omega_t^2 \otimes\Xi_t)[[\nu]],\\[6pt]
\ric^\f_{t}=\ric_{t}\in(\Omega \ot\Omega)[[\nu]], \quad &
\mathfrak{R}^{\f}_t=\mathfrak{R}_t\in\X.
\ea
\ee
Hence \ $\gm_{t\star} = \left\la \,\cdot \ots \cdot\,,\gm_t^\f\right\ra_\star \colon \Xi_{t\star}\ots\Xi_{t\star}\to\X_\star$, \
$II^\f_\star = \left\la \,\cdot \ots \cdot\,,II^\f\right\ra_\star
     \colon  \Xi_{t\star}\ots\Xi_{t\star}\to\Xi_{\scriptscriptstyle{\perp}\star}$, \
$\rR^\f_{t\star}\!=\!
\left\la \,\cdot \ots \cdot\ots \cdot\,,\rR^\f_{t}\right\ra_\star\colon\Xi_{t\star}\ot_\star\Xi_{t\star}\ot_\star\Xi_{t\star}\to\Xi_{t\star}$, \
$\ric^\f_{t\star}= \left\la \,\cdot \ots \cdot\,,\ric^\f_{t}\right\ra_\star\colon\Xi_{t\star}\ot_\star\Xi_{t\star}\to\X_\star$,
\begin{comment}
\bea
&&\gm_{t\star}=\left\la \,\cdot \ots \cdot\,,\gm_t^\f\right\ra_\star
\:     \colon \: \Xi_{t\star}\ots\Xi_{t\star}\to\X_\star\nn[4pt]
&&   II_\star:=\mathrm{pr}_{\scriptscriptstyle{\perp}\star}\circ\nabla^\f=
\left\la \,\cdot \ots \cdot\,,II^\f\right\ra_\star
\:     \colon \: \Xi_{t\star}\ots\Xi_{t\star}\to\Xi_{\scriptscriptstyle{\perp}\star}\\[4pt]
&& \rR_{t\star}\\[4pt]
&&\ric^\f_{t\star}\colon\Xi_{t\star}\ot_\star\Xi_{t\star}\to\X_\star
\eea
\end{comment}
%
are $U\k^\f$-equivariant   maps,
and for all $X,Y,Z\in\Xi_{t\star}$ they actually reduce to
\begin{equation}\label{II^F}
\ba{ll}
\gm_{t\star}(X,Y)=\gm_t\!\left(\bF_1\trc X,\bF_2\trc Y\right), \quad   & 
\rR^\f_{t\star}(X,Y,Z)=\rR_{t}\!\left(\bF_1\trc X,\bF_2\trc Y,\bF_3\trc Z\right) ,\\[8pt]
II^\f_\star(X,Y)=
II\!\left(\bF_1\trc X,\bF_2\trc Y\right),\quad & \ric^\f_{t\star}(X,Y)=\ric_{t}\!\left(\bF_1\trc X,\bF_2\trc Y\right),
\ea
\end{equation}
where $\bF_1\ot \bF_2\ot \bF_3$ is the inverse of 
%the intertwiner between $\Delta^{(2)},\Delta^{(2)}_\f$, defined at either side of
 (\ref{cocycle}); these maps are 
 left (resp. right) $\X_\star$-linear in the first (resp. last) argument, `middle' $\X_\star$-linear otherwise, in the sense $\gm_{t\star}(X\star h,Y)=\gm_{t\star}(X,h\star Y)$, etc. \
Furthermore, the following {\it twisted Gauss equation} holds for all $X,Y,Z,W\in\Xi_{t\star}$
\begin{equation} \label{GaussQuantum}
\begin{split}
    \gm_\star\left(\rR^\f_\star(X,Y,Z),W\right)
    =&\gm_\star\left(\rR^\f_{t\star}(X,Y,Z),W\right)
    +\gm_\star\big(II^\f_\star(X,\bR_1\trc Z),II^\f_\star(\bR_2\trc Y,W)\big)\\[2pt]
    &-\gm_\star\big(II^\f_\star(\bR_{1\widehat{(1)}}\trc Y,\bR_{1\widehat{(2)}}\trc Z),
    II^\f_\star(\bR_2\trc X,W)\big).
\end{split}
\end{equation}
The twisted first and second fundamental forms,  Levi-Civita connection, 
curvature tensor, Ricci tensor, Ricci scalar {\it on $M$} are 
finally obtained from the above
by applying the further projection $\X_\star\to\XM_\star$, which amounts to choosing the $c=0$ manifold $M$ out of the $M_c$ family. 
Of course, one can do the same on any other $M_c$.

\medskip
{\bf Decompositions  (\ref{eq14}) in terms of bases or complete sets.} \  
%Fix a (non-degenerate) metric $\gm$ on $\RR^n$.
%Let $(x^1,\ldots,x^n)$ be the Cartesian coordinates of $\RR^n$, $\{\mathrm{d}x^i\}_{i=1}^n$
%the induced basis of $\Omega$ and $\{\partial_i\}_{i=1}^n$ the dual basis of vector fields
%$\Xi$. In particular every vector field decomposes as $X=X^i\partial_i\in\Xi$ and every
%differential $1$-form decomposes as $\omega=\omega_i\mathrm{d}x^i\in\Omega$ with
%$X^i,\omega_i\in\X$. Here and in the following we sum over repeated upper and lower indices (Einstein sum convention).
In terms of Cartesian coordinates $(x^1,\ldots,x^n)$  of $\RR^n$
the components of the metric and of the inverse metric on $\RR^n$
are denoted by $g_{ij}=\gm(\partial_i,\partial_j)$ and  $g^{ij}=\gm^{-1}(\mathrm{d}x^i,\mathrm{d}x^j)$
(as before $\partial_i\equiv \partial/\partial x^i$). Using
them we lower and raise indices: 
$\mathrm{d}x_i:=g_{ij}\mathrm{d}x^j$, $Y_i:=g_{ij}Y^j$, $\partial^i:=g^{ij}\partial_j$,
etc. In particular
\begin{equation}
\begin{split}
    \gm
    =&\mathrm{d}x^i\ot\mathrm{d}x_i,\\
    \gm^{-1}
    =&\partial^i\ot\partial_i,
\end{split}
\hspace{2cm}
\begin{split}
    \gm(X,Y)
    =&X^iY_i,\\
    \gm^{-1}(\omega,\eta)
    =&\omega^i\eta_i.
\end{split}
\end{equation}
Let  $\E:=f^if_i$ ($f_i\equiv \partial_i f$), ${\cal D}_f'\!\subset\!{\cal D}_f\!\subset\!\RR^n$ be the subset where $\E\neq 0$, and $K:=\E^{-1}$ on ${\cal D}_f'$.
If  $\gm$ is {\it Riemannian} then ${\cal D}_f'\!=\!{\cal D}_f$, because
$\E>0$ on all of ${\cal D}_f$ (as  $\gm^{-1}$ is positive-definite).  Let
\be
\Vp :=\gm^{-1}(df  ,dx^i)\:\partial_i=f^{i}\partial_i,\qquad 
 \Up:=\sqrt{|K|}\Vp, \qquad \theta:=\sqrt{|K|} df ;
\label{DefNpa}
\ee
$\Vp,\,\Np\!:=\!%K \: \gm^{-1}(df  ,dx^i)\:\partial_i%=K^{ab}f^{bi}\partial_i=
K \Vp\!= K \!\star\!\Vp$  or $\Up$ spans $\Xip$ (and $\Xi_{{\scriptscriptstyle \perp}\star}$), while $df$
or $\theta$ spans $\Omp$ (and $\Omega_{{\scriptscriptstyle \perp}\star}$). \ All
are $U\mathfrak{k}$-invariant. 
$\Np,df$ are $\star$-dual,   $\la \Np ,df \ra_\star=1$,  but  
$\gm^{-1}_\star(df ,df )=\E $,  $\gm_\star(\Np ,\Np)=K$,   while
\be
\la \Up ,\theta \ra_\star=1, \qquad 
\gm_\star(\Up ,\Up )=\epsilon, \qquad \gm^{-1}_\star(\theta,\theta)=\epsilon,           \qquad \quad\epsilon:=\mbox{sign}(\E) \label{perpdualstar}
\ee 
(see Proposition 8  in \cite{FioreWeber2020}); these relations hold also without 
$\star$. %, because $\E,K$ are $\k$-invariant.
The projection $\Pps$ ($\CC[[\nu]]$-linear extension  of $\Pp$) on  $X\in\Xi_\star$,  $\omega\in\Omega_\star$   can be equivalently expressed as
\bea
\ba{l}
\Pps(\omega)=\omega_{\scriptscriptstyle \perp}= \epsilon\: \theta \star   \gm^{-1}_\star(\theta,\omega)=   df  \star K  \star 
\gm^{-1}_\star(df  ,\omega)
= \gm^{-1}_\star(\omega,df ) \star K\star df , \\[8pt]
\Pps(X)=X_{\scriptscriptstyle \perp}= \epsilon\: \gm_\star(X ,\Up ) \star\Up
=\gm_\star(X ,\Vp) \star K\star\Vp
=\Vp \star K\star\gm_\star(\Vp,X)   
\ea\label{XiOmegaPerpstar}
\eea 
(see Proposition 14  in \cite{FioreWeber2020}).
By the $\star$-bilinearity of $\gm_\star$ these equations imply in particular
\bea
\ba{l}
\omega_{\scriptscriptstyle \perp}= df \star K \star 
\gm^{-1}_\star(df ,dx^i)\star\check\omega_i
=\hat\omega_i \star  \gm^{-1}_\star(dx^i,df) \star K \star df, \\[8pt]
X_{\scriptscriptstyle \perp}=\hat X^i\star\gm_\star(\partial_i ,\Vp) \star K \star\Vp
=\Vp \star K \star\gm_\star(\Vp,\partial_i) \star\check X^i,   
\ea\label{XiOmegaPerpstar'}
\eea 
in terms of the left  and right decompositions  \ $\omega=\hat\omega_i \star dx^i=dx^i\star\check\omega_i\in\Omega_\star$, \ $X=\hat X^i\star\partial_i=\partial_i \star\check X^i\in\Xi_\star$ 
  in the bases  $\{dx^i\}_{i=1}^n$, $\{\partial_i\}_{i=1}^n$. 
One can decompose  $df,\Np,\theta,\Up$ themselves in the same way,
if one wishes.  
If the metric is Euclidean  ($g_{ij}=\delta_{ij}$) or Minkowski 
[$g_{ij}=g^{ij}= \eta_{ij}=\mbox{diag}(1,...,1,-1)$] one makes (\ref{XiOmegaPerpstar'})
 more explicit replacing
\bea
\ba{ll} \gm^{-1}_\star(dx^i,df)= \gm^{-1}_\star(df,dx^i)=\gm^{-1}(dx^i,df)=f^{i},\qquad
&,\\[8pt]
\gm_\star(\partial_i ,\Np) =\gm_\star(\Np^a,\partial_i)=
\gm(\partial_i ,\Np )= Kf_i= K\star f_i.
\ea  
\eea 
Finally, we can express the  tangent projection acting on  $X\in\Xi_\star$,  $\omega\in\Omega_\star$  simply as \ $\Pts(X)=X_t :=X-X_{\scriptscriptstyle \perp}$,  \  $\Pts(\omega)= \omega_t:=\omega-\omega_{\scriptscriptstyle \perp}$. 
All the above formulae hold also if we drop all $\star$.

\smallskip
Having determined bases of $ \Xips,\Omps$ we now consider 
$\Xi_{t\star},\Omega_{t\star}$.  The globally  defined sets
$\Theta_t:=\left\{\vartheta^j\right\}^n_{j=1}$,
$S_W:=\left\{W_j\right\}^n_{j=1}$, where $\vartheta^j\!:=\Pt(dx^j)$,
$W_j\!:=\Pt(\partial_j)=:K\,V_j$,
are respectively complete in $\Omega_t$, $\Xi_t$; they are not bases, because of the linear dependence relations $\vartheta^jf_j=0$, $f^{j}W_j=0$. 
An alternative complete set (of globally  defined vector fields) in $\Xi_t$ is
\bea
%S_V:=\left\{V_j\right\}^n_{j=1},\quad
S_L:=\left\{L_{ij}\right\}_{i,j=1}^n, \qquad\quad \mbox{where }\qquad
%V_j:=(f^i\!f_i)\partial_j\!-\! f_j\Vp,\quad
L_{ij}:=f_i\partial_j\!-\!f_j\partial_i.
\eea
In fact,  $L_{ij}$ manifestly annihilate $f$, and $S_L$  is complete because 
%one recovers the $V_j$ from the $L_{ij}$ through 
the combinations $K f^iL_{ij}=W_j$ make up $S_W$.  
Clearly $L_{ij}=-L_{ji}$, so at most  $n(n\!-\!1)/2$ \  $L_{ij}$ (e.g. those
with  $i<j$) are linearly independent over $\RR$ (or $\CC$), while $S_L$
is of rank $n\!-\!1$ over $\X$ because of the dependence relations  
%(over $\X$)
\be
%f^iV_i=0, \qquad\qquad 
f_{[i}L_{jk]}=0                          \label{DepRel}
\ee
(square brackets enclosing indices mean a complete  antisymmetrization of the latter).
Contrary to the $W_j$, the $L_{ij}$ are anti-Hermitian under the $*$-structure 
%(\ref{natural*}), namely 
$L_{ij}^*=-L_{ij}$ and {\it do not involve $\gm$}, so they can be used even if we introduce no metric. Setting $f_{ih}=\partial_i\partial_h f$, their Lie brackets are
\bea
[L_{ij},L_{hk}]
%& = & [f_i\partial_j-f_j\partial_i,f_h\partial_k-f_k \partial_h]\nn
%& = & f_i[ \partial_j(f_h )\partial_k-\partial_j(f_k )\partial_h]
%-f_j[\partial_i(f_h )\partial_k-\partial_i(f_k )\partial_h] -(ij\leftrightarrow hk)\nn
& = & f_{jh}L_{ik}-f_{ih}L_{jk}-
f_{jk}L_{ih}+f_{ik}L_{jh}.       \label{comm}
\eea

By the mentioned propositions, every complete set of  $\Omega_t$, e.g. $\Theta_t$, is also a complete set  of $\Omega_{t\star}$; similarly, every complete set of $\Xi_t$, e.g.  $S_W$ or $S_L$, is also a complete set  of   $\Xi_{t\star}$.

\section{Twisted algebraic submanifolds of \texorpdfstring{$\RR^n$}{Rn}: the quadrics}
\label{TwistDiffGeomAlgSubman}

%\subsection{General features}
%\label{GenFeat}

We can apply the whole machinery developed in the previous chapter to twist deform 
algebraic manifolds of codimension 1 embedded in $\RR^n$ provided we adopt
$\X=\mbox{Pol}^\bullet(\RR^n)$, etc. everywhere.
We can assume without loss of generality that the $f$ be an irreducible
polynomial function\footnote{If $f(x)=g(x)h(x)$, we find
$$
L_{ij}=h(x)[g_i\partial_j-g_j\partial_i]+g(x)[h_i\partial_j-h_j\partial_i];
$$
on $M_g$ the second term vanishes and the first is tangent to $M_g$, as it must be;
and similarly on $M_h$. Having assumed the 
Jacobian everywhere of maximal rank  $M_g, M_h$ have empty intersection
and can be analyzed separately. Otherwise  $L_{ij}$ vanishes on
$M_g\cap M_h\neq\emptyset$ (the singular part of $M$),
so that  on the latter a twist built using the $L_{ij}$ will reduce to the identity, and the $\star$-product to the pointwise product (see the conclusions).}.
%As mentioned before,
It is interesting to ask for which algebraic submanifolds $M_c\subset\RR^n$ the infinite-dimensional Lie algebra  $\Xi_t$ 
admits a nontrivial finite-dimensional subalgebra $\g$ over $\RR$ (or $\CC$),
so that we can build concrete examples
of twisted $M_c$ by choosing a twist $\F\in (U\g\otimes U\g)[[\nu]]$ of a known type. 
%As said, 
If $M_c$ are manifestly symmetric under a % finite-dimensional  
Lie group\footnote{For instance, the sphere $S^{n-1}$ is 
$SO(n)$ invariant; a cylinder in $\RR^3$ is invariant under $SO(2)\times \RR$; 
the hyperellipsoid of equation $(x^1)^2\!+\!(x^2)^2\!+\!2[(x^3)^2\!+\!(x^4)^2] =1$  is invariant under $SO(2)\times SO(2)$; etc.} $\mathfrak{K}$,
 then such a $\g$ exists and contains the Lie algebra $\mathfrak{k}$ 
of $\mathfrak{K}$ (if $M$ is maximally symmetric  then $\mathfrak{k}$ is even complete - over $\X$ - in $\Xi_t$).
In general, given any set $S$ of vector fields that is complete in $\Xi_t$  the question is whether there are combinations  of them   (with coefficients in $\X$)
that close a finite-dimensional Lie algebra $\g$.

Here we answer this question in the simple situation where the $L_{ij}$ themselves close a finite-dimensional Lie algebra $\g$. This means that  in   (\ref{comm}) $f_{ij}=$const, hence $f(x)$ is a quadratic polynomial,
and $M$ is either a quadric or the union of two hyperplanes (reducible case); moreover $\g$ is a  Lie subalgebra of the affine Lie algebra ${\sf aff}(n)$ of $\RR^n$. In the next subsection we find some results valid for all $n\ge 3$ drawing some general consequences from the only assumptions $\X=\mbox{Pol}^\bullet(\RR^n)$ and  $\g\subset {\sf aff}(n)$; in particular,
in sections \ref{Qstar}, \ref{QMstar} we show that the {\it global} description of differential geometry on $\RR^n,M_c$ in terms of generators and relations extends to their
twist deformations, in such a way to preserve the  spaces 
consisting of polynomials of any fixed degrees in the coordinates $x^i$, differential $dx^i$ and  vector fields chosen as generators. 
%Moreover such deformations are not formal, but {\it strict} in the parameter $\nu$, namely make sense for all $\nu\in\RR$.
In section \ref{quadricsR^3} we shall analyze in detail the twisted quadrics embedded in $\RR^3$.

\medskip
If $f$ is  of degree two then there are real
constants $a_{\mu\nu}\!=\!a_{\nu\mu}$ ($\mu,\nu=0,1,...,n$) such that
\be
f(x)\equiv \frac 12 a_{ij}x^ix^j+a_{0i}x^i+\frac 12 a_{00}=0;  \label{quadricsn}
\ee
hence  $f_i=a_{ij}x^j\!+\!a_{i0}$,  all $f_{ij} =a_{ij}$ are constant, and  (\ref{comm}) has already the desired form 
\begin{equation}\label{eq06}
[L_{ij},L_{hk}]=a_{jh}L_{ik}-a_{ih}L_{jk}-
a_{jk}L_{ih}+a_{ik}L_{jh},    
\end{equation}
i.e. the $L_{ij}$ span a finite-dimensional Lie algebra $\g$ over $\RR$.
This is a Lie subalgebra of the affine Lie algebra of $\RR^n$, because all
$L_{ih}\trc $ act as linear transformations of the coordinates $x^k$:
\be
L_{ij}\trc x^h%=[L_{ij}, x^h]
=(a_{ix}x^k\!+\!a_{0i})\delta^h_j-(a_{jx}x^k\!+\!a_{j0})\delta^h_i
 \label{L_ij_su_x^h}
\ee
Let \ $r:=\mbox{rank}(a_{\mu\nu})$.
To identify  $\g$  for irreducible $f$'s ($r>2)$\footnote{If all $a_{ij}=0$ vanish, but $a_{0i}\neq0$ for some $i$ then $r=1$,  $M$ is a (hyper)plane, 
and rhs(\ref{eq06}) vanishes; one can express  all $L_{ij}$ (or $V_j$) as   combinations with constant coefficients of $(n\!-\!1)$
independent ones: i.e. $\g\sim \RR^{n-1}$ is the abelian group of translations in the  (hyper)plane.
$r=2$ corresponds to a reducible $f$, i.e. two (hyper)planes.}
we note that by a suitable Euclidean transformation %of the frame 
(this will be also an affine one)  one can always 
make the $x^i$ canonical coordinates for the quadric, so 
that $a_{ij}=a_i\delta_{ij}$ (no sum over $i$), $b_i:=a_{0i}=0$ if  $a_i\neq 0$, and coordinates are ordered  so that 
\be
a_1> 0,\quad ...,\quad a_l>0, \quad a_{l+1}<0,\quad  ...,\quad a_{m}<0,\quad  
\left\{\!\!\ba{l}a_{m+1}= 0, \\ b_{m+1}<0,\ea\right.  \quad...,\quad  
\left\{\!\!\ba{l}a_n= 0, \\ b_n<0,\ea\right. 
\ee
with $l\le m\le n$; moreover, if $m\!<\!n$ one can  make $%c\!\equiv\!\frac 12 
a_{00}=0$ by translation of a $x^j$ with $j\!>\!m$. The associated new $L_{ij}$ (which are related to the old by a linear transformation) fulfill 
\bea
[L_{ij},L_{hk}]=a_j[\delta_{jh}L_{ik}\!-\!\delta_{jk}L_{ih}]-
a_i[\delta_{ih}L_{jk}\!-\!\delta_{ik}L_{jh}].                                             \label{comm0}
\eea
It is easy to check that $r=n\!+\!1$ if $m\!=\!n$, $r=m\!+\!2$ if $m\!<\!n$. %(?)
One can always make $a_1=1$
by replacing $f\mapsto f/a_1$; one can make also 
the other nonzero $a_i$'s in (\ref{comm0}) be $\pm 1$ by the rescalings
$x^i\mapsto y^i:=|a_i|^{1/2}x^i$ of the corresponding coordinates 
(another affine transformation). So the associated new $L_{ij}$ fulfill (\ref{comm0})
with the $a_i\in\{-1,0,1\}$. Then:

\begin{itemize}
\item If $k>j>m$ (what is possible only if $m<n\!-\!1$), then \ $[L_{jk},L_{hi}]=0$. \ 
Hence the center  $\Z(\g)$
of $\g$  is trivial if $m\!=n,n\!-\!1$; otherwise it contains all such $L_{jk}=a_{0j}\partial_k\!-\!a_{0k}\partial_j$, and $\Z(\g)\!\simeq\!\RR^{n-m-1}$;
a basis of $\Z(\g)$ is $\B=\{L_{(m+1)(m+2)},L_{(m+2)(m+3)},...,L_{(n-1)n}\}$.

\item The $L_{ij}$ with   $j\!>\!m$  span an ideal  $\I(\g)\supset\Z(\g)$ of $\g$,
because (\ref{comm0}) becomes $[L_{ij},L_{hk}]=
a_i[\delta_{ih}L_{kj}\!-\!\delta_{ik}L_{hj}]$; adding the $m(n\!-\!m)$ elements $L_{ij}$  with   $i\le m\!<\!j$ to $\B$ one obtains a basis of $\I(\g)$, hence dim$[\I(\g)]=m(n\!-\!m)+ (n\!-\!m\!-\!1)\theta(n\!-\!m\!-\!1)$.  
$\I(\g)$ is a nilpotent Lie subalgebra, the radical $\R(\g)$ (the largest solvable ideal) of $\g$.

\item Finally,  the $L_{ij}$ with   $i<j\le m$  make up a basis of a $m(m\!-\!1)/2$-dimensional simple Lie-subalgebra  $\g_s\simeq so(l,m\!-\!l)$, in view of the signs of $a_i,a_j$.

\end{itemize}
\noindent 
Summing up, 
%\ dim$( \g)=m(m\!-\!1)/2+m(n\!-\!m)+ (n\!-\!m\!-\!1)\theta(n\!-\!m\!-\!1)$, \ and 
the Levi decomposition of $\g$ becomes
$\g\simeq so(l,m\!-\!l)\:\cross\,\R$.

The cones, which in the $y$ coordinates are represented by the
homogeneous equations
$$
f(y):=(y^1)^2+...+(y^l)^2-(y^{l+1})^2-...-(y^n)^2= 0,
$$
strictly speaking are not encompassed in the above analysis because the Jacobian 
matrix $(f_i)(y)$ vanishes at the apex $y=0$ (the only singular point). They  are algebraic varieties that are limits of the hyperboloids $f_c(y)=0$ as $c\to 0$. If we omit the apex, a cone becomes a disconnected
union of  two nappes (which are open in $\RR^n$), and $\g$ is spanned
not only by the $L_{ij}$, but also by the central anti-Hermitian element $\eta=x^i\partial_i+n/2$ %(strictly speaking, $\eta$ belongs to the central extension $\Xi_t\oplus\CC$, due to the constant term $n/2$)
 generating dilatations; note that all of them vanish on the apex.
Hence \ $\g\simeq so(l,n\!-\!l)\:\times\RR$ in this case.

\medskip
If we endow $\RR^n$ with the Euclidean metric, the metric matrix $g_{ij}=\delta_{ij}$ is not changed by the above Euclidean
changes of coordinates, because
the Euclidean group is the isometry group $\mathfrak{H}$ of  $\RR^n$, whereas its
nonzero (diagonal) elements are rescaled if we rescale  $x^i\mapsto |a_i|^{1/2}x^i$. Similarly,
if we endow $\RR^n$ with the Minkowski metric, Euclidean
changes of coordinates involving only the space ones,
or a translation of the time coordinate, do not alter the metric matrix $g_{ij}=\eta_{ij}$. 

\subsection{Twisted differential calculus on \texorpdfstring{$\RR^n$}{Rn}
by generators, relations}
\label{Qstar}

Let us  abbreviate $\xi^i:=dx^i$.
We  name {\it differential calculus algebra  on $\RR^n$ %${\cal D}_f$
} 
 the unital associative  
$*$-algebra $\Q^\bullet$ over $\CC$ generated by  Hermitian elements
$\{\1,x^i,\xi^i,\mathrm{i}\partial_i\}_{i=1}^n$ fulfilling 
\bea 
&&\ba{l}
%&&\label{1etarel}
 \1\eta^i-\eta^i= \eta^i\1-\eta^i=0, \qquad \mbox{for }\:\eta^i=x^i,\xi^i,\partial_i\\[4pt]
%\label{xxrel}
x^ix^j-x^jx^i=0, \\[4pt] 
%&& \label{xxirel} 
\xi^ix^j-x^j\xi^i=0,
%&& \label{ddrel}
\ea\label{DCcomrel1}\\[10pt]  
&&\ba{l}
% \1\partial_i-\partial_i= \partial_i\1-\partial_i=0, \\[4pt]
\partial_i\partial_j-\partial_j\partial_i=0, \\[4pt]  
%&& \label{dxirel}
\partial_j\xi^i-\xi^i\partial_j=0,\\[4pt] 
%&& \label{xixirel}
\xi^i\xi^j+\xi^j\xi^i=0,                 \\[4pt] 
%&& \label{dxrel}
\partial_i x^j - \delta^j_i\1-x^j\partial_i=0. 
\ea\label{DCcomrel2}
\eea
The $x^0\equiv\1,x^i,\xi^i,\partial_i$ play respectively the role of the unit, of Cartesian  coordinate functions on $\RR^n$, of differentials $dx^i$ of $x^i$,
of partial derivatives $\partial/\partial x^i$ with respect to $x^i$.
This is the adaptation of the definition of $\Q^\bullet$ 
in the smooth context (sections 3.1.3,  3.2.3 in \cite{FioreWeber2020})  to the polynomial one: 
the relations in the first two lines define the algebra structure of $\X$,
the other ones determine 
the relations (113-114) of \cite{FioreWeber2020} for the current choice
of $\X$ and of the pair  $\{\xi^i\}$, $\{\partial_i\}$ of dual frames.
The $x^\mu$ ($\mu=0,...,n$)
span the fundamental module $(\check\M,\tau)$  
of $U {\sf aff}(n)$ (the invariant element $\1$ itself spans a 1-dim, non-faithful  submodule), 
the $\xi^i$ span a related module $(\M ,\tau )$,
the $\partial_i$   the contragredient one $(\M^\vee ,\tau^\vee)$. 
More precisely they are related by
\bea
\ba{l}
g\trc\1=\varepsilon(g)\1,\\[4pt]
g\trc x^i=x^\mu\check\tau^{\mu i}(g)=:x^j\tau^{ji}(g)+\1\check\tau^{0 i}(g), \\[4pt]
g\trc \xi^i=\xi^j\tau^{ji}(g), \\[4pt]
 g\trc \partial_i=\tau^\vee{}^{ji}(g)\partial_j 
=\tau^{ij}(Sg)\partial_j;
\ea                  \label{lintransf1}
\eea
the first relation and $g\trc x^0=x^\mu\check\tau^{\mu 0}(g)$ imply
$\check\tau^{\mu 0}(g)=\varepsilon(g)\delta^{\mu 0}$. \
We encompass these $U {\sf aff}(n)$-modules into a single one $(\widetilde{\M},\rho)$ 
spanned by $(a^0\!,a^1\!,\!..., a^{3n})\equiv(\1,x^1,...,x^n,\xi^1,...,\xi^n,\partial_1,...,\partial_n)$. All are trivially also $U\g$-modules; 
also $\g$ is, under the adjoint action. 
%, and (\ref{DCcomrel}) can be formulated as a set of polynomial relations $f^J(a)=0$.
Of course, this $U {\sf aff}(n)$  action  is compatible with the
relations (\ref{DCcomrel1}-\ref{DCcomrel2});  the ideal $\I$ generated by 
their left-hand sides in the free $*$-algebra $\A^f$ generated by 
$\{a^0,a^1,..., a^{3n}\}$ is $U {\sf aff}(n)$-invariant. 
The $U {\sf aff}(n)$-action is also compatible with the invariance
of the exterior derivative, because $g\trc \xi^i=d(g\trc x^i)$. 

In the $\Q^\bullet$ framework $Xh=hX+X(h)$ is the inhomogeneous first order differential operator sum of a first order part (the vector field  $hX$) and a zero order part  (the multiplication operator by $X(h)$); it must not be confused with the  product of  $X$ by $h$ from the right, which  is equal to  $hX$ and so far has been denoted in the same way.  In the $\Q^\bullet$ framework we denote the latter by $X \ltlc  h$ (of course
$(X \ltlc  h)(h')=X(h') h=hX(h')$,  $X \ltlc  (hh')= hh'X$ remain valid).

When choosing a basis $\B$ of $\Q^\bullet$ made out of monomials in these generators, relations (\ref{DCcomrel1}-\ref{DCcomrel2}) allow to order them in any prescribed way; 
in particular we may choose 
$$
\B:=\left\{\beta^{\vec{p},\vec{q},\vec{r}}:=(\xi^1)^{p_1}...(\xi^n)^{p_n}(x^1)^{q_1}...(x^n)^{q_n}\partial_1^{r_1}...\partial_n^{r_n}\:\: |
\:\: \vec{p}\in\{0,1\}^n,\: \vec{q},\vec{r}\in\NN_0^n\right\}
$$
(we define $\beta^{\vec{0},\vec{0},\vec{0}}:=\1$).
The $*$-algebra structure of $\Q^\bullet$ is compatible with the 
form grading $\natural$
\be
\natural\left(\beta^{\vec{p},\vec{q},\vec{r}}\right)=p,\qquad
 \qquad p:=\sum_{i=1}^np_i,\quad q:=\sum_{i=1}^nq_i, \quad r:=\sum_{i=1}^nr_i
\ee
and the one $\sharp$ defined by \ $\sharp\!\left(\beta^{\vec{p},\vec{q},\vec{r}}\right)\!=q-r$ \ ($p,q,r$ are the total degrees   in $\xi^i,x^i,\partial_i$ respectively).
Fixing part or all  of $p,q,r$ we obtain the various relevant $U{\sf aff}(n)$ modules or module subalgebras or $\X$-bimodules: \ $\Lambda^\bullet,\Lambda^p,\Omega^\bullet,\Omega^p,...$.
For instance the exterior algebra $\Lambda^\bullet$ is generated by the $\xi^i$ alone
 ($q=r=0$) and     its $\natural\!=\!p$ component is the $U{\sf aff}(n)$-submodule of exterior 
$p$-forms $\Lambda^p$; by (\ref{DCcomrel2})$_3$  dim$\left(\Lambda^p\right)
=\binom{n}{p}$; in particular this is zero for $p>n$, \ 1 for $p=n$, and 
$\Lambda^\bullet=\bigoplus_{p=0}^n\Lambda^p$.
Let $\X^q$ be the  component of $\X$ of degree $q$, and 
$\widetilde{\X}^q:=\bigoplus_{h=0}^q\X^q$ (i.e.
 $\X^q,\widetilde{\X}^q$ consist resp. of homogeneous and inhomogenous polynomials in $x^i$ of degree $q$);   $\X=\bigoplus_{q=0}^\infty\X^q$ is trivially a filtered algebra
$\X=\biguplus_{q=0}^\infty\widetilde{\X}^q$. 
Let $\D$ be the unital subalgebra generated by the $\partial_i$ alone,  $\D^r$  its  component 
of degree $r$, and $\widetilde{\D}^r:=\bigoplus_{h=0}^r\D^r$; then
$\D=\bigoplus_{r=0}^\infty\D^r$  is trivially a filtered algebra 
$\D=\bigoplus_{r=0}^\infty\widetilde{\D}^r$. 
\ Finally, let 
\be
\Q^{pqr}:= \Lambda^p\widetilde{\X}^q\widetilde{\D}^r.
\ee
By (\ref{lintransf1}) the $U{\sf aff}(n)$ action 
 maps $\Lambda^p,\widetilde{\X}^q,\D^r$ into themselves, and 
all $\Q^{pqr}$ are $U{\sf aff}(n)$-$*$-modules.  By (\ref{DCcomrel1}-\ref{DCcomrel2}), 
$\widetilde{\D}^r\widetilde{\X}^{q'}=\widetilde{\X}^{q'}\widetilde{\D}^r$, whence
\be
\Q^{pqr}\Q^{p'q'r'}\subseteq \Q^{(p+p')(q+q')(r+r')}                  \label{Qproduct}
\ee
(this multiplication rule would not hold if we had defined 
$\Q^{pqr}\!:=\! \Lambda^p\X^q\D^r$,
because, $\D^r\X^{q'}\neq\X^{q'}\D^r$). \
A basis of $\Q^{pqr}$ is \
$\B^{pqr}:=\{\beta^{\vec{p},\vec{q},\vec{r}}\: \:|\:\:  p=\sum\limits_{i=1}^np_i,\:\:  \sum\limits_{i=1}^nq_i\le q, \:\:  \sum\limits_{i=1}^nr_i\le r\}$. \ $\Q^\bullet$ is graded by $p$ and filtered by both $q,r$; it  decomposes as
\be
\Q^\bullet=\bigoplus_{p=0}^n\biguplus_{q=0}^\infty\biguplus_{r=0}^\infty\Q^{pqr}.
 \label{Qdeco}
\ee

Choosing a twist $\F$ based on $U{\sf aff}(n)$ 
(in particular, on $U\g$) and setting  (\ref{starprod}) for all $a,b\in\Q^\bullet$
one makes  $\Q^\bullet$ into a $U{\sf aff}(n)^\f$-module (resp. $U\g^\f$-module)
algebra $\Q^\bullet_\star$
with grading $\natural$ (whereas the grading $\sharp$ is not preserved).
In the appendix we prove

\begin{prop} The vector fields 
$\partial_i'\!:=\!S(\beta)\trc\partial_i \!=\!\tau^{ij}(\beta)\partial_j$ are
the $\star$-dual ones  to the $\xi^i=dx^i$;
under the $U{\sf aff}(n)$ (and $U\g$) action they transform according to
$g\trc \partial_i'=\tau^{ij}[S_\f(g)]$.
The polynomials relations (\ref{DCcomrel1}-\ref{DCcomrel2}) are deformed into the ones 
\bea 
&&\ba{l}
%&&\label{1etarel}
 \1\star\eta^i-\eta^i= \eta^i\star\1-\eta^i=0, \qquad  \mbox{for }\:\eta^i=x^i,\xi^i,
\\[4pt]
%\label{xxrel}
x^i\star x^j- x^\nu \star x^\mu R^{\mu\nu}_{ij}=0, \\[4pt] 
%&& \label{xxirel} 
\xi^i\star x^j-x^\nu \star \xi^h R^{h\nu}_{ij}=0,\\[4pt] 
%&& \label{xixirel}
\xi^i\star\xi^j+ \xi^k\star \xi^h R^{hk}_{ij}=0,
\ea\label{DCcomrelstar1}\\[8pt]  
&&\ba{l}
 \1\star\partial_i'-\partial_i'= \partial_i'\star\1-\partial_i'=0, \\[4pt]
%&& \label{ddrel}
\partial_i'\star\partial_j'-R^{ij}_{hk}\partial_k'\star\partial'_h=0, \\[4pt]  
%&& \label{dxirel}
\partial_i'\star\xi^j-R^{hi}_{jk}\xi^h\star\partial_k'=0,                 \\[4pt] 
%&& \label{dxrel}
\partial_i' \star x^j - \delta^j_i\1-R^{\mu i}_{jk}x^\mu \star\partial_k'=0. 
\ea\label{DCcomrelstar2}
\eea
where $R^{\mu\nu}_{ij}:=(\tau^{\mu i}\ot\tau^{\nu j})(\R)$. 
Defining 
$\Q^{pqr}_\star:= \Lambda^p_\star\widetilde{\X}^q_\star\widetilde{\D}^r_\star$,
we find not only \ $\Q^\bullet_\star=\Q^\bullet[[\nu]]$,  \ but
that   for all $p,q,r\in\NN_0$ also
\be
\Q^{pqr}_\star=\Q^{pqr}[[\nu]]                          \label{Qpqr=Qpqrstar}
\ee 
hold as equalities of $\CC[[\nu]]$-modules. A basis $\B^{pqr}_\star$ of $\Q^{pqr}_\star$ is obtained replacing all products in the definition of $\B^{pqr}$   by $\star$-products. %Again, 
$\Q^\bullet_\star$ is graded by $p$, filtered by both $q,r$, and
%(\ref{Qproduct}-\ref{Qdeco}) remain true: 
\be
\Q^\bullet_\star=\bigoplus_{p=0}^n\biguplus_{q=0}^\infty\biguplus_{r=0}^\infty\Q^{pqr}_\star,\qquad \Q^{pqr}_\star \star\Q^{p'q'r'}_\star\subseteq \Q^{(p+p')(q+q')(r+r')}_\star.
 \label{Qproduct+decostar}
\ee
$\Q^\bullet_\star$  is  a $U\g^\f$-module $*$-algebra with the
$\Q^{pqr}_\star$ as $*$-submodules, \ if $\F$ is either real or unitary; 
correspondingly the involution is the undeformed one $*$, respectively is given by
(\ref{eq03}), i.e.
\bea
\1^{*_\star}=\1,
\quad x^i\,{}^{*_\star} =x^\mu\check\tau^{\mu i}\! \left[ S(\beta)\right],
\quad \xi^i\,{}^{*_\star} =\xi^k\tau^{ki}\! \left[ S(\beta)\right],
\quad \partial_i'\,{}^{*_\star} =- \tau^{ik}\!\left(\beta^{-1}\right)\hat \partial_k'.
\label{*_FR^n}
\eea
\label{propQR^nstar}
\end{prop} 
In the $\Q^\bullet_\star$ framework 
$X\star h=(\bR_1\trc h)\star (\bR_2\trc X) +X_\star(h)$,  while so far
it stood just for the $\star$-product of the vector field $X$
by the function $h$ from the right, i.e. for the first term at the rhs; denoting the latter by  \ $X \ltlc_\star  h:=(\bR_1\trc h)\star (\bR_2\trc X)$, \   we can abbreviate
$X\star h=X \ltlc_\star  h+X_\star(h)$. \
Of course  \ $(X \ltlc_\star  h)_\star(h')=[X_\star(\bR_1\trc h')]\star (\bR_2\trc h)$, \ 
 $(X \ltlc_\star  h)\ltlc_\star  h'=X\ltlc_\star  (h\star h')$ remain valid.

These results are the strict analogues of their untwisted counterparts.
Relation (\ref{Qpqr=Qpqrstar})  is much stronger than  the equality of infinite-dimensional $\CC[[\nu]]$-modules $\Q^\bullet_\star=\Q^\bullet[[\nu]]$; it  implies \ $\mbox{dim}\big(\Q^{pqr}_\star\big)=\mbox{dim}\big(\Q^{pqr}\big)$ over $\CC[[\nu]]$, \ so that
the Hilbert-Poincar\'e series of the $p$-graded
and $(q,r)$-filtered algebras \ $\Q^\bullet_\star,\Q^\bullet[[\nu]]$ coincide.
In particular,  $p\!=\!r\!=\!0$ yields   $\mbox{dim}\big(\widetilde{\X}^q_\star\big)=\mbox{dim}\big(\widetilde{\X}^q\big)$. 

The $U{\sf aff}^\f$-equivariant relations (\ref{DCcomrelstar1}-\ref{DCcomrelstar2}) defining $\Q^\bullet_\star$ have the same form (see e.g. formulae (1.10-15)
in \cite{Fio04JPA}) as the quantum group equivariant ones defining the differential calculus
algebras on the celebrated `quantum spaces' introduced in \cite{FRT}.
The  relations, among (\ref{DCcomrelstar1}-\ref{DCcomrelstar2}), that involve
only the generators $x^i,\partial_j'$ of the twisted Heisenberg algebra on $\RR^n$ (the $p=0$ component
of $\Q_\star^\bullet$) were already determined in \cite{Fio98JMP,Fio00RMP}, while (\ref{Qpqr=Qpqrstar}) extends results of \cite{Fiore2010}.

\subsection{Twisted differential calculus on $M$ by generators, relations}
\label{QMstar}

%Assume now that  $M\subset\RR^n$ is the algebraic manifold defined by 
%a single polynomial equation (\ref{quadricsn}) $f(x)=0$.   
Chosen a basis $\{e_1,...,e_B\}$ of $\g$ (e.g. consisting of $L_{ij}$), on ${\cal D}_f\subseteq\RR^n$ one can use  
$S'\equiv\{e_1,...,e_B,e_{B+1}=\Vp\}$,  instead of
$S\equiv\{\partial_1,...,\partial_n\}$, as a complete set of vector fields in $\Xi$. They fulfill  the following commutation relations with the coordinates
\be
\Vp x^h-x^h\Vp -f^i=0,\qquad \quad
e_\alpha x^h- x^he_\alpha-x^\mu\check\tau^{\mu h}(e_\alpha)=0, \quad\alpha=1,...,B
                    \label{DCMcrel1'}
\ee
and the remaining relations of the type eq. (113) in  \cite{FioreWeber2020}, i.e.
\be
\ba{l}
% \1 e_\alpha-e_\alpha= e_\alpha\1-e_\alpha=0,\\[4pt]
\sum_{\alpha=1}^A t_l^\alpha\, e_\alpha=0, \qquad l=1,...,B+1-n,\\[4pt]
e_\alpha e_\beta- e_\beta e_\alpha-  C^\gamma_{\alpha\beta}\,e_\gamma=0,\\[4pt]  
%&& \label{dxirel}
e_\alpha\xi^i-\xi^ie_\alpha=0, 
%\\[4pt] && \label{xixirel}
%\xi^i\xi^j+\xi^j\xi^i=0,                            
\ea \label{DCrel2}
\ee
with suitable $t_l^a,C^\gamma_{\alpha\beta}\in\X$. For instance,
if $S\equiv\{L_{ij},\Vp\}$ then 
 the dependence relations in the first line  amount to  (\ref{DepRel}), while the commutation relations
 in the second line have   constant $C^\gamma_{\alpha\beta}$ 
%$e_\alpha\neq\Vp\neq e_\beta$
and amount to (\ref{eq06})  for $\alpha,\beta\leq B$. 
%Thus (\ref{DCMcrel1'}) and (\ref{DCcomrel1})$_2$ replace (\ref{DCrel1}).
%The  $L_{ij}$ span the adjoint module of $U\g$. %also a module of $U {\sf aff}(n)$?
%$E(x):=f_i(x)f_i(x)$ is positive definite everywhere, and so is $K(x):=1/E(x)$. 
We collectively rename $\1,x^1,...,x^n,\xi^1,...,\xi^n,e_1,...,e_B$
as $a^0,a^1,..., a^N$; we denote as $\A'{}^\bullet$  the free algebra generated by $a^0,..., a^N$, and as $\A'{}^{pqr}$ the subspace consisting of polynomials in the $a^A$
of degree $q$ in the $x^i$, of degree $r$ in the $e_\alpha$ and homogeneous of degree $p$ in the $\xi^i$. Clearly
\be
\A'{}^{pqr}\A'{}^{p'q'r'}\subseteq \A'{}^{(p+p')(q+q')(r+r')}.                  \label{A'product}
\ee 
%A basis of $\QM^{pqr}$ is \ $\B^{pqr}:=\{\beta^{\vec{p},\vec{q},\vec{r}}\: \:|\:\:  p=\sum\limits_{i=1}^np_i,\:\:  \sum\limits_{i=1}^nq_i\le q, \:\:  \sum\limits_{i=1}^nr_i\le r\}$.
 \ $\A'{}^\bullet$ is graded by $p$ and filtered by both $q,r$; it  decomposes as
\be
\A'{}^\bullet=\bigoplus_{p=0}^\infty\biguplus_{q=0}^\infty\biguplus_{r=0}^\infty\A'{}^{pqr}.
 \label{A'deco}
\ee
For all $c\in\RR$ denote as $\{f_c^J(a^0,...,a^N)\}_{J\in \J}$  the set of polynomial functions at the lhs of   (\ref{DCrel2}), (\ref{DCcomrel1}) , (\ref{DCMcrel1'}) involving only
$e_\alpha$ with $\alpha\le B$, together with
\be
\ba{l}
f_c\equiv f(x)\!-\!c=0,\\[6pt] 
df(x)\equiv\xi^hf_h=0 ,     
\ea                    \label{DCMcrel'}
\ee
which are (\ref{quadricsn}) and its exterior derivative. Let $\IM$ be the ideal generated by all the  $f_c^J(a)$ in $\A'{}^\bullet$. We define the 
{\it differential calculus algebra on $M_c$}  as the quotient 
\be
\Q^\bullet_{\scriptscriptstyle M_c}:=\A'{}^\bullet/\IM.
\ee
$\IM^{pqr}:=\IM\cap\A'{}^{pqr}$ is  a subspace of $\A'{}^{pqr}$.
The %(linear space) 
quotient subspaces
$\QM^{pqr}:= \A'{}^{pqr}/\IM^{pqr}$ fulfill
%$\E$ as the unital subalgebra generated by the $e_\alpha$ with $\alpha\le B$ (alone),   
%as $\widetilde{\E}^r$ its subspace consisting in polynomials of degree $r$ in the $e_\alpha$; 
% $\E$  is a filtered algebra $\E=\biguplus_{r=0}^\infty\widetilde{\E}^r$. \
%We also define $\Lambda^p,\widetilde{\X}^q\subset\QM^\bullet$ as before
%[but clearly now their dimensions may be smaller due to (\ref{DCMcrel'})], 
%and $\QM^{pqr}:= \Lambda^p\widetilde{\X}^q\widetilde{\E}^r$. \ 
\be
\QM^{pqr}\QM^{p'q'r'}\subseteq \QM^{(p+p')(q+q')(r+r')} \label{QMproduct}
\ee 
because of the equations $f_c^J(a)=0$,  in particular because  $x^\mu\check\tau^{\mu h}(e_\alpha)$
in (\ref{DCMcrel1'}) are polynomial functions of first degree in $x^i$.
%A basis of $\QM^{pqr}$ is \ $\B^{pqr}:=\{\beta^{\vec{p},\vec{q},\vec{r}}\: \:|\:\:  p=\sum\limits_{i=1}^np_i,\:\:  \sum\limits_{i=1}^nq_i\le q, \:\:  \sum\limits_{i=1}^nr_i\le r\}$.
 \ $\QM^\bullet$ is graded by $p$ and filtered by both $q,r$; it  decomposes as
\be
\QM^\bullet=\bigoplus_{p=0}^{n-1}\biguplus_{q=0}^\infty\biguplus_{r=0}^\infty\QM^{pqr}.
 \label{QMdeco}
\ee
By (\ref{lintransf1}), (\ref{DCrel2})$_2$ the $a^i$ span 
a (reducible) $U\g$-$*$-module. Hence  $\A'{}^\bullet$, 
which is generated by them, is a 
$U\g$-module $*$-algebra, and the $\A'{}^{pqr}$ are $U\g$-$*$-submodules.
It is immediate to check that also the  $f_c^J(a)$ span a (reducible)  $U\g$-$*$-module,
\be
[f_c^J(a)]^*=f_c^J(a), \qquad
g\trc f_c^J(a)=\sum_{J'\in \J} f_c^{J'}(a){\bf \tau}^J_{J'}(g) ,      \label{f_c^JModule}
\ee 
%(${\bf \tau}^J_{J'}(g)\in \CC$), 
more precisely $g\trc f^c =\varepsilon(g)f^c$, \
%$g\trc e_\alpha=e_{\alpha g}^i(x)\partial_i$ with $e_{\alpha g}^i(x)$
%a polynomial function of the same degree as $e_{\alpha}^i(x)$ (at most), 
while more generally $g\trc f_c^J(a)$ is a numerical combination of the $f_c^{J'}(a)$
appearing in the same equation 
where $ f_c^J(a)$ appears, e.g. $g\trc(\xi^i\xi^j+\xi^j\xi^i)=(\xi^h\xi^k+\xi^k\xi^h)\tau^{hi}(g_{(1)})\tau^{kj}(g_{(2)})$. 
Therefore  $\IM$  is a  $U\g$-$*$-module, and $\Q^\bullet_{\scriptscriptstyle M_c}$ is a $U\g$-module $*$-algebra as well; moreover $\IM^{pqr}\subset\IM$ and $\QM^{pqr}\subset\Q^\bullet_{\scriptscriptstyle M_c}$ are $U\g$ $*$-submodules as well.
% $\Q^\bullet_{\scriptscriptstyle M_c}$  is thus a $U\g$-module $*$-algebra with the
%$\QM^{pqr}$ as $*$-submodules. 

\bigskip
Eq. (\ref{lintransf1}) and (\ref{f_c^JModule})  with a twist $\F\!\in\! U\g\otimes U\g[[\nu]]$   imply that:

\begin{enumerate} 
\item $\A'{}^\bullet_\star$ is a
$U\g^\f$-module $*_\star$-algebra;   each component $\A'{}^{pqr}_\star$
consisting of polynomials in the $a^A$
of degree $q$ in the $x^i$, of degree $r$ in the $e_\alpha$ and homogeneous of degree $p$ in the $\xi^i$ is a $U\g^\f$-$*_\star$-submodule;  
$\A'{}^{pqr}_\star=\A'{}^{pqr}[[\nu]]$, $\A'{}^\bullet_\star=\A'{}^\bullet[[\nu]]$
hold as equalities of $\CC[[\nu]]$-modules.

\item For all $J\!\in\! \J\!$, $\alpha,\alpha'\!\in\! \A'{}^\bullet[[\nu]]$, $\beta,\beta'\!\in\!\IM[[\nu]]$,  also
%\bea
$$
f_c^J(a)\star\alpha,\quad 
 \alpha\star f_c^J(a) ,\quad \beta\star \alpha,\quad  \alpha\star \beta,\quad  
(\alpha+\beta)\star(\alpha'+\beta')-\alpha\star \alpha'
$$
%\nonumber\eea
belong to $\IM[[\nu]]$; if the twist $\F$ is either real or unitary then also
$[f_c^J(a)]^{*_\star}$, $\beta^{*_\star}$ do. 
Therefore \ $\IMst:=\IM[[\nu]]$ \ is a two-sided ($*_\star$-)ideal of  $\A'{}^\bullet_\star$.  For each component \ $\IMst^{pqr}:=\IMst\cap\A'{}^{pqr}_\star$
we find \ $\IMst^{pqr}=\IM^{pqr}[[\nu]]$. \
$\IMst$ and  $\IMst^{pqr}$ \ are  $U\g^\f$-$*_\star$-submodules.

\end{enumerate} 

This leads to the following

\begin{prop}    For all $c\in{\cal D}_f$ \
$\QMst^\bullet:=\A'{}^\bullet_\star/\IMst$ \ defines a $U\g^\f$-module $*_\star$-algebra,
which we shall name {\rm twisted differential calculus algebra on $M_c$}; taking the quotient commutes with deforming the product: 
\be
\QMst^\bullet:=\A'{}^\bullet_\star/\IMst=(\A'{}^\bullet/\IM)_\star.
\ee
All components $\QMst^{pqr}:=\A'{}^{pqr}_\star/\IMst^{pqr}$ ($p,q,r\in\NN_0$) are $U\g^\f$-$*_\star$-submodules. 
 $\QMst^\bullet$ is graded by $p$, filtered by both $q,r$, and
%(\ref{Qproduct}-\ref{Qdeco}) remain true: 
\be
\QMst^\bullet=\bigoplus_{p=0}^n\biguplus_{q=0}^\infty\biguplus_{r=0}^\infty\QMst^{pqr},\qquad \QMst^{pqr}\star\QMst^{p'q'r'}\subseteq \Q^{(p+p')(q+q')(r+r')}_\star.
 \label{QMproduct+decostar}
\ee
$\QMst^\bullet=\QM^\bullet[[\nu]]$ \ and  
\be
\QMst^{pqr}=\QM^{pqr}[[\nu]]             \label{QMpqr=QMpqrstar}
\ee 
hold for all $p,q,r\in\NN_0$ as equalities of $\CC[[\nu]]$-modules. 
The set of characterizing polynomial relations $f_c^J(a)=0$ is  equivalent to the  set of relations $\hat f_c^J(a\star)=0$
consisting of  (\ref{DCcomrelstar1}) and other relations of the same degrees in 
$x^i,\xi^i,e_\alpha$ ($\alpha\le B$) as their undeformed counterparts.
From any basis $\BM^{pqr}$ of $\QM^{pqr}$ consisting
of polynomials in $x^i,\xi^i,e_\alpha$ one can obtain a basis $\BMst^{pqr}$ of 
$\QMst^{pqr}$ consisting of $\star$-polynomials of the same degrees.
If $\F$ is either real or unitary,  $\QMst^\bullet$  is  a $U\g^\f$-module $*$-algebra with the
$\QMst^{pqr}$ as $*$-submodules. If $\F$ is real 
the involution is  undeformed $*$. If $\F$ is unitary the involution is given by
(\ref{eq03}), i.e. on $\xi^i,x^i$ $*_\star$ acts as in (\ref{*_FR^n}), while $L_{ij}^{*_\star} =-\tau^{ih}\left(\beta_{(1)}\right)\tau^{jk}\left(\beta_{(2)}\right)L_{hk}$ 
(this differs from $L_{ij}^* =-L_{ij}$).  
\label{propQM_cstar}
\end{prop}

These results are the strict analogue of their undeformed counterparts.
Relation  (\ref{QMpqr=QMpqrstar})  is much stronger than  the equality of infinite-dimensional $\CC[[\nu]]$-modules $\QMst^\bullet=\QM^\bullet[[\nu]]$; it  implies \ $\mbox{dim}\big(\QMst^{pqr}\big)=\mbox{dim}\big(\QM^{pqr}\big)$ over $\CC[[\nu]]$, \ so that
the Hilbert-Poincar\'e series of 
\ $\QMst^\bullet,\QM^\bullet[[\nu]]$ coincide.
In particular, setting $p\!=\!r\!=\!0$, we find   $\mbox{dim}\big(\widetilde{\X}^q_\star\big)=\mbox{dim}\big(\widetilde{\X}^q\big)$.

%%%%%%%

In section \ref{quadricsR^3} we explictly determine all of the relations $\hat f_c^J(a\star)=0$  in the specific case of some deformed quadrics in $\RR^3$.

%%%%%%5

%Perhaps we should in particular add:
%why the only admissible polynomial metrics on an algebraic submanifold
%of $\RR^n$ are constant

\section{The quadrics in \texorpdfstring{$\RR^3$}{Rn}}
\label{quadricsR^3}

\begin{figure}[!htbp]
\begin{subfigure}{.48\textwidth}
\centering
    \includegraphics[width=0.46\textwidth, angle=0]{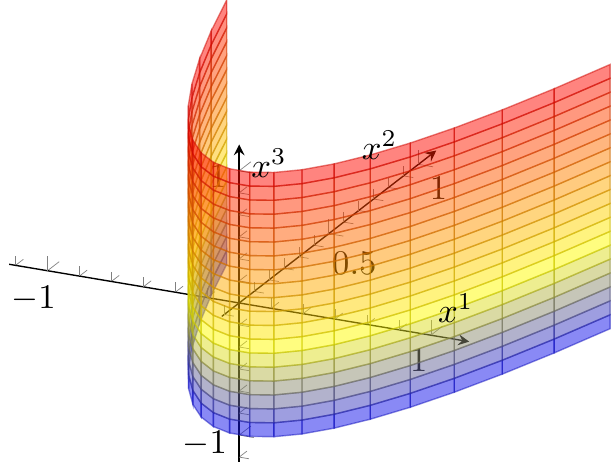}
    \caption{Parabolic cylinder with $a_{02}=-1$}
    \label{Parabolic cylinder}
\end{subfigure}
\begin{subfigure}{.48\textwidth}
\centering
    \includegraphics[width=0.46\textwidth, angle=0]{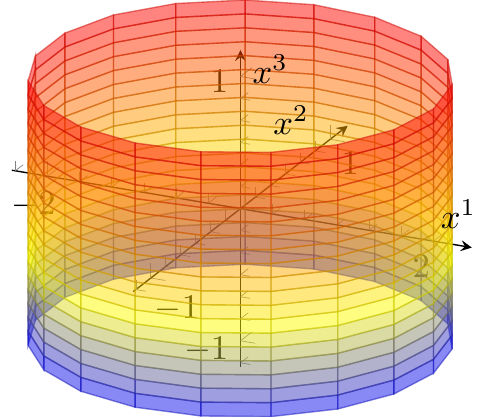}
    \caption{Elliptic cylinder with $a_1=\frac{1}{2}$, $a_2=2$} 
    \label{Elliptic cylinder}
\end{subfigure}
\vskip.25cm
\begin{subfigure}{.48\textwidth}
\centering
    \includegraphics[width=0.46\textwidth, angle=0]{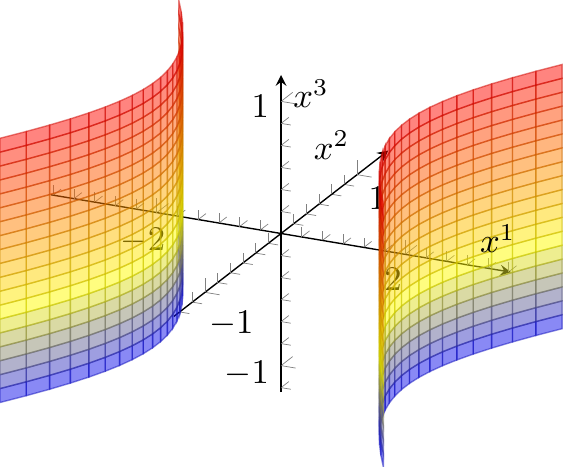}
    \caption{Hyperbolic cylinder with $a_1=\frac{1}{2}$, $a_2=-2$} 
    \label{Hyperbolic cylinder}
\end{subfigure}
\begin{subfigure}{.48\textwidth}
\centering
    \includegraphics[width=0.46\textwidth, angle=0]{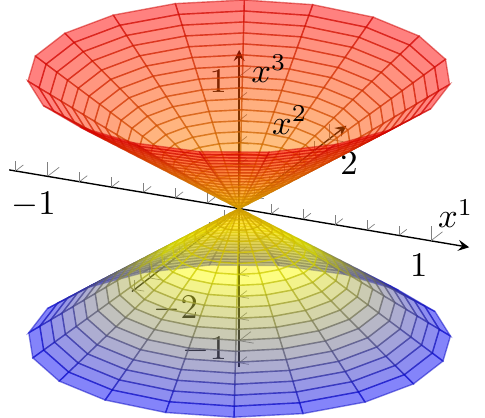}
    \caption{Elliptic cone with $a_1=-a_3=2$, $a_2=\frac{1}{2}$} 
    \label{Elliptic cone}
\end{subfigure}
\vskip.25cm
\begin{subfigure}{.48\textwidth}
\centering
    \includegraphics[width=0.46\textwidth, angle=0]{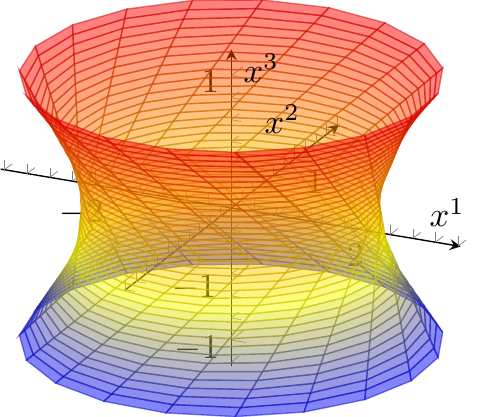}
    \caption{One-sheet elliptic hyperboloid with $a_1=\frac{1}{2}$, $a_2=-a_3=2$} 
    \label{Elliptic hyperboloid of one sheet}
\end{subfigure}
\begin{subfigure}{.48\textwidth}
\centering
    \includegraphics[width=0.46\textwidth, angle=0]{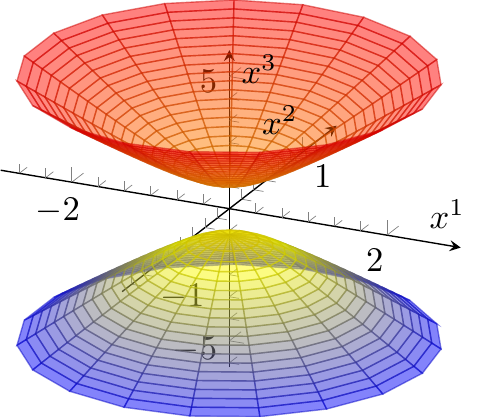}
    \caption{Two-sheet elliptic hyperboloid  with $a_1=8$, $a_2=32$, $a_3=-2$} 
    \label{Elliptic hyperboloid of two sheets}
\end{subfigure}
\vskip.25cm
\begin{subfigure}{.48\textwidth}
\centering
    \includegraphics[width=0.46\textwidth, angle=0]{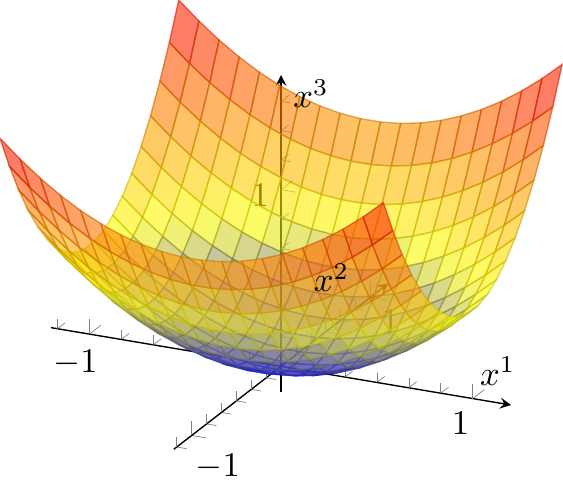}
    \caption{Elliptic paraboloid with $a_1=1$, $a_2=2$} 
    \label{Elliptic paraboloid}
\end{subfigure}
\begin{subfigure}{.48\textwidth}
\centering
    \includegraphics[width=0.46\textwidth, angle=0]{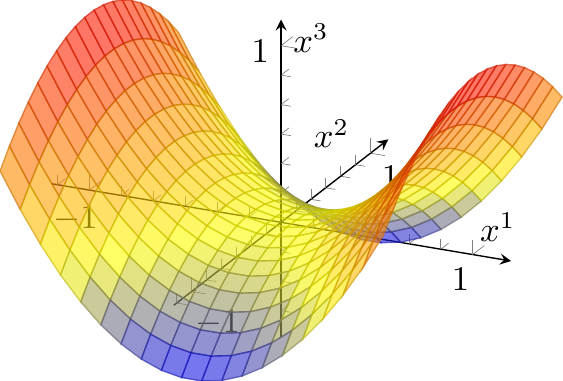}
    \caption{Hyperbolic paraboloid with $a_1=2$, $a_2=-1$}
    \label{Hyperbolic paraboloid}
\end{subfigure}
\vskip.25cm
\begin{subfigure}{.48\textwidth}
\centering
    \includegraphics[width=0.46\textwidth, angle=0]{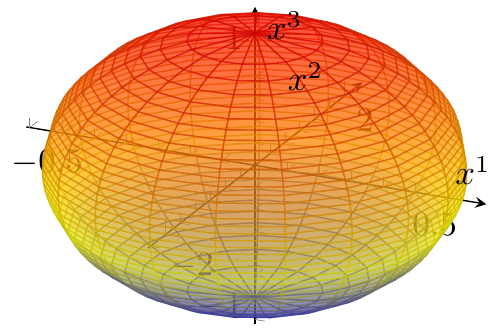}
    \caption{Ellipsoid with $a_1=8$, $a_2=\frac{1}{2}$, $a_3=2$}
    \label{Ellipsoid}
\end{subfigure}
\caption{The  irreducible quadric surfaces of $\mathbb{R}^3$}
\label{QuadricSurfaces}
\end{figure}

Using the notions and results presented in the previous sections, here
we study in detail  twist deformations of the  quadric surfaces in $\RR^3$.
As usual, we identify two quadric surfaces if they can be translated into each other via an
Euclidean transformation. This leads to nine classes of  quadrics, 
identified by their equations in canonical (i.e. simplest)
form. These are summarized in Fig.~\ref{QuadricSummary}, together with their
rank, the associated symmetry Lie algebra $\g$, and the type of twist deformation
we perform.
A plot of each class is given in Fig.~\ref{QuadricSurfaces}.
These classes make up 7 families of submanifolds, differing by the value of $c$.
In fact  classes (f), (g), (h)  altogether give a single family: (f) consists of connected manifolds,
the 1-sheeted hyperboloids;   (g), (h) of two-component manifolds, the
2-sheeted hyperboloids and the cone, which has two nappes separated by the apex (a singular point); all are closed, except the cone. For all families, except   (i) (consisting of ellipsoids), we succeed in   building  $U\g$-based Drinfel'd twists of 
either abelian (\ref{abeliantwist}) or Jordanian (\ref{Jordaniantwist}) type (depending on the coefficients of
the normal form) and through the latter in creating explicit twist deformations. 
%In particular, we are interested in the deformed relations of coordinate functions 
%and coordinate vector fields. We explicitly formulate them in all possible cases. 
Those twists are the simplest ones
resp. based on an abelian or $``ax+b"$ Lie subalgebra of the symmetry
Lie algebras.
Note that there are other choices of Drinfel'd twists on the $``ax+b"$-Lie algebra.
In particular we like to mention the twist of Theorem~2.10 of \cite{GiZh98},
which is the real  (i.e. $\F^{*\ot
*}=(S\ot S)[\F_{21}]$) counterpart  of the unitary Jordanian twist we utilize;
both twists  lead to the same commutation
relations. 
Since we are especially interested in describing the deformed spaces in terms of
deformed generators and relations, i.e. we intend to explicitly
calculate $\star$-commutators and the twisted Hopf algebra structures, we use
abelian and Jordanian twists, which admit an explicit exponential formulation.
Furthermore, all of the considered symmetry Lie algebras (except the one of the ellipsoids)
contain an abelian or $"ax+b"$ Lie subalgebra, which allows us to perform
a homogeneous deformation approach for all quadric surfaces.
We devote a subsection to each of the remaining six families of quadrics,
and a proposition to each twist deformation; propositions are proved in
the appendix. Throughout this section the star product $X\star h$
of a vector field $X$ by a function $h$ from the right is understood in the
$\Q_\star,\QM{}_\star$ sense   (see section \ref{Qstar}) \ $X\star h=X \ltlc_\star  h+X_\star(h)\equiv (\bR_1\trc h)\star (\bR_2\trc X) +X_\star(h)$.

\begin{figure}[h!]
\begin{center}
\begin{tabular}{|c|c|c|c|c|c|c|c|c|c|c|}
\hline
& $a_1$ & $a_2$ & $a_3$ & $a_{03}$ & $a_{00}$ & $r$ &  quadric  &$\mathfrak{g}\simeq$ & Abelian & Jordanian \\
\hline
(a) & $+$ & 0 & 0 & $-$ & & 3 &  parabolic cylinder & $\mathfrak{h}(1)$ & Yes & No \\
\hline
(b) & $+$ & $+$ & 0  & $-$ & & 4 &
elliptic paraboloid & $\mathfrak{so}(2)\ltimes\RR^2$ & Yes & No \\
\hline
(c) & $+$ & $+$ & 0  & 0 &  $-$ & 3 & elliptic cylinder & 
\begin{tabular}{c}
$\mathfrak{so}(2) \cross \RR^2$\\ 
$\mathfrak{so}(2) \times \RR$ 
\end{tabular}
& \begin{tabular}{c}
Yes\\ 
Yes
\end{tabular}
 &  \begin{tabular}{c}
No\\ 
No
\end{tabular} \\
\hline
(d) & $+$ & $-$ & 0 & $-$ &  & 4 & hyperbolic paraboloid  & $\mathfrak{so}(1,\!1)\!\ltimes\!\RR^2 $ %(?)
& Yes & Yes \\
\hline
(e) & $+$ & $-$ & 0 & 0 & $-$  & 3 & hyperbolic cylinder  & 
\begin{tabular}{c}
$\mathfrak{so}(1,\!1) \cross \RR^2$\\ 
$\mathfrak{so}(1,\!1) \!\times\! \RR$ 
\end{tabular}
& \begin{tabular}{c}
Yes\\ 
Yes
\end{tabular}
 &  \begin{tabular}{c}
Yes\\ 
No
\end{tabular} \\
\hline
(f) & $+$ & $+$ & $-$  & 0 & $-$ & 4 &  1-sheet  hyperboloid
& $\mathfrak{so}(2,1)$ & No & Yes \\
\hline
(g) & $+$ & $+$ & $-$  & 0 &+& 4 & 2-sheet hyperboloid
& $\mathfrak{so}(2,1)$ & No & Yes \\
\hline
(h) & $+$ & $+$ & $-$  & 0 & 0 & 3 &  elliptic cone$^\dagger$
& $\mathfrak{so}(2,\!1)\!\times\!\RR$ & Yes$^\dagger$ & Yes \\
\hline
(i) & $+$ & $+$ & $+$ & 0 & $-$ & 4 & ellipsoid & $\mathfrak{so}(3)$ & No & No \\
\hline
\end{tabular}
\end{center}
\caption{Overview of the quadrics in $\RR^3$: signs of the coefficients of the equations in canonical form
(if not specified, all $a_{00}\in\RR$ are possible), 
rank,  associated symmetry Lie algebra $\g$, type of twist deformation;
$\mathfrak{h}(1)$  stands for the Heisenberg algebra.
%: \  $[L_{23},\mathfrak{h}(1)]=0$,  \ $[L_{13},  L_{12}]=L_{23}$. 
For fixed $a_i$ each class gives a family of submanifolds $M_c$ parametrized by $c$,
except classes (f), (g), (h), which altogether give a single family; 
so there are 7 families of submanifolds. We can always make $a_1=1$
by a rescaling of $f$. The $\dagger$ reminds  that the
cone (e) is not a single closed manifold, due to the singularity in the apex;
we build an abelian twist for it using also the generator of dilatations.}
\label{QuadricSummary}
\end{figure}

\subsection{(a)   Family of parabolic cylinders\texorpdfstring{: $a_2\!=\!a_3\!=\!a_{01}\!=\!a_{02}\!=\!0$}{}}
 
Their equations in canonical form are parametrized by \
$c,\, b\!\equiv\! a_{03}\in\mathbb{R}$ and read
\be\label{ParCyleq}
    f_c(x):=\frac{1}{2}(x^1)^2-bx^3-c=0.
\ee
For every fixed $b$, $\{M_c\}_{c\in\RR}$ is a foliation of $\RR^3$.
The Lie algebra $\g$ is spanned by the  vector fields $L_{12}=x^1\partial_2$, 
$L_{13}=x^1\partial_3+b\partial_1$,
$L_{23}=b\partial_2$, which  fulfill
\bea
[L_{23},\g]=0, \qquad [L_{13},  L_{12}]=L_{23}. 
\eea
Clearly, $\g\simeq \mathfrak{h}(1)$, the Heisenberg algebra. 
The actions of the $L_{ij}$ on the $x^h,\xi^h,\partial_h$ are 
\bea
\ba{lll}
  L_{12}\trc x^i=   \delta^i_2 x^1,\qquad
& L_{13}\trc x^i=    \delta^i_1b+\delta^i_3x^1,
\qquad   & L_{23}\trc x^i=    \delta^i_2b , \\[8pt]
  L_{12}\trc \xi^i=   \delta^i_2 \xi^1,\qquad
& L_{13}\trc \xi^i=    \delta^i_3\xi^1,
\qquad   & L_{23}\trc \xi^i=    0 , \\[8pt]
L_{12}\trc\partial_i=
    -\delta_{i1}\partial_2,\qquad &L_{13}\trc \partial_i=
    -\delta_{i1}\partial_3,\qquad
  &  L_{23}\trc\partial_i=0;
\ea     \label{ParCyl-gaction}
\eea
%\bea\ba{lll}
%  [L_{12}, x^i]=   \delta^i_2 x^1,\qquad& [L_{13}, x^i]=    \delta^i_1b+\delta^i_3x^1,\qquad   & [L_{23},x^j]=    \delta^j_2b , \\[8pt]
%[L_{12},\partial_k]=    -\delta_{k1}\partial_2,\qquad &[L_{13}, \partial_k]=
 %   -\delta_{k1}\partial_3,\qquad   &  [L_{23},\partial_j]=0.
%\ea \eea
the commutation relations \
 $[L_{ij},x^h]=L_{ij}\trc x^h$, \ $[L_{ij},\partial_h]=L_{ij}\trc \partial_h$, \
$[L_{ij},\xi^h]=0$ \ hold in $\Q^\bullet$.
\begin{prop}\label{prop01'}
$\mathcal{F}=\exp(i\nu L_{13}\otimes L_{23})$ is a unitary abelian twist 
 inducing the following twisted deformations of $U\g$,
 of $\Q^\bullet$ on $\RR^3$
and of $\QM^\bullet$ on the parabolic cylinders (\ref{ParCyleq}). 
%$\beta=S(\beta)=\exp(-i\nu L_{13}L_{23})$.
The $U\g^\f$ counit, coproduct, antipode on the 
$\{L_{ij}\}_{1\leq i<j\leq 3}$ coincide with the undeformed
ones, except
\bea          \label{DeltaSParCyl}
\Delta_{\f}(L_{12})=L_{12}\otimes \1+\1\otimes L_{12} +i\nu L_{23}\otimes L_{23},
\qquad S_\f(L_{12})=-L_{23}+i\nu L_{23}^2.
\eea
The twisted star products %$L_{ij}\star L_{hk}=L_{ij}L_{hk}$ 
and Lie brackets of the $L_{ij}$ coincide with the untwisted ones. 
The twisted star products of the $L_{ij}$ with the %coordinate functions 
$x^i,\xi^i\equiv dx^i,\partial_i$, and those among the 
latter, equal their undeformed counterparts, except \
\ \  $L_{12}\star x^2=L_{12}  x^2-i\nu b\, L_{23}$, 
$$
\ba{ll}
    x^1\star x^2=x^1x^2-i\nu b^2,\qquad
      &x^3\star x^2    =x^2x^3-i\nu b x^1,\\[4pt]
\xi^3\star x^2    =\xi^3x^2-i\nu b \,\xi^1,\qquad
&\partial_1\star x^2=\partial_1x^2+ i\nu b\partial_3.
\ea
$$
Hence  the $\star$-commutation relations of the $U\g^\f$-equivariant 
$*$-algebra  $\Q^\bullet_\star$  read
\bea\ba{l}
 x^2\star x^1= x^1\star x^2+i\nu b^2,\qquad 
  x^3\star x^1=x^1\star x^3 ,\qquad
 x^3\star x^2= x^2\star x^3 -i\nu b\, x^1 ,\\[8pt]
 x^i\star \xi^j=\xi^j\star x^i+ \delta^i_2\,\delta_3^j\, i\nu\, b\, \xi^1,
\qquad \xi^i\star\xi^j+\xi^j\star\xi^i=0,\qquad 
\partial_i\star \xi^j=\xi^j\star \partial_i,\\[8pt]
\partial_j\star x^i=\delta^i_j\1+x^i\star  \partial_j+
\delta_{1j}\delta^i_2\: i\nu b\,\partial_3,\qquad \qquad 
\partial_i\star \partial_j=\partial_j\star \partial_i,\\[8pt]
L_{12}\star x^2=x^2\star L_{12}-i\nu b\, L_{23},\qquad
L_{ij}\star x^h = x^h\star L_{ij} +L_{ij}\trc x^h \quad\mbox{otherwise},\\[8pt]
 L_{ij}\star \partial_h =\partial_h\star  L_{ij}+ L_{ij}\trc \partial_h,\qquad  L_{ij}\star\xi^h= L_{ij}\star\xi^h.
%[L_{12}, x^2]_\star=x^1-i\nu b\, L_{23},\qquad
%[L_{ij},x^h]_\star=L_{ij}\trc x^h \quad\mbox{otherwise},\\[8pt]
% [L_{ij},\partial_h]_\star=L_{ij}\trc \partial_h,\qquad\qquad [L_{ij},\xi^h]_\star=0,
\ea\eea
In terms of star products \ $L_{12}=x^1\star\partial_2$, \
$L_{13}=x^1\star\partial_3+b\partial_1$,\
$L_{23}=b\partial_2$. \ Also  the relations  characterizing the $U\g^\f$-equivariant 
$*$-algebra  $\QMst^\bullet$, i.e. equation  (\ref{ParCyleq}), %of the parabolic cylinder 
 its differential   and the linear dependence relations,
 %(\ref{DepRel})$_2$,
%$\epsilon^{ijk} f_i L_{jk}=0$  (sum over repeated indices)
  keep the same form:
\bea
f_c(x)\equiv\frac{1}{2}x^1\star x^1-bx^3-c=0,\qquad 
d f_c \equiv x^1\star\xi^1-b \,\xi^3=0,\qquad
%,\\[4pt] &&L_{12}=x^1\star\partial_2,\qquad L_{13}=x^1\star\partial_3+b\partial_1,\qquad 
%L_{23}=b\partial_2\\[6pt] &&
\epsilon^{ijk} f_i \star L_{jk}=0. 
\eea
The $*$-structures on $U\g^\f$, $\Q^\bullet_\star,\QMst^\bullet$ remain 
undeformed.
\end{prop}

\noindent
Alternatively,  one could twist everything by the unitary abelian twist $\mathcal{F}=\exp(i\nu L_{12}\otimes L_{23})$. 
%$\star$-relations are simpler in the latter

%
%\begin{align*}     0=&    b x^1\star\frac{\partial}{\partial x^2}\\
%    &    -b x^1\star\frac{\partial}{\partial x^2}\end{align*}
%

\subsection{(b)   Family of elliptic paraboloids\texorpdfstring{: $a_2>0$, $a_3=0$, $a_{03}<0$}{}}

Their equations in canonical form are parametrized by 
$a=a_2,c=-a_{00}\in\RR$, $b=-a_3\in\RR^+$ and read
\be\label{eq18}
    f_c(x):=\frac{1}{2}\big[(x^1)^2+a(x^2)^2\big]-bx^3-c=0.
\ee
For every fixed $a,b$, $\{M_c\}_{c\in\RR}$ is a foliation of $\RR^3$.
The  vector fields $L_{12}=x^1\partial_2-ax^2\partial_1$, 
$L_{13}=x^1\partial_3+b\partial_1$,
$L_{23}=a x^2\partial_3+b\partial_2$  fulfill

\bea\label{eq04'}
[L_{12},L_{13}]=-L_{23}, \qquad [L_{12},L_{23}]=aL_{13}, 
\qquad  [L_{13},  L_{23}]=0. 
\eea
Clearly,  \ $\g\simeq \mathfrak{so}(2)\cross \RR^2$. \
The actions of the $L_{ij}$ on the $x^h,\xi^h,\partial_h$ are given by
\bea
&&\ba{l}
L_{12}\trc \partial_i=\delta_{i2}a\,\partial_1-\delta_{i1}\partial_2,\qquad
L_{12}\trc  u^i=   \delta^i_2 u^1 -\delta^i_1 a\, u^2,\qquad\: \mbox{for } 
u^i=x^i,\xi^i,\ea\label{EllParL12}\qquad\\ [8pt]
&&\ba{lll}
L_{13}\trc  \partial_i=    -\delta_{i1}\partial_3, \qquad\quad &
L_{13}\trc  x^i=  \delta^i_3x^1+b\delta^i_1,
  \qquad &L_{13}\trc  \xi^i=  \delta^i_3\xi^1,\\ [6pt]
 L_{23}\trc \partial_i=-\delta_{i2}a\partial_3, \qquad\qquad 
& L_{23}\trc x^i=\delta^i_3 a x^2+b\delta^i_2 \qquad\:\: & L_{23}\trc \xi^i=\delta^i_3 a \xi^2;
\ea\label{EllParL13L23}
 \eea
the commutation relations \
 $[L_{ij},x^h]=L_{ij}\trc x^h$, \ $[L_{ij},\partial_h]=L_{ij}\trc \partial_h$, \
$[L_{ij},\xi^h]=0$ \ hold in $\Q^\bullet$.

%%%%%%%%%%%%%
\begin{comment}
%\bea 
%&&  [L_{12}, x^i]=   \delta^i_2 x^1 -\delta^i_1 a x^2,\qquad\qquad 
%[L_{12},\partial_k]=\delta_{k2}a\partial_1-\delta_{k1}\partial_2,\label{EllCylL12}\\[8pt]
%&&[L_{13}, x^i]=  \delta^i_3x^1,\quad    [L_{23},x^i]=    \delta^i_3 a x^2 , 
%\quad [L_{13}, \partial_k]=    -\delta_{k1}\partial_3,\quad
%    [L_{23},\partial_j]=-\delta_{k2}a\partial_3.\qquad \quad 
% \eea
\bea
\ba{l}
L_{12}\trc  x^i=   \delta^i_2 x^1 -\delta^i_1 a\, x^2,\\ [4pt]
L_{12}\trc  \xi^i=   \delta^i_2 \xi^1 -\delta^i_1 a\, \xi^2,\\ [4pt]
L_{12}\trc \partial_i=\delta_{i2}a\,\partial_1-\delta_{i1}\partial_2,
\ea
\eea
\bea
\ba{ll}
L_{13}\trc  x^i=  \delta^i_3x^1+b\delta^i_1
, \qquad\qquad  & L_{23}\trc x^i=\delta^i_3 a x^2+b\delta^i_2, \\ [6pt]
L_{13}\trc  \xi^i= \delta^i_3\xi^1, \qquad\qquad & L_{23}\trc \xi^i=\delta^i_3 a \xi^2, \\ [6pt]
L_{13}\trc  \partial_i=    -\delta_{i1}\partial_3,\qquad\qquad
  &  L_{23}\trc \partial_i=-\delta_{i2}a\partial_3;
\ea
 \eea
the commutation relations \
 $[L_{ij},x^h]=L_{ij}\trc x^h$, \ $[L_{ij},\partial_h]=L_{ij}\trc \partial_h$, \
$[L_{ij},\xi^h]=0$ \ hold in $\Q^\bullet$.
%\bea 
%&&  [L_{12}, x^i]=   \delta^i_2 x^1 -\delta^i_1 a x^2,\qquad\qquad 
%[L_{12},\partial_k]=\delta_{k2}a\partial_1-\delta_{k1}\partial_2,\label{EllCylL12}\\[8pt]
%&&[L_{13}, x^i]=  \delta^i_3x^1,\quad    [L_{23},x^i]=    \delta^i_3 a x^2 , 
%\quad [L_{13}, \partial_k]=    -\delta_{k1}\partial_3,\quad
%    [L_{23},\partial_j]=-\delta_{k2}a\partial_3.\qquad \quad 
% \eea
\end{comment}
%%%%%%%%%%%%%%%%%%%%%%%%%%%%%%

\begin{prop}\label{prop07}
$\F=\exp(i\nu L_{13}\otimes L_{23})$  is a unitary abelian twist 
 inducing the following twisted deformation of $U\g$,
 of $\Q^\bullet$ on $\RR^3$
and of $\QM^\bullet$  on the elliptic paraboloids (\ref{eq18}). 
%$\beta=S(\beta)=\exp(-i\nu L_{13}L_{23})$.
The $U\g^\f$ counit, coproduct, antipode on the 
$\{L_{ij}\}_{1\leq i<j\leq 3}$ coincide with the undeformed
ones, except
\bea
\ba{l}
\Delta_{\f}(L_{12})=L_{12}\otimes \1+\1\otimes L_{12} +i\nu \left(L_{23}\otimes L_{23}-aL_{13}\otimes L_{13}\right),\\[8pt]
S_\f(L_{12})=-L_{12}+i\nu \left(L_{23}^2-a L_{13}^2\right).
\ea        \label{DSEllypticParab}
\eea
The twisted star products %$L_{ij}\star L_{hk}=L_{ij}L_{hk}$ 
and Lie brackets of the $\{L_{ij}\}_{1\leq i<j\leq 3}$ coincide with the untwisted ones except
$L_{12}\star L_{12}=L_{12}^2+i\nu a L_{23}L_{13}$.
The twisted star products of the $L_{ij}$ with the %coordinate functions 
$x^i,\xi^i\equiv dx^i,\partial_i$, and those among the 
latter, equal their undeformed counterparts, except 
\bea
\ba{ll}
L_{12}\star u^3=L_{12}  u^3-i\nu a\, L_{23} u^2,\qquad\qquad
& u^3\star L_{12}=u^3  L_{12}+i\nu a\, u^1 L_{13} ,\\[6pt]
  L_{12}\star x^2 =L_{12}x^2-i\nu bL_{23} 
\qquad &  x^1\star L_{12}    =x^1L_{12}+i\nu abL_{13}, \\[6pt]
L_{12}\star \partial_2=L_{12}  \partial_2+i\nu a\, L_{23} \partial_3,\qquad
& \partial_1\star L_{12}=\partial_1 L_{12}-i\nu a\, \partial_3 L_{13},\\[6pt]
  x^1\star x^2 =x^1x^2-i\nu b^2,\qquad\qquad &
   x^1\star x^3 =x^1x^3-i\nu abx^2,\\[6pt]
    x^3\star x^3=x^3x^3-i\nu a  x^1 x^2-ab^2\frac{\nu^2}{2},\qquad &
     x^3\star x^2     =x^3x^2-i\nu bx^1,\\[6pt]
  x^1\star\xi^3=x^1\xi^3-i\nu ab\xi^2,\qquad &
 x^3\star \xi^3=x^3\xi^3-i\nu a\,  x^1 \xi^2,\\[6pt]
  \xi^3\star x^2 =\xi^3x^2-i\nu b\xi^1,\qquad &
\xi^3\star x^3=\xi^3x^3-i\nu a\,  \xi^1 x^2,\\[6pt]
    \xi^3\star\xi^3=-i\nu a\xi^1%\wedge
\xi^2,\qquad &
\partial_1\star\partial_2=\partial_1\partial_2-i\nu a\,\partial_3\partial_3,\\[6pt]
    x^1\star\partial_2 =x^1\partial_2+i\nu ab\partial_3,\qquad &
 x^3\star\partial_2 =x^3\partial_2+ i\nu a\,x^1\partial_3,\\[6pt]
  \partial_1\star x^2 =\partial_1x^2+i\nu b\partial_3, \qquad &
\partial_1\star x^3=\partial_1x^3+ i\nu a\,\partial_3x^2,
\ea                \label{starprodEllypticParab}
\eea
%%%%%%%%%%%%%%%%%%%%%%
\begin{comment}
$$
\ba{ll}
    x^3\star x^3=x^3x^3-i\nu a  x^1 x^2-ab^2\frac{\nu^2}{2},\qquad
      &  \xi^3\star x^3=\xi^3x^3-i\nu a\,  \xi^1 x^2,\\[6pt]
 x^3\star \xi^3=x^3\xi^3-i\nu a\,  x^1 \xi^2,\qquad
&\partial_1\star\partial_2=\partial_1\partial_2-i\nu a\,\partial_3\partial_3,\\[6pt]
\partial_1\star x^3=\partial_1x^3+ i\nu a\,\partial_3x^2,
\qquad & x^3\star\partial_2 =x^3\partial_2+ i\nu a\,x^1\partial_3,\\[6pt]
L_{12}\star x^3=L_{12}  x^3-i\nu a\, L_{23} x^2,\qquad\qquad
& x^3\star L_{12}=x^3  L_{12}+i\nu a\, x^1 L_{13} ,\\[6pt]
L_{12}\star \xi^3=L_{12}  \xi^3-i\nu a\, L_{23} \xi^2,\qquad
& \xi^3\star L_{12}=\xi^3  L_{12}+i\nu a\, L_{13} \xi^1,\\[6pt]
L_{12}\star \partial_2=L_{12}  \partial_2+i\nu a\, L_{23} \partial_3,\qquad
& \partial_1\star L_{12}=\partial_1 L_{12}-i\nu a\, \partial_3 L_{13},
\ea
$$
and
%
\begin{align*}
\begin{split}
    x^1\star x^3
    =&x^1x^3-i\nu abx^2,\\
    x^3\star x^2
    =&x^3x^2-i\nu bx^1,\\
    x^1\star x^2
    =&x^1x^2-i\nu b^2,\\
    x^1\star\xi^3
    =&x^1\xi^3-i\nu ab\xi^2,\\
    x^1\star L_{12}
    =&x^1L_{12}+i\nu abL_{13},
\end{split}
\begin{split}
    \xi^3\star x^2
    =&\xi^3x^2-i\nu b\xi^1,\\
    x^1\star\partial_2
    =&x^1\partial_2+i\nu ab\partial_3,\\
    \partial_1\star x^2
    =&\partial_1x^2+i\nu b\partial_3,\\
    \xi^3\star\xi^3
    =&-i\nu a\xi^1\wedge\xi^2,\\
    L_{12}\star x^2
    =&L_{12}x^2-i\nu bL_{23}.
\end{split}
\end{align*}
\end{comment}
%%%%%%%%%%%%%%%%%%%%%%%%%
where $u^i=x^i,\xi^i$. Hence the $\star $-commutation relations of the $U\g^\f$-equivariant 
algebra $\Q_\star$ read
\bea\ba{l}
% x^i\star x^j = x^j\star x^i +i\nu\bigg( (a\delta^i_3x^2+b\delta^i_2)\star
% (\delta^j_3x^1+b\delta^j_1) -(a\delta^j_3x^2+b\delta^j_2)\star
% (\delta^i_3x^1+b\delta^i_1) \bigg),\\[6pt]
 x^1\star x^2=x^2\star x^1 -i\nu b^2,\quad
x^1\star x^3=x^3\star x^1 -i\nu ab \, x^2,\quad
x^2\star x^3 =x^3\star x^2 + i\nu b\, x^1,\\[6pt]
x^i\star \xi^j=
\xi^j\star x^i
+i\nu\delta^j_3\bigg(
\xi^1\star(a\delta^i_3x^2+b\delta^i_2)
-\xi^2\star (\delta^i_3x^1+b\delta^i_1) a
\bigg),\\[6pt]
 \xi^i\star \xi^j+\xi^j\star \xi^i=-\delta_3^j\,\delta^i_3\,i2\nu a\,  
\xi^1\star  \xi^2, \qquad
\partial_i\star \partial_j=\partial_j\star \partial_i-\delta_{1i}\,\delta_{2j}\,i\nu a\,
\partial_3\star\partial_3,\\[6pt]
\partial_j\star x^i=\delta^i_j\1+x^i\star  \partial_j
+i\nu\bigg(
\delta_{j1}(a\delta^i_3x^2+b\delta^i_2)
-a\delta_{j2}(\delta^i_3x^1+b\delta^i_1) 
\bigg)\star\partial_3,\\[6pt]
\partial_i\star \xi^j=\xi^j\star \partial_i+
\delta^i_3 i\nu a\,(\delta_{1j} \xi^2-\delta_{2j} \xi^1)\star\partial_3,
\ea                    \label{starcomEllypticParab1}
\eea
%except
%\bea\ba{l}  x^3\star \xi^3=\xi^3\star x^3+i\nu a\,  (\xi^1\star  x^2
%-\xi^2 \star x^1),\qquad  \xi^3\star \xi^3=-i\nu a\,  \xi^1\star  \xi^2, \\[6pt]
%\partial_1\star \partial_2=\partial_2\star \partial_1-i\nu a\,
%\partial_3\star\partial_3,\\[6pt]
%\partial_j\star x^i=\delta^i_j\1+x^i\star  \partial_j+
%\delta^i_3 i\nu a\,(\delta_{1j} x^2-\delta_{2j} x^1)\star\partial_3,\\[6pt]
%\partial_i\star \xi^j=\xi^j\star \partial_i+
%\delta^i_3 i\nu a\,(\delta_{1j} \xi^2-\delta_{2j} \xi^1)\star\partial_3.
%\ea\eea
while those among the tangent vectors $L_{ij}$ and the generators 
$x^i,\xi^i,\partial_i$ read
\bea
\ba{l}
L_{12}\star x^i=L_{12}\trc x^i+x^i\star L_{12} -i\nu b \, 
(a\delta^i_1 L_{13}+\delta^i_2 L_{23}+a\delta^i_3)-i\nu a \delta^i_3 \, 
(x^1\star L_{13}+x^2\star L_{23}),\\[6pt]
L_{12}\star \xi^i=\xi^i\star L_{12}-i\nu a \delta^i_3 \, 
(\xi^1\star L_{13}+\xi^2\star L_{23}),\\[6pt]
 L_{12}\star \partial_i=L_{12} \trc \partial_i+\partial_i\star L_{12}
+i\nu a\, \partial_3\star L_{i3} ,\\[6pt]
L_{j3}\star x^i=L_{j3}\trc x^i+x^i\star L_{j3},\qquad   
L_{j3}\star \xi^i=\xi^i\star L_{j3}, \qquad j=1,2,\\[6pt]
L_{j3}\star \partial_i=L_{j3}\trc \partial_i+\partial_i\star L_{j3},
\qquad \qquad \qquad \qquad\qquad \qquad j=1,2.
\ea  \label{starcomEllypticParab2}
\eea
In terms of star products \ $L_{12}=\partial_2\star x^1-ax^2\star\partial_1$, \
$L_{13}=x^1\star\partial_3+b\partial_1$,\
$L_{23}=a x^2\star\partial_3+b\partial_2$. \ Also the relations  characterizing the $U\g^\f$-equivariant 
$*$-algebra  $\QMst^\bullet$, i.e. equation  (\ref{eq18}), %of the parabolic cylinder 
 its differential   and the linear dependence relations 
 %(\ref{DepRel})$_2$,
  keep the same form
\bea \label{characterizingEllPar}
f_c(x)\equiv\frac{1}{2}(x^1\!\star\! x^1\!+\!a x^2\!\star\! x^2)\!-\!bx^3\!-\!c=0,
df\equiv \xi^1\!\star\! x^1\!+\!a\, \xi^2\!\star\! x^2\!-\!b\xi^3=0,
 \epsilon^{ijk} f_i \star L_{jk}=0.
\eea
The $*$-structures on $U\g^\f$, $\Q^\bullet_\star, \QMst^\bullet$ remain undeformed.
\end{prop}

\subsection{(c)   Family of elliptic cylinders\texorpdfstring{:  $a_2\!>0$, $a_3\!=\!a_{0i}\!=\!0$, $a_{00}\!<\!0$}{}}

Their equations in canonical form are parametrized by \
$c,\, a\equiv a_2\in\mathbb{R}^+$ and read
\be\label{EllCyleq}
    f_c(x):=\frac{1}{2}\big[(x^1)^2+a(x^2)^2\big] -c=0.
\ee
For every $a>0$, $\{M_c\}_{c\in\RR^+}$ is a foliation of $\RR^3\setminus \vec{z}$,
where $\vec{z}$ is the axis $x^1\!=\!x^2\!=\!0$.
Eq. (\ref{EllCyleq}) can be obtained from the one (\ref{eq18}) characterizing the
elliptic paraboloids (b) setting $b=0$. Hence also
the tangent vector fields $L_{ij}$, their commutation relations, 
their actions on the $x^h,\xi^h,\partial_h$, 
the commutation relations of the $L_{ij}$ with the $x^h,\xi^h,\partial_h$
can be obtained from the ones of case (b)
by setting $b=0$. The $L_{ij}$   fulfill again (\ref{eq04'}), so that
 \ $\g\simeq \mathfrak{so}(2)\cross \RR^2$. \
Hence we can deform all objects with the same abelian
twist as in (b), and obtain the corresponding results:

\begin{prop}
$\F=\exp(i\nu L_{13}\otimes L_{23})$ is a unitary abelian twist 
 inducing  the twisted deformation of $U\g$,  of $\Q^\bullet$ on $\RR^3$
and of $\QM^\bullet$  on the elliptic cylinders (\ref{EllCyleq}) which is
obtained by setting $b=0$ in Proposition \ref{prop07}. 
\end{prop}

This is essentially the same as Proposition 15 in \cite{FioreWeber2020}.
Alternatively, as a complete set in $\Xi_t$ instead of $\{L_{12},L_{13},L_{23}\}$
we can use $S_t=\{L_{12},\partial_3\}$, which is actually a basis of $\Xi_t$;
the Lie algebra $\g\simeq \mathfrak{so}(2)\times \RR$ generated by the latter is abelian; 
the relevant  relations are (\ref{EllParL12})$_{b=0}$,
$$
L_{12}\trc \partial_i=\delta_{i2}a\,\partial_1-\delta_{i1}\partial_2,\qquad\quad
L_{12}\trc  u^i=   \delta^i_2 u^1 -\delta^i_1 a\, u^2,\quad\: \mbox{for } u^i\in\{x^i,\xi^i\}, \qquad \eqno{(\ref{EllParL12})_{b=0} }
%\label{EllCylL12}
$$
and
\be
%[\partial_3,x^i]=
\partial_3\trc x^i\equiv \partial_3(x^i)=   \delta^i_3\1 , \qquad  \partial_3\trc\partial_i=[\partial_3,\partial_i]=0, \qquad  \partial_3\trc L_{12}=[\partial_3,L_{12}]=0.
\ee
We correspondingly adopt   the unitary abelian twist $\F=\exp(i\nu \partial_3\otimes  L_{12})$.

\bigskip
\noindent
%\begin{prop}\label{prop03}
{\bf Proposition 16 in \cite{FioreWeber2020}}. \
{\it $\F=\exp(i\nu \partial_3\otimes L_{12})$   is a unitary abelian twist
 inducing  the following twist  deformation of $U\g$,
 of $\Q^\bullet$ on $\RR^3$
and of $\QM^\bullet$  on the elliptic cylinders (\ref{EllCyleq}). 
%$\beta=S(\beta)=\exp(-i\nu L_{13}L_{23})$.
The $U\g^\f$ counit, coproduct, antipode on $\{\partial_3,L_{12}\}$
coincide with the undeformed ones.
The twisted star products %$L_{ij}\star L_{hk}=L_{ij}L_{hk}$ 
and Lie brackets of $\{\partial_3,L_{12}\}$ coincide with the untwisted ones. 
The twisted star products of $\partial_3,L_{12}$ with  %coordinate functions 
$x^i,\xi^i\equiv dx^i,\partial_i$, and those among the 
latter, equal the untwisted ones, except 
\begin{align*}
\begin{split}
    x^3\star x^1=&x^1x^3+i\nu a x^2,\\
    x^3\star\xi^1=&x^3\xi^1+i\nu a\xi^2,\\
    x^3\star\partial_1=&x^3\partial_1+i\nu\partial_2,
\end{split}
\begin{split}
    x^3\star x^2=&x^2x^3-i\nu x^1,\\
    x^3\star\xi^2=&x^3\xi^2-i\nu\xi^1,\\
    x^3\star\partial_2=&x^3\partial_2-i\nu a\partial_1.
\end{split}
\end{align*}
Hence the $\star $-commutation relations of the $U\g^\f$-equivariant 
algebra $\Q_\star$ read
\begin{equation}\label{eq07}
\begin{split}
    x^i\star x^j
    =&x^j\star x^i+i\nu\delta^i_3(\delta^j_1ax^2-\delta^j_2x^1)
    -i\nu\delta^j_3(\delta^i_1ax^2-\delta^i_2x^1),\\
    x^i\star\xi^j
    =&\xi^j\star x^i+i\nu\delta^i_3(\delta^j_1a\xi^2-\delta^j_2\xi^1),\\
    x^i\star\partial_j
    =&-\delta^i_j\1
    +\partial_j\star x^i
    +i\nu\delta^i_3(\delta^j_1\partial_2-\delta^j_2a\partial_1),\\
    \xi^i\star\xi^j
    =&-\xi^j\star\xi^i,\\
    \xi^i\star\partial_j
    =&\partial_j\star\xi^i,\\
    \partial_i\star\partial_j
    =&\partial_j\star\partial_i.
\end{split}
\end{equation}
%
%except
%\bea\ba{l}  x^3\star \xi^3=\xi^3\star x^3+i\nu a\,  (\xi^1\star  x^2
%-\xi^2 \star x^1),\qquad  \xi^3\star \xi^3=-i\nu a\,  \xi^1\star  \xi^2, \\[6pt]
%\partial_1\star \partial_2=\partial_2\star \partial_1-i\nu a\,
%\partial_3\star\partial_3,\\[6pt]
%\partial_j\star x^i=\delta^i_j\1+x^i\star  \partial_j+
%\delta^i_3 i\nu a\,(\delta_{1j} x^2-\delta_{2j} x^1)\star\partial_3,\\[6pt]
%\partial_i\star \xi^j=\xi^j\star \partial_i+
%\delta^i_3 i\nu a\,(\delta_{1j} \xi^2-\delta_{2j} \xi^1)\star\partial_3.
%\ea\eea
In terms of star products \ $L_{12}=x^1\star\partial_2-ax^2\star\partial_1$. 
 \ Also the relations  characterizing the $U\g^\f$-equivariant 
$*$-algebra  $\QMst^\bullet$, i.e. eq.  (\ref{EllCyleq}), 
 its differential   and %the linear dependence relations,
 (\ref{DepRel}),
  keep the same form:
\bea\label{eq08}
f_c(x)\equiv\frac{1}{2}(x^1\!\star\! x^1+a x^2\!\star\! x^2)-c=0,\quad
df_c\equiv \xi^1\!\star\! x^1+a\, \xi^2\!\star\! x^2=0, \quad \epsilon^{ijk} f_i \star L_{jk}=0.
\eea
The $*$-structures on $U\g^\f$, $\Q^\bullet_\star, \QMst^\bullet$ remain undeformed.
\label{prop03}}
%\end{prop}

\subsubsection{Circular cylinders embedded in Euclidean $\RR^3$}

If  $a_1\!=\!a_2\!=\!1$, i.e. \ $f_c(x)=\frac{1}{2}\big[(x^1)^2+(x^2)^2\big]-c=0$ \ 
and we endow $\RR^3$ with the Euclidean metric
(circular cylinder of radius $R=\sqrt{2c}$), then
$S:=\{L,\partial_3,\Np\}$ is an orthonormal basis of $\Xi$  alternative to $S'\!:=\!\{\partial_1,\partial_2,\partial_3\}$
and such that $S_t\!:=\!\{L,\partial_3\}$, $S_{\scriptscriptstyle \perp}\!:=\!\{\Np\}$
%$\gm\left(\Vp,\Vp\right)=(x^1)^2\!+\!(x^2)^2=2c$,
are orthonormal bases of $\Xi_t$, $\Xi_{\scriptscriptstyle \perp}$  respectively;
here  $L:=L_{12}/R$, $\Np=f^i\partial_i/R=(x^1\partial_1+x^2\partial_2)/R$
({\it outward} normal). 
The Killing Lie algebra $\k$ is abelian and spanned (over $\RR$) by $S_t$. \ 
 $\nabla_XY=0$ for all $X,Y\!\in\! S'$, whereas the only non-zero$\nabla_XY$, with $X,Y\in S$ are
\be
\nabla_LL=-\frac 1 R \Np,\qquad\nabla_L\Np=\frac 1 R L,\qquad
\nabla_{\Np}L=\frac 1 R L,\qquad\nabla_{\Np}\Np=\frac 1 R\Np.
\ee
%In terms of the decomposition $ Z=\tilde Z L+Z^3\partial_3$ of the generic $Z\in \Xi_t$ 
The second fundamental form \ $II(X,Y)=\left( \nabla_X Y\right)_{\perp}$, \  $X,Y\in \Xi_t$, \ 
is thus explicitly given by
\bea
II(X,Y)&=&
%\gm\left( \nabla_X Y,\Np\right)\Np  =
%\gm\left(X^i\partial_i(Y^j)\partial_j,f_h\partial_h\right) \frac{\Vp}{2c}=
% X^i\partial_i(Y^j)f_j \frac{\Vp}{2c} \nn &\!\!\stackrel{Y^jf_j=0}{=}\!\! &
%-X^iY^j \partial_i (f_j) \frac{\Vp}{2c}=-\left( X^1Y^1 +X^2Y^2\right)  \frac{\Vp}{2c}=
-\frac{\tilde X\,\tilde Y }{R}\,\Np;
\eea
here we are using the decomposition 
$ Z=\tilde Z L+Z^3\partial_3$ of a generic $Z\in \Xi_t$.
%, which implies $Z^1=-\tilde Z\frac{x^2}{R}$, $Z^2=\tilde Z \frac{x^1}{R}$, 
% $X^1Y^1\!+\!X^2Y^2=\tilde X\tilde Y\big[(x^1)^2\!+\!(x^2)^2\big]/R^2=\tilde X\tilde Y$.
Thus $II$ is diagonal in the basis $S_t$,
with diagonal elements (i.e. principal curvatures) \ $\kappa_1=0, \kappa_2=-1/R$. \
Hence the Gauss (i.e. intrinsic) curvature $K=\kappa_1\kappa_2$ vanishes; 
$\rR_{t}=0$ easily follows also from $\rR=0$ using the Gauss theorem. The mean (i.e. extrinsic) curvature is $H=(\kappa_1\!+\!\kappa_2)/2=-1/2R$. \ 
%Signs ok?
The Levi-Civita covariant derivative $\nabla\!_t$ on $M_c$ is the tangent projection of $\nabla$ 
$$
\nabla\!_{t,X}Y=\mathrm{pr}_t(\nabla_XY)=\nabla_XY-II(X,Y)=\nabla_XY +\tilde X\,\tilde Y \,\Np/R.
$$
The deformation via the abelian twist $\F=\exp(i\nu \partial_3\otimes  L_{12})\in U\k\otimes U\k[[\nu]]$  yields
\bea
    &&\nabla^\f_{X} =\nabla^{}_X\qquad\qquad  \forall\, X\in S\cup S'=
\{\partial_1,\partial_2,\partial_3,L,\Np\}, \label{eq09}\\[6pt]
   &&\nabla\!_{t,X}^{\,\f}Y = \mathrm{pr}_t(\nabla_XY)=\nabla\!_{t,X}Y\qquad  \forall\, X,Y\in S_t=
\{\partial_3,L\},  \label{eq09'}
\eea
because $\partial_3$ commutes with all such $X$,
so that $\bF_1\trc X\otimes\bF_2=X\otimes\1$, \ and the projections $\mathrm{pr}_{\scriptscriptstyle \perp},\mathrm{pr}_t,$ stay undeformed, as shown in 
Proposition \ref{prop08}. 
\ Eq. (\ref{eq09}-\ref{eq09'}) determine $ \nabla^\f_{X}Y$
for all $X,Y\in\Xi_\star$ and  $\nabla\!_{t,X}^{\,\f}Y=\nabla\!_{t,X}Y$ for all $X,Y\in\Xi_{t\star}$ via  the function left $\star$-linearity
in $X$ and the deformed Leibniz rule for $Y$.  The twisted curvatures $\rR^{\f},\rR^{\f}_t$ vanish, by Theorem 7 in \cite{AschieriCastellani2009}.
Furthermore, 
\begin{equation}\label{II^F=II}
    II^\f_\star(X,Y)
    \stackrel{(\ref{II^F})}{=}II(\mathcal{F}_1^{-1}\rhd X,\mathcal{F}_2^{-1}\rhd Y)
   % =\gm(\nabla_{\mathcal{F}_1^{-1}\rhd X}(\mathcal{F}_2^{-1}\rhd Y),\Np)
    \Np    =II(X,Y)
\end{equation}
for all $X,Y\in S_t$, leading to the same principal curvatures  
\ $\kappa_1=0, \kappa_2=1/R$, \
Gauss and mean curvatures as in the undeformed case.

\subsection{(d) Family of   hyperbolic paraboloids\texorpdfstring{: $a_2,a_{03}<0$, $a_3=0$}{}}

Their equations in canonical form are parametrized by 
$a=-a_2,b=-a_{03}>0$, $c=-a_{00}\in\RR$ and read
\be\label{HypPareq}
    f_c(x):=\frac{1}{2}\big[(x^1)^2-a(x^2)^2\big] -bx^3-c=0.
\ee
For all fixed $a,b>0$, $\{M_c\}_{c\in\RR}$ is a foliation of $\RR^3$.
The Lie algebra $\g$ is spanned by the  vector fields
 $L_{12}=x^1\partial_2+ax^2\partial_1$, 
$L_{13}=x^1\partial_3+b\partial_1$,
$L_{23}=b\partial_2-a x^2\partial_3$, which  fulfill
\bea\label{HypParg}
[L_{12},L_{13}]=-L_{23}, \qquad [L_{12},L_{23}]=-aL_{13}, 
\qquad  [L_{13},  L_{23}]=0, 
\eea
whence 
 \ $\g\simeq \mathfrak{so}(1,1)\cross \RR^2$. \  
The abelian twist deformation is entirely similar to the one of (b): just
replace $a$ by $-a$ in the equations of Proposition~\ref{prop07}.

In addition, there is also a Jordanian twist deformation on the hyperbolic paraboloid
which we are going to discuss in detail. The tangent vector fields
$H=-\frac{2}{\sqrt{a}}L_{12}$, $E=L_{13}+\frac{1}{\sqrt{a}}L_{23}$,
$E'=L_{13}-\frac{1}{\sqrt{a}}L_{23}$ fulfill the commutation relations
\begin{equation}\label{eq11}
    [H,E]=2E,\quad
    [H,E']=-2E',\quad
    [E,E']=0.
\end{equation} 
%Remark that $L_{13}=\frac{1}{2}(E+E')$ and $L_{23}=\frac{\sqrt{a}}{2}(E-E')$. 
To compute the action of $\mathcal{F}$ on  functions it is convenient 
to adopt the eigenvectors of $H$
\begin{equation} \label{ycoord}
    y^1=x^1-\sqrt{a}x^2,\qquad
    y^2=x^1+\sqrt{a}x^2,\qquad
    y^3=x^3,
\end{equation}
as new coordinates.  In fact, $H\rhd y^i=\lambda_iy^i$ with $\lambda_1=2$,
$\lambda_2=-2$ and $\lambda_3=0$.
Abbreviating  $\tilde{\partial}_i=\frac{\partial}{\partial y^i}$, 
%$\tilde{\partial}^1=\tilde{\partial}_2$, $\tilde{\partial}^2=\tilde{\partial}_1$, 
the inverse
coordinate and  the partial derivatives transformations read%
\be
\ba{lll}\label{ycoord'}
      x^1=\frac{1}{2}(y^1+y^2) ,\qquad
     &  \tilde{\partial}_1
     =\frac{1}{2}(\partial_1-\frac{1}{\sqrt{a}}\partial_2) ,\qquad
     &  \partial_1=\tilde{\partial}_1+\tilde{\partial}_2 , \\[6pt]
      x^2=\frac{1}{2\sqrt{a}}(y^2-y^1) ,\qquad
     &  \tilde{\partial}_2
     =\frac{1}{2}(\partial_1+\frac{1}{\sqrt{a}}\partial_2) ,\qquad
     &  \partial_2=\sqrt{a}(\tilde{\partial}_2-\tilde{\partial}_1) , \\[6pt]
      x^3=y^3 ,\qquad
     &  \tilde{\partial}_3=\partial_3 ,\qquad
     &  \partial_3=\tilde{\partial}_3 .
\ea\ee
In the new coordinates $f_c(y)=\frac{1}{2}y^1y^2-by^3-c$  and 
%$$ L_{12}=\sqrt{a}(y^2\tilde{\partial}_2-y^1\tilde{\partial}_1),~
%L_{13}=\frac{1}{2}(y^1+y^2)\tilde{\partial}_3
%+b(\tilde{\partial}_1+\tilde{\partial}_2),~
%L_{23}=\frac{\sqrt{a}}{2}(y^1-y^2)\tilde{\partial}_3
%+b\sqrt{a}(\tilde{\partial}_2-\tilde{\partial}_1),$$
%
$$
H=2(y^1\tilde{\partial}_1-y^2\tilde{\partial}_2), \qquad
E=y^1\tilde{\partial}_3+2b\tilde{\partial}_2, \qquad
E'=y^2\tilde{\partial}_3+2b\tilde{\partial}_1.
$$ 
%
%\begin{equation}
%    [L_{12},L_{23}]=-aL_{13},~
%    [L_{12},L_{13}]=-L_{23},~
%    [L_{13},L_{23}]=0,
%\end{equation}
%
%and
%
The actions of $H,E,E'$ on coordinate functions,
differential forms $\eta^i=dy^i$ and vector fields are given by
for all $1\leq i\leq 3$
\begin{equation} \label{gQaction}
\begin{split}
    H\rhd y^i =&\lambda_iy^i,\\[6pt]
    H\rhd\eta^i = &\lambda_i\eta^i,\\[6pt]
    H\rhd\tilde{\partial}_i = &-\lambda_i\tilde{\partial}_i,
\end{split}
\hspace{1cm}
\begin{split}
    E\rhd y^i
    =&\delta^i_3y^1
    +2b\delta^i_2,\\[6pt]
    E\rhd\eta^i
    =&\delta^i_3\eta^1,\\[6pt]
    E\rhd\tilde{\partial}_i
    =&-\delta_{i1}\tilde{\partial}_3,
\end{split}
\hspace{1cm}
\begin{split}
    E'\rhd y^i =&\delta^i_3y^2
    +2b\delta^i_1,\\[6pt]
    E'\rhd\eta^i =&\delta^i_3\eta^2,\\[6pt]
    E'\rhd\tilde{\partial}_i =&-\delta_{i2}\tilde{\partial}_3.
\end{split}
\end{equation}
\begin{prop}\label{prop08}
$\F=\exp[H/2\otimes\log(\1+i\nu E)]$ is a unitary Jordanian twist 
 inducing  the following twisted deformation of $U\g$,
 of $\Q^\bullet$ on $\RR^3$
and of $\QM^\bullet$  on the hyperbolic paraboloid. 
%$\beta=S(\beta)=\exp(-i\nu L_{13}L_{23})$.
The $U\g^\f$  coproduct, antipode on $\{H,E,E'\}$
read
\bea  \label{DeltaSHypPar}
\ba{ll}
%    \Delta_\mathcal{F}(L_{12})
%    =&\Delta(L_{12})
%    +\frac{i\nu\sqrt{a}}{2}\bigg(H\otimes\frac{E}{1+i\nu E}\bigg),\\
%    \Delta_\mathcal{F}(L_{13})
%    =&\Delta(L_{13})
%    +\frac{i\nu}{2}\bigg(E\otimes E
%    -E'\otimes\frac{E}{1+i\nu E}\bigg),\\
%    \Delta_\mathcal{F}(L_{23})
%    =&\Delta(L_{23})
%    +\frac{i\nu\sqrt{a}}{2}\bigg(E\otimes E
%    +E'\otimes\frac{E}{1+i\nu E}\bigg),\\
\displaystyle    \Delta_\f(H)   =\Delta(H) 
    -i\nu\, H\otimes\frac{E}{\1+i\nu E}, 
\qquad     & S_\f(H)     = S(H)-i\nu HE,\\[8pt]
\displaystyle       \Delta_\f(E)   =\Delta(E)+i\nu E\otimes E,\qquad
 & \displaystyle   S_\f(E) = \frac{S(E)}{\1+i\nu E},\\[8pt]
\displaystyle       \Delta_\f(E')
    =\Delta(E')-i\nu E'\otimes\frac{E}{\1+i\nu E},\qquad  
& S_\f(E') = S(E')-i\nu EE'.
\ea\eea
%
%
%\bea
%    S_\mathcal{F}(L_{12})
%    =&S(L_{12})-i\nu L_{12}E,\\
%    S_\mathcal{F}(L_{13})
%    =&S(L_{13})+\frac{i\nu}{2}E\bigg(\frac{E}{1+i\nu E}-E'\bigg),\\
%    S_\mathcal{F}(L_{23})
%    =&S(L_{23})+\frac{i\nu\sqrt{a}}{2}E\bigg(\frac{E}{1+i\nu E}+E'\bigg),\\
%\eea
%
The $*$-structures on $U\g^\f$, $\Q^\bullet_\star, \QMst^\bullet$ remain undeformed apart from
$(y^2)^{*_\star}
=y^2+2i\nu b$
and
$(\tilde{\partial}_1)^{*_\star}
=-\tilde{\partial}_1+i\nu\tilde{\partial}_3$. \
The twisted star products %$L_{ij}\star L_{hk}=L_{ij}L_{hk}$ and Lie brackets 
of $\{H,E,E'\}$ coincide with the untwisted ones, except
\be  \label{starprodgHypPar}
    E\star H
    =EH+2i\nu E^2,\qquad\qquad
    E'\star H
    =E' H+2i\nu E^2.
\ee
The twisted star products of $H,E,E'$ with  
$y^i,\eta^i,\tilde{\partial}_i$ equal the untwisted ones, except 
\be
\ba{lll}
    E\star y^i     & = & Ey^i-i\nu E(2b\delta^i_2+y^1\delta^i_3),\\[6pt]
       E\star\eta^3  & = & E\eta^3-i\nu E\eta^1,\\[6pt]
     E\star\tilde{\partial}_1  & = & E\tilde{\partial}_1
    +i\nu E\tilde{\partial}_3,\\[6pt]
 E'\star y^i     & = & E'y^i+i\nu E'(2b\delta^i_2+y^1\delta^i_3),\\[6pt]
 E'\star\eta^3     & = & E'\eta^3+i\nu E'\eta^1,\\[6pt]
     E'\star\tilde{\partial}_1     & = & E'\tilde{\partial}_1
    -i\nu E'\tilde{\partial}_3,\\[6pt]
    y^i\star H     & = & y^iH-2i\nu(\delta^i_2-\delta^i_1)y^iE,\\[6pt]
    \eta^i\star H      & = & \eta^iH-2i\nu(\delta^i_2-\delta^i_1)\eta^iE,\\[6pt]
    \tilde{\partial}_i\star H      & = & \tilde{\partial}_iH
    -2i\nu(\delta_{i1}-\delta_{i2})\tilde{\partial}_iE;
\ea\ee
the twisted star products among 
$y^i,\eta^i,\tilde{\partial}_i$ equal the untwisted ones, except
\be
\ba{lll}
    y^i\star y^j     &=& y^iy^j
    +i\nu(\delta^i_2-\delta^i_1)y^i(2b\delta^j_2+\delta^j_3y^1),\\[6pt]
    y^i\star\tilde{\partial}_1
    &=&y^i\tilde{\partial}_1
    +i\nu(\delta^i_1-\delta^i_2)y^i\tilde{\partial}_3,\nonumber%\\[6pt]
\ea\ee
\be   \label{starprodgQHypPar}
\ba{lll}
    \tilde{\partial}_i\star y^j
    &=&\tilde{\partial}_iy^j
    +i\nu(\delta_i^1-\delta_i^2)\tilde{\partial}_i
    (2b\delta^j_2+\delta^j_3y^1),\\[6pt]
    \tilde{\partial}_1\star\tilde{\partial}_1
    &=&\tilde{\partial}_1\tilde{\partial}_1
    -i\nu\tilde{\partial}_1\tilde{\partial}_3,\\[6pt]
    \eta^2\star\eta^3
    &=&\eta^2 %\wedge
\eta^3+i\nu \eta^2 %\wedge
\eta^1,\\[6pt]
    y^i\star\eta^3
    &=&y^i\eta^3+i\nu(\delta^i_2-\delta^i_1)y^i\eta^1,\\[6pt]
    \eta^i\star\tilde{\partial}_1
    &=&\eta^i\tilde{\partial}_1
    +i\nu(\delta^i_1-\delta^i_2)\eta^i\tilde{\partial}_3,\\[6pt]
    \tilde{\partial}_i\star\eta^3
    &=&\tilde{\partial}_i\eta^3
    +i\nu(\delta_{i1}-\delta_{i2})\tilde{\partial}_i\eta^1,\\[6pt]
    \tilde{\partial}_2\star\tilde{\partial}_1
    &=&\tilde{\partial}_1\tilde{\partial}_2
    +i\nu\tilde{\partial}_2\tilde{\partial}_3,\\[6pt]
    \eta^i\star y^j    &=&\eta^iy^j+i\nu(\delta^i_2-\delta^i_1)\eta^i
    (2b\delta^j_2+\delta^j_3y^1).
\ea\ee
Hence the $\star $-commutation relations of the $U\g^\f$-equivariant 
algebra $\Q_\star$ read
\be \label{starcomrelQHypPar}
\ba{lll}
    y^1\star y^2    &=&y^2\star y^1-2bi\nu y^1,\\[6pt]
    y^i\star y^3    &=&y^3\star y^i+
    i\nu (\delta^i_2-\delta^i_1) \, y^i\star y^1,\qquad\mbox{for }i=1,2\\[6pt]
    y^1\star\eta^j    &=&\eta^j\star y^1
    -i\nu\delta^j_3\eta^1\star y^1,\\[6pt]
    y^2\star\eta^j
    &=&\eta^j\star y^2+2i\nu b(\delta^j_1-\delta^j_2)\eta^j
    +i\nu\delta^j_3\eta^1\star  (y^2+2i\nu b\1),\\[6pt]
    y^3\star\eta^j    &=&\eta^j\star y^3
    +i\nu(\delta^j_1-\delta^j_2)\eta^j \star y^1,\\[6pt]
% y^3\star\eta^i    &=&\eta^i\star y^3+
% i\nu(\delta^i_1-\delta^i_2)y^2\star\eta^i
 %   +2b\nu^2\delta^i_1\eta^1,\\[6pt]
    \tilde{\partial}_i\star y^1&=&\delta_i^1\1+
y^1\star\tilde{\partial}_i    -i\nu\delta_i^1 y^1\star\tilde{\partial}_3,\\[6pt]
 \tilde{\partial}_i\star y^2    &=&\delta_i^2\1+ y^2\star\tilde{\partial}_i
    +i\nu\delta_i^1y^2\star\tilde{\partial}_3
    +2i\nu b(\delta_{i1}-\delta_{i2})\tilde{\partial}_i,\\[6pt]
    \tilde{\partial}_i\star y^3 &=&\delta_i^3\1+
y^3\star\tilde{\partial}_i    +i\nu (\delta_i^1-\delta_i^2) \, y^1\star\tilde{\partial}_i
+i\nu \delta_i^1+\nu^2 \delta_i^1\, y^1\star\tilde{\partial}_3,\\[6pt]
    \eta^i\star\eta^j    &=&-\eta^j\star\eta^i    +i\nu(\delta^i_3\delta^j_1
    -\delta^j_3\delta^i_1)\eta^1\star\eta^2,\\[6pt]
\tilde{\partial}_j\star\eta^i&=& \eta^i\star\tilde{\partial}_j
    +i\nu[\delta_j^1(\delta^i_2-\delta^i_1)\eta^i\star\tilde{\partial}_3
 +\delta^i_3(\delta_j^1-\delta_j^2)\eta^1\star\tilde{\partial}_j
   -i\nu \delta^i_3\delta_j^1\,\eta^1\star\tilde{\partial}_3]\\[6pt]
%  \eta^i\star\tilde{\partial}_j     &=&\tilde{\partial}_j\star\eta^i
%    +i\nu(\delta_{j2}(\delta^i_1\eta^1-\delta^i_2\eta^2)
%    \star\tilde{\partial}_3     -\delta^i_3(\delta_{j1}\tilde{\partial}_1
%    -\delta_{j2}\tilde{\partial}_2)\star\eta^2),\\[6pt]
    \tilde{\partial}_i\star\tilde{\partial}_j    &=&\tilde{\partial}_j\star\tilde{\partial}_i
    +i\nu(\delta_{i2}\delta_{j1}    -\delta_{j2}\delta_{i1}
    )\tilde{\partial}_2\star\tilde{\partial}_3
\ea\ee
and
\be \label{starcomrelQgHypPar}
\ba{lll}
H\star y^i &=& y^i\star H+\lambda_i y^i+2i\nu(\delta^i_2-\delta^i_1)\, y^i\star E,\\[6pt]
       H\star\eta^i &=&  \eta^i\star H
    +2i\nu(\delta^i_2-\delta^i_1)\,\eta^i\star E,\\[6pt]
H\star\tilde{\partial}_i  &=&     \tilde{\partial}_i\star H  - \lambda_i \tilde{\partial}_i+
    2i\nu(\delta_i^1-\delta_i^2)\,\tilde{\partial}_i\star E,\\[6pt]
%    L_{13}\star y^i
%    &=&y^i\star L_{13}
%    -\frac{i\nu}{\sqrt{a}}(2b\delta^i_2+\delta^i_3y^2)\star L_{23}
%    -\nu^2b\delta^i_3E,\\[6pt]
%    L_{13}\star\eta^i
%    &=&\eta^i\star L_{13}
%    -\frac{i\nu}{\sqrt{a}}(2b\delta^i_2+\delta^i_3\eta^2)\star L_{23},\\[6pt]
%    L_{13}\star\tilde{\partial}_i
%    &=&\tilde{\partial}_i\star L_{13}
%    +\frac{i\nu}{\sqrt{a}}\delta_{i1}\tilde{\partial}_3\star L_{23},\\[6pt]
%    L_{23}\star y^i
%    &=&y^i\star L_{23}
%    -i\nu\sqrt{a}L_{12}\star(2b\delta^i_2+\delta^i_3y^2)
%    -\nu^2\sqrt{a}b\delta^i_3E,\\[6pt]
%    L_{23}\star\eta^i
%    &=&\eta^i\star L_{23}
%    -i\nu\sqrt{a}L_{12}\star(2b\delta^i_2+\delta^i_3\eta^2),\\[6pt]
%    L_{23}\star\tilde{\partial}_i
%    &=&\tilde{\partial}_i\star L_{23}
%    +i\nu\sqrt{a}\delta_{i1}L_{12}\star\tilde{\partial}_3,
    E\star y^i
    &=& E\rhd y^i+y^i\star E    -i\nu(2b\delta^i_2+y^1\delta^i_3)\star E,\\[6pt]
      E\star\eta^i &=&\eta^i\star E -i\nu\delta_3^i\,\eta^1\star E,\\[6pt]
      E\star\tilde{\partial}_i    &=&E\rhd \tilde{\partial}_i +
\tilde{\partial}_i\star E
    +i\nu\delta_i^1\,\tilde{\partial}_3\star E,\\[6pt]
     E'\star y^i     &=& E'\rhd y^i+ y^i\star E'
    +i\nu[(2b\delta^i_2+y^1\delta^i_3)\star E'+ 2b\delta^i_3],\\[6pt]
 E'\star\eta^i&=&\eta^i\star E'    +i\nu\delta^i_3\,\eta^1\star E',\\[6pt]
  E'\star\tilde{\partial}_i    &=& E'\rhd \tilde{\partial}_i+\tilde{\partial}_i\star E'
    -i\nu\delta_i^1\,\tilde{\partial}_3\star E'.
\ea\ee
%
%except
%\bea\ba{l}  x^3\star \xi^3=\xi^3\star x^3+i\nu a\,  (\xi^1\star  x^2
%-\xi^2 \star x^1),\qquad  \xi^3\star \xi^3=-i\nu a\,  \xi^1\star  \xi^2, \\[6pt]
%\partial_1\star \partial_2=\partial_2\star \partial_1-i\nu a\,
%\partial_3\star\partial_3,\\[6pt]
%\partial_j\star x^i=\delta^i_j\1+x^i\star  \partial_j+
%\delta^i_3 i\nu a\,(\delta_{1j} x^2-\delta_{2j} x^1)\star\partial_3,\\[6pt]
%\partial_i\star \xi^j=\xi^j\star \partial_i+
%\delta^i_3 i\nu a\,(\delta_{1j} \xi^2-\delta_{2j} \xi^1)\star\partial_3.
%\ea\eea
In terms of star products
\begin{align*}
    H=&2	\big(y^1\star\tilde{\partial}_1-y^2\star\tilde{\partial}_2-
i\nu y^1\star\tilde{\partial}_3\big),\qquad
    E    = y^1\star\tilde{\partial}_3
    +2b\tilde{\partial}_2,\qquad
    E'=y^2\star\tilde{\partial}_3+2b\tilde{\partial}_1
\end{align*}
and the relations  characterizing the $U\g^\f$-equivariant 
$*$-algebra  $\QMst^\bullet$ become
\begin{equation} \label{starcomrelQMHypPar}
    f(y)
    =\frac{1}{2}y^2\star y^1-by^3,\qquad
    \mathrm{d}f
    =\frac{1}{2}(y^2\star\eta^1+y^1\star\eta^2),\qquad
    \epsilon^{ijk}f_i\star L_{jk}=0.
\end{equation}
\end{prop}

\subsection{(e)   Family of hyperbolic cylinders\texorpdfstring{:  $a_2\!<0$, $a_3\!=\!a_{0\mu}\!=\!0$}{}}

Their equations in canonical form are parametrized by \
$c,\, a\equiv -a_2\in\mathbb{R}^+$ and read
\be\label{HyperbolicCylinder}
    f_c(x):=\frac{1}{2}\big[(x^1)^2-a(x^2)^2\big] -c=0.
\ee
For every $a>0$,  this equation with  $c=0$
singles out a variety $\pi$ consisting of two planes intersecting along
the $\vec{z}$-axis; $\{M_c\}_{c\in\RR^+}$ is a foliation of $\RR^3\setminus \pi$.
The case $c<0$ is reduced to the
case $c>0$ by a $\pi/2$ rotation around the $\vec{z}$-axis.
Eq. (\ref{HyperbolicCylinder}) can be obtained from the one (\ref{HypPareq}) characterizing the
 hyperbolic paraboloids (d) setting $b=0$. Hence also
the tangent vector fields $L_{ij}$  (or equivalently 
$H,E,E'$), their commutation relations, 
their actions on the $x^h,\xi^h,\partial_h$ (or equivalently on the
$y^h,\eta^h=dy^h,\tilde\partial_h$ defined by (\ref{ycoord}-\ref{ycoord'})),
the commutation relations of the $L_{ij}$ with the $x^h,\xi^h,\partial_h$
can be obtained from the ones of case (d)
by setting $b=0$. The $L_{ij}$   fulfill again (\ref{HypParg}), or
equivalently (\ref{eq11}), \ so that
 \ $\g\simeq \mathfrak{so}(1,1)\cross \RR^2$.

\begin{prop}\label{prop03'}
$\mathcal{F}=\exp(i\nu L_{13}\otimes L_{23})$ is a unitary abelian 
twist  inducing the twisted deformation of $U\g$,
 of $\Q^\bullet$ on $\RR^3$
and of $\QM^\bullet$  on the hyperbolic cylinders (\ref{HyperbolicCylinder}) that is
obtained by replacing $a\mapsto -a$ in Proposition 16 in \cite{FioreWeber2020}, section \ref{prop03}. 
\end{prop}

We can also deform everything with the same Jordanian
twist as in (d). We find

\begin{prop} Setting $b=0$ in Proposition \ref{prop08} one obtains 
the deformed $U\g$,  $\Q^\bullet$ on $\RR^3$
and $\QM^\bullet$  on the hyperbolic cylinders (\ref{HyperbolicCylinder}) 
induced by the  unitary twist $\F=\exp\left[\frac H2\otimes\log(\1\!+\!i\nu E)\right]$. 
\end{prop}

\subsection{(f-g-h)  Family of  hyperboloids and cone: 
$a_2,-a_3>0$}

Their equations in canonical form are parametrized by
$a=a_2,b=-a_3>0,c=-a_{00}$ ($c>0$, $c<0$ resp. for the  $1$-sheet  
and the $2$-sheet hyperboloids, $c=0$ for the cone) and read
\begin{equation}\label{eq19}
    f_c(x) :=\ba{l}\!\frac{1}{2}\!\ea \!
[(x^1)^2+a(x^2)^2-b(x^3)^2]-c=0.
\end{equation}
For all $a,b>0$, $\{M_c\}_{c\in\RR\setminus\{0\}}$ is a foliation of $\RR^3\setminus M_0$,
where $M_0$ is the cone of equation $f_0=0$ (see  section \ref{Cone}).
The Lie algebra  $\g$ is spanned by  $L_{12}=x^1\partial_2-ax^2\partial_1$, 
$L_{13}=x^1\partial_3+bx^3\partial_1$,
$L_{23}=ax^2\partial_3+bx^3 \partial_2$, which  fulfill \
%\begin{equation}
%    [L_{12},L_{13}]=-L_{23},\quad     [L_{12},L_{23}]=aL_{13},\quad
%    [L_{13},L_{23}]    =bL_{12},
%\end{equation}
$[L_{12},L_{13}]=-L_{23}$, \ $ [L_{12},L_{23}]=aL_{13}$, \ 
$[L_{13},L_{23}]    =bL_{12}$. \
Setting $H:=\frac{2}{\sqrt{b}}L_{13}$,
$E:=\frac{1}{\sqrt{a}}L_{12}+\frac{1}{\sqrt{ab}}L_{23}$
and $E':=\frac{1}{\sqrt{a}}L_{12}-\frac{1}{\sqrt{ab}}L_{23}$,
we obtain 
\be
[H,E]=2E,\qquad [H,E']=-2E',\qquad [E,E']=-H,       \label{so(2,1)}
\ee 
showing that the corresponding symmetry Lie algebra is
$ \mathfrak{g}\simeq\mathfrak{so}(2,1)$. 
The commutation relations \
 $[L_{ij},x^h]=L_{ij}\trc x^h$, \ $[L_{ij},\partial_h]=L_{ij}\trc \partial_h$, \
$[L_{ij},\xi^h]=0$ \ hold in $\Q^\bullet$.
To compute the action of $\mathcal{F}$ on  functions it is convenient 
to adopt the eigenvectors of $H$
\be
    y^1    =x^1+\sqrt{b}x^3,\qquad
    y^2=x^2,\qquad
    y^3=x^1-\sqrt{b}x^3,
\ee
as new coordinates; the eigenvalues are $\lambda_1=2$, $\lambda_2=0$ and $\lambda_3=-2$. \ Abbreviating 
$$
\eta^i:=dy^i, \qquad\tilde{\partial}_i:=%\frac{\partial}{\partial y^i}
\partial/\partial y^i,\qquad 
\tilde{\partial}^2:=\tilde{\partial}_2, \quad \tilde{\partial}^1:=2a\,\tilde{\partial}_3, \quad \tilde{\partial}^3:=2a\,\tilde{\partial}_1
$$
the inverse
coordinate  and  the partial derivative transformations read
\be
\ba{lll}
       x^1=\frac{1}{2}(y^1+y^3),  & 
     \tilde{\partial}_1
    =\frac{1}{2}\left(\partial_1
    +\frac{1}{\sqrt{b}}\partial_3\right)= \frac 1{2a}\tilde{\partial}^3,  &
     \partial_1
    =\tilde{\partial}_1+\tilde{\partial}_3, \\
     x^2=y^2,  & 
     \tilde{\partial}_2=\partial_2= \tilde{\partial}^2,  &
     \partial_2=\tilde{\partial}_2,  \\
     x^3=\frac{1}{2}\frac{1}{\sqrt{b}}(y^1-y^3),\qquad  & 
     \tilde{\partial}_3    =\frac{1}{2}\left(\partial_1
    -\frac{1}{\sqrt{b}}\partial_3\right)= \frac 1{2a}\tilde{\partial}^1,\qquad  &
     \partial_3
    =\sqrt{b}\left(\tilde{\partial}_1
    -\tilde{\partial}_3\right) .
\ea
\ee
In the new coordinates, \ $\big(\tilde{\partial}_i\big)^*
=-\tilde{\partial}_i$, \ $ f_c(y)
    =\frac{1}{2}y^1y^3+\frac{a}{2}(y^2)^2-c$ \  and%
\begin{equation}
    H=2y^1\tilde{\partial}_1-2y^3\tilde{\partial}_3,\quad
    E=\frac{1}{\sqrt{a}}y^1\tilde{\partial}_2
    -2\sqrt{a}y^2\tilde{\partial}_3,\quad
    E'=\frac{1}{\sqrt{a}}y^3\tilde{\partial}_2
    -2\sqrt{a}y^2\tilde{\partial}_1.
\end{equation}
The actions of $H,E,E'$ on any \  $u^i\in\{y^i,\tilde{\partial}^i, \eta^i\}$ \ read
\bea   H\rhd u^i=\lambda_iu^i,
\qquad     E\rhd u^i
    =\delta^i_2\frac{1}{\sqrt{a}}u^1
    -2\delta^i_3\sqrt{a}u^2,
\quad    E'\rhd u^i
    =\delta^i_2\frac{1}{\sqrt{a}}u^3
    -2\delta^i_1\sqrt{a}u^2.\qquad
\eea

\bigskip
\noindent
%\begin{prop}
{\bf Proposition 17 in \cite{FioreWeber2020}}. \
{\it $\F=\exp(H/2\otimes\log(\1+i\nu E))$ is a unitary twist
 inducing the following twisted deformation of $U\g$,
 of $\Q^\bullet$ on $\RR^3$
and of $\QM^\bullet$  on the hyperboloids or cone
(\ref{eq19}). 
%$\beta=S(\beta)=\exp(-i\nu L_{13}L_{23})$.
The $U\g^\f$  coproduct, antipode on $\{H,E,E'\}$
are given by
\be
\ba{l}
    \Delta_\f(E)
    =\displaystyle \Delta(E)+i\nu E\otimes E,\qquad
    \Delta_\f(H)
    =\Delta(H)    -i\nu H\otimes\frac{E}{\1+i\nu E},\\
    \Delta_\f(E')
    =\displaystyle
\Delta(E')     -\frac{i\nu}{2}H\otimes\bigg(\!H\!+\!\frac{i\nu E}{\1\!+\!i\nu E}\!\bigg)
    \frac{\1}{\1\!+\!i\nu E}\\ 
 \qquad\qquad\displaystyle -i\nu E'\otimes\frac{E}{\1\!+\!i\nu E}-\frac{\nu^2}{4}
    H^2\otimes\frac{E}{(\1\!+\!i\nu E)^2},
\ea  \label{copr}
\ee
\begin{equation}
\begin{split}
    S_\f(H)
    =&S(H)(\1+i\nu E),\qquad\qquad
    S_\f(E)
    =\frac{S(E)}{\1+i\nu E},\\
    S_\f(E')
    =&S(E')(\1\!+\!i\nu E)
    \!-\!\frac{i\nu}{2}H(\1\!+\!i\nu E)\bigg(\!H\!+\!\frac{i\nu E}{\1\!+\!i\nu E}\!\bigg)\!+\!\frac{\nu^2}{4}H(\1\!+\!i\nu E)HE.
\end{split}
\end{equation}
The twisted star products %$L_{ij}\star L_{hk}=L_{ij}L_{hk}$ 
%and Lie brackets 
of $\{H,E,E'\}$ coincide with the untwisted ones, except
\be
\ba{llllll} \label{gstargEllHyp}
    E\star H
     &=& EH+2i\nu E^2,\qquad
    &E'\star H
    &=& E' H-2i\nu E'E,\\[6pt]
    E\star E'
    &=& EE'+i\nu EH-2\nu^2 E^2,\qquad
    & E'\star E'   &=& (E')^2-i\nu E'  H.
%,\\    [E',H]_\star     =&[E',H]-2i\nu[E',E],\\
% [E,E']_\star =&[E,E']+i\nu[E,H],\\ [E',E']_\star  =&-i\nu[E',H].
\ea
\ee
The twisted star products of $u^i=y^i,\eta^i,\tilde{\partial}^i$ with
$v^j= y^j,\eta^j,\tilde{\partial}^j$  and with  $H,E,E'$
are given by
\bea       \label{starEllHyp}
\ba{lll}
    u^i\star v^j    &=& u^iv^j+i\nu(\delta^i_3-\delta^i_1)u^i
    \left(\frac{1}{\sqrt{a}}\delta^j_2v^1-2\sqrt{a}\delta^j_3v^2\right)+\delta^i_1\delta^j_3 2\nu^2u^1v^1,\\[8pt]
  H\star u^i  &=& Hu^i,\qquad
    u^i\star H     =  u^iH
    +2i\nu\left(\delta^i_1-\delta^i_3\right)u^iE,\\[8pt]
u^i\star E &=&u^iE,    \qquad   E\star u^i     =Eu^i
    +i\nu E\left(2\delta^i_3\sqrt{a}u^2-\frac{1}{\sqrt{a}}\delta^i_2u^1\right)
    +2\nu^2\delta^i_3Eu^1,\\[8pt]
        E'\star u^i     &=&E'u^i
    +i\nu\left(\frac{1}{\sqrt{a}}\delta^i_2E'u^1-2\sqrt{a}\delta^i_3E'u^2\right),\\[8pt]
     u^i\star E'   &=&u^iE'+i\nu\left(\delta^i_1-\delta^i_3\right)u^iH
    -2i\nu\delta^i_1u^1 E.
\ea
\eea
Hence the $\star $-commutation relations of the $U\g^\f$-equivariant 
algebra $\Q_\star$ read as follows: 
\bea
\ba{l}
 u^1\!\star\! u^2=u^2\!\star\! u^1\!-\!\frac{i\nu}{\sqrt{a}}u^1\!\star\! u^1, 
\qquad  u^1\!\star\! u^3=u^3\!\star\! u^1\!+\!2 i\nu \sqrt{a}\, u^2\!\star\! u^1
\!+\! 2\nu^2u^1\!\star\! u^1, \\[8pt]
 u^2\!\star\! u^3=u^3\!\star\! u^2\!-\!\frac{i\nu}{\sqrt{a}}u^3\!\star\! u^1, 
\qquad u^1\!\star\! \eta^1=\eta^1\!\star\! u^1, \quad 
u^1\!\star\! \eta^2=\eta^2\!\star\! u^1-\frac{i\nu}{\sqrt{a}}\eta^1\!\star\! u^1,\\[8pt]
u^1\!\star\! \eta^3=\eta^3\!\star\! u^1+2 i\nu \sqrt{a}\, \eta^2\!\star\! u^1\!+\! 2\nu^2\eta^1\!\star\! u^1,
\qquad  u^2\!\star\! \eta^1=\eta^1\!\star\! u^2+\frac{i\nu}{\sqrt{a}}\eta^1\!\star\! u^1 ,  \\[8pt]
u^2\!\star\! \eta^2=\eta^2\!\star\! u^2, \quad 
u^2\!\star\! \eta^3=\eta^3\!\star\! u^2-\frac{i\nu}{\sqrt{a}}\eta^3\!\star\! u^1,\quad
u^3\!\star\! \eta^1=\eta^1\!\star\! u^3-2i\nu \sqrt{a}\,\eta^1\!\star\! u^2,\\[8pt] 
u^3\!\star\! \eta^2=\eta^2\!\star\! u^3+\frac{i\nu}{\sqrt{a}}\eta^1\!\star\! u^3
+2\nu^2\eta^1\!\star\! u^2,\\[8pt] 
%u^3\!\star\! \eta^2=\eta^2\!\star\! u^3+\frac{i\nu}{\sqrt{a}}\eta^1\!\star\! \left(u^3
%-2\nu \sqrt{a} u^2-2i\nu^2\sqrt{a} u^1\right),\\[8pt] 
u^3\!\star\! \eta^3=\eta^3\!\star\! u^3+2i \nu \sqrt{a}\big(\eta^3\!\star\! u^2-\eta^2\!\star\! u^3\big)+2\nu^2 \,\eta^3\!\star\! u^1
\ea\eea
for $u^i=y^i,\tilde{\partial}^i$; the twisted Leibniz rule for the derivatives read
\bea
\ba{l}
\tilde{\partial}^1\!\star\! y^1=y^1\!\star\! \tilde{\partial}^1, \quad  \tilde{\partial}^2\!\star\! y^1
=y^1\!\star\! \tilde{\partial}^2+\frac{i\nu}{\sqrt{a}}y^1\!\star\! \tilde{\partial}^1 , \quad
\tilde{\partial}^3\!\star\! y^1=2a+y^1\!\star\! \tilde{\partial}^3-i2\nu \sqrt{a}y^1\!\star\! \tilde{\partial}^2,\\[8pt] 
\tilde{\partial}^1\!\star\! y^2=y^2\!\star\! \tilde{\partial}^1-\frac{i\nu}{\sqrt{a}}y^1\!\star\! \tilde{\partial}^1,\qquad 
\tilde{\partial}^3\!\star\! y^2=y^2\!\star\! \tilde{\partial}^3+i2\nu\sqrt{a}+
\frac{i\nu}{\sqrt{a}}y^1\!\star\! \tilde{\partial}^3
+2\nu^2y^1\!\star\! \tilde{\partial}^2,\\[8pt] 
 \tilde{\partial}^2\!\star\! y^2=1+y^2\!\star\! \tilde{\partial}^2, \qquad \qquad\quad \:
\tilde{\partial}^1\!\star\! y^3=2a+y^3\!\star\! \tilde{\partial}^1+i2 \nu \sqrt{a}\, y^2\!\star\! \tilde{\partial}^1
+2 \nu^2 y^1\!\star\! \tilde{\partial}^1,\\[8pt]
\tilde{\partial}^2\!\star\! y^3=y^3\!\star\! \tilde{\partial}^2-\frac{i\nu}{\sqrt{a}}y^3\!\star\! \tilde{\partial}^1, 
\qquad
\tilde{\partial}^3\!\star\! y^3=y^3\!\star\! \tilde{\partial}^3+i2 \nu \sqrt{a}\big(y^3\!\star\! \tilde{\partial}^2-y^2\!\star\! \tilde{\partial}^3\big)+2\nu^2 \,y^3\!\star\! \tilde{\partial}^1,
\ea\eea
while  the twisted wedge products fulfill
\bea
\ba{lll}
\eta^1\!\star\! \eta^1=0, \qquad 
&\eta^2\!\star\! \eta^2=0,\qquad 
&\eta^3\!\star\! \eta^3=2i \nu \sqrt{a}\, \eta^2\!\star\! \eta^3,\\[8pt]
\eta^1\!\star\! \eta^2+\eta^2\!\star\! \eta^1=0,\quad 
&\eta^1\!\star\! \eta^3+\eta^3\!\star\! \eta^1=2i\nu\sqrt{a}\,\eta^1\!\star\! \eta^2 ,\quad &
\eta^2\!\star\! \eta^3+\eta^3\!\star\! \eta^2=\frac{i\nu}{\sqrt{a}}\eta^3\!\star\! \eta^1.
\ea\eea
The $\star $-commutation relations between  generators of $\Q_\star$  and the tangent  vectors $H,E,E'$ are
\bea
\ba{l}
u^i\star H= H\star u^i- \vartheta \lambda_i u^i 
+2i\nu\left(\delta^i_1-\delta^i_3\right)u^i\star E,\\[6pt]
    u^1\star E=  E\star u^1, \qquad
     u^2\star E=  E\star u^2 -\frac{\vartheta}{\sqrt{a}}u^1+
\frac{i\nu}{\sqrt{a}} E\star u^1, \\[6pt]
      u^3\star E=  E\star u^3 +2\vartheta \sqrt{a} u^2
-2i\nu\sqrt{a}E\star u^2,\\[6pt]
u^1\star E'=  E'\star u^1+2\vartheta \big(\sqrt{a} u^2-i\nu u^1\big)+
i\nu H\star u^1-2i\nu E\star u^1\\[6pt]
u^2\star E'=  E'\star u^2-\frac{\vartheta}{\sqrt{a}} u^3-\frac{i\nu}{\sqrt{a}} E'\star u^1,\\[6pt]
u^3\star E'=  E'\star u^3-2\vartheta i\nu u^3+2i\nu\sqrt{a} E'\star u^2-
i\nu H\star u^3+2\nu^2  E'\star u^1,
\ea
\eea
where $\vartheta=1$ if $u^i=y^i,\tilde{\partial}^i$,
$\vartheta=0$ if $u^i=\eta^i$. In terms of star products
%
%\begin{split}
%    L_{13}
%    =&\sqrt{b}(\tilde{\partial}_1\star y^1
%    -y^3\star\tilde{\partial}_3),\\
%    L_{12}
%    =&\frac{1}{2}\tilde{\partial}_2\star(y^1+y^3)
%    -ay^2\star(\tilde{\partial}_1+\tilde{\partial}_3),\\
%    L_{23}
%    =&a\sqrt{b}
%    y^2\star(\tilde{\partial}_1-\tilde{\partial}_3)
%    +\frac{\sqrt{b}}{2}\tilde{\partial}_2\star(y^1-y^3),
%\end{split}
%\hspace{0.5cm}
\be   \label{gstarEllHyp}
\ba{lll}  
    H    &=&2(\tilde{\partial}_1\star y^1-1
    -y^3\star\tilde{\partial}_3),\\[6pt]
    E    &=& \displaystyle\frac{1}{\sqrt{a}}\tilde{\partial}_2\star y^1
    -2\sqrt{a}y^2\star\tilde{\partial}_3,\\[6pt]
    E'    &=& \displaystyle\frac{1}{\sqrt{a}}\tilde{\partial}_2\star y^3
    -2\sqrt{a}y^2\star\tilde{\partial}_1.
\ea\ee
The relations  characterizing the $U\g^\f$-equivariant 
$*$-algebra  $\QMst^\bullet$ become
\be  \label{QMstarEllHyp}
\ba{lll}
    0 &=&f_c(y)\equiv
    \frac{1}{2}y^3\star y^1+\frac{a}{2}y^2\star y^2-c,\\[6pt]
  0&=& \mathrm{d}f_c =\frac{1}{2}(y^3\star\eta^1+\eta^3\star y^1)    +ay^2\star\eta^2,\\[6pt]
  0&= &y^3\star E-y^1\star E'-\sqrt{a}\,y^2\star H+i\nu y^1\star H-2i\nu (1+i\nu) y^1\star E.
% 0 =&\frac{1}{2}(y^3\star L_{23}-i\nu\sqrt{a}L_{13}\star y^3)
 %   -aL_{13}\star y^2     +\frac{1}{2}L_{12}\star y^1.
\ea\ee
The $*$-structures on $U\g^\f$, $\Q^\bullet_\star, \QMst^\bullet$  remain undeformed except
$(u^3)^{*_\star}
=(u^3)^*-2i\nu\sqrt{a}(u^2)^*$ for $u^i=y^i,\eta^i,\tilde{\partial}^i$.
\label{prop05}
}
%\end{prop}
%

\subsubsection{Circular hyperboloids and cone embedded  in Minkowski $\RR^3$}
\label{CircHyperb}

We now focus on the case $1=a_1=a=b$, i.e.
$f_c(x)=\frac{1}{2}[(x^1)^2+(x^2)^2-(x^3)^2]-c$. This covers the
circular cone and hyperboloids of one and two sheets.
We endow $\RR^3$ with the Minkowski metric $\gm:=\eta_{ij}\mathrm{d}x^i\otimes\mathrm{d}x^j=\mathrm{d}x^1\otimes\mathrm{d}x^1
+\mathrm{d}x^2\otimes\mathrm{d}x^2-\mathrm{d}x^3\otimes\mathrm{d}x^3$,
whence $\gm(\partial_i,\partial_j)=\eta_{ij}$. \ $\gm$
is equivariant with respect to $U\mathfrak{g}$, where
$\mathfrak{g}\simeq\mathfrak{so}(2,1)$ is the Lie $^*$-algebra spanned by the vector fields $L_{ij}$, tangent to $M_c=f_c^{-1}(\{0\})$. 
%As known,  the induced metric on $M_c$ (or {\it first fundamental form}) 
The first fundamental form $\gm_t:=\gm\circ(\Pt\ot\Pt)$ makes 
$M_c$ Riemannian if $c<0$, Lorentzian if $c>0$,  whereas  is degenerate on the cone $M_0$. Moreover,  
\bea\label{IIFF}
II(X,Y)=-\frac {1}{2c}\,\gm(X,Y)\,\Vp \qquad \forall X,Y\in\Xi_t 
\eea
where $\Vp=f_j\eta^{ji}\partial_i=x^i\partial_i$ ({\it outward} normal); in particular, this implies the proportionality
relation \ $II(v_\alpha,v_\beta)=-\frac {1}{2c}\,g_{\alpha\beta}\,\Vp$ \ 
(here $g_{\alpha\beta}:=\gm(v_\alpha,v_\beta)$) \
between the matrix elements of $II,\gm_t$ in any basis $S_t:=\{v_1,v_2\}$ of $\Xi_t$,
and, applying the Gauss theorem, one finds the following components of the 
curvature and  Ricci tensors,  Ricci scalar   (or {\it  Gauss 
curvature}) on $M_c$:
\bea
\rR_{t}{\,}^{\delta}_{\alpha\beta\gamma}=
\frac{g_{\alpha\gamma}\delta_\beta^\delta-g_{\beta\gamma}\delta_\alpha^\delta}{2c},\qquad \ric_{t}{}_{\beta\gamma}=\rR_{t}{\,}_{\alpha\beta\gamma}^{\alpha}
=-\frac {g_{\beta\gamma}}{2c},\qquad \mathfrak{R}_{t}
=\ric_{t}{}_{\beta}^{\beta}=-\frac 1{c}  \label{curvatures}
\eea
[we recall that by the Bianchi identity one can express the whole curvature tensor on a 
(pseudo)Riemanian surface  in terms of the Ricci scalar in this way, and that $\rR_{t}{\,}^{\delta}_{\alpha\beta\gamma}v_\delta=\rR_{t}(v_\alpha,v_\beta,v_\gamma)$]. All diverge as $c\to 0$ (i.e. in  the cone $M_0$ limit). $M_c$ is therefore
de Sitter space $dS_2$ if $c>0$, the union of two copies of anti-de Sitter space
$AdS_2$ (the hyperbolic plane) if $c<0$. In appendix \ref{ClassicalCircularHyperboloids} we recall
how these results can be derived.
In terms of the $y^i$ coordinates and the tangent vector fields $H,E,E'$
(\ref{eq19}), (\ref{DepRel})  become the linear dependence relations \
$y^1y^3+(y^2)^2=2c$ \ and \  $y^3E-y^1E'-y^2H=0$, i.e. (\ref{QMstarEllHyp}) for $a=1$, $\nu=0$.
%\be
%y^1y^3+(y^2)^2=2c,\qquad
%%0=2(x^1L_{23}-x^2L_{13}-x^3L_{12})=...=
%y^3E-y^1E'-y^2H=0.   \label{lindip}
%\ee
At all points of $M_c$ at least two out of  $E,E',H$ are non-zero (in the case $c\!=\! 0$
we have already excluded the only point where this does not occur, the apex)
and make up another basis $S_t'=\{\epsilon_1,\epsilon_2\}$ of $\Xi_t$.
More precisely, we can choose $\epsilon_1:=E$, $\epsilon_2:=E'$ in a chart where $y^2\neq 0$, 
$\epsilon_1:=E$, $\epsilon_2:=H$ in a chart where $y^1\neq 0$, $\epsilon_1:=E'$, $\epsilon_2:=H$ in a chart where $y^3\neq 0$.  
One can use (\ref{IIFF}), (\ref{curvatures})  with each basis $S_t'$; 
$g_{\alpha\beta}$ stands for $g_{\alpha\beta}\equiv\gm(\epsilon_{\alpha},\epsilon_{\beta})$, and these matrix elements are given in (\ref{metricHEE'}). Alternatively,
we can use the complete set $S_t^c=\{E,E',H\}$ on all of $M_c$, keeping in mind the mentioned linear dependence relations.

\medskip
We now analyze the effects on the geometry of the twist deformation of Proposition 17 in \cite{FioreWeber2020} restated above. The curvature (and Ricci) tensor on $\RR^3$ remain zero.
Moreover, eq. (\ref{tangentgIIR}),  (\ref{II^F}) apply; namely, on $M_c$  
the first and second fundamental forms, as well as the
 curvature and Ricci tensor, remain undeformed as elements of the corresponding
tensor spaces; only the associated multilinear maps of twisted tensor products 
$\gm_{t\star}:\Xi_{t\star}\ots \Xi_{t\star}\to\X_\star$, ...,
 `feel' the twist 
(compare also to \cite{AschieriCastellani2009}~Theorem~7 and eq.~6.138).
Also the Ricci scalar (or Gauss curvature) $\mathfrak{R}_{t}^\f$ remains
the undeformed one $-1/c$. By (\ref{II^F})  the twisted counterpart of (\ref{IIFF}) becomes
\bea \label{IIfstar}
II^\f_\star(X,Y)=-\frac {1}{2c}\,\gm_{t\star}(X,Y)\,\Vp=-\frac {1}{2c}\,\gm_{t\star}(X,Y)\star\Vp;
\eea
the second equality holds because $\Vp$ is $U\k$-invariant. 
Similarly, by (\ref{II^F}), (\ref{Killing})  
\bea
\rR^\f_{t\star}(X,\!Y,\!Z)=
\frac{\left(\bR_1\!\trc\! Y\right) \ltlc_\star\gm_{t\star}\!\left(\bR_2\!\trc\!  X,\!Z\right)-X\ltlc_\star \gm_{t\star}(Y,\!Z)}{2c},\quad \ric^\f_{t\star}(Y,\!Z)
=-\frac {\gm_{t\star}(Y,Z)}{2c}  \label{curvaturesstar}
\eea
for all $X,Y,Z\in\Xi_{t\star}$;
the twisted counterpart of (\ref{curvatures}) is obtained  choosing $(X,Y,Z)=(v_\alpha,v_\beta,v_\gamma)$.
Hence the matrix elements of $II^\f_\star,\rR^\f_{t\star}, \ric^\f_{t\star}$ in any basis $S_t'$ are obtained from those of the twisted metric $\gm_{t\star}$ on $M_c$. In the appendix we  sketchily prove that
on $E,E',H$
%the same results as in (\ref{metricHEE'}), apart from
\bea \label{gmtstar}
\ba{l}
\gm_{t\star}(H,H)=-8y^1y^3, \qquad\gm_{t\star}(H,E)=-2y^1y^2, \qquad\gm_{t\star}(H,E')=-2y^2y^3\\[6pt]
\gm_{t\star}(E,E)=(y^1)^2, \qquad\quad  \gm_{t\star}(E,E')=2c+(y^2)^2
-2i\nu y^1y^2-2\nu^2(y^1)^2, \\[6pt]
\gm_{t\star}(E,H)=-2y^1y^2+2i\nu (y^1)^2, \qquad\qquad \gm_{t\star}(E',E)=2c+(y^2)^2,\\[6pt] 
\gm_{t\star}(E',E')=(y^3)^2 ,\quad \gm_{t\star}(E',H)=-2y^2y^3-2i\nu[2c+(y^2)^2]
+2i\nu y^2y^3.
\ea
\eea
Finally, we also show that the twisted Levi-Civita connection on $E,E',H$ gives
\bea\label{TwistedConnection}
\ba{l}
   \nabla^\f_EE%=\nabla_E E
    =-2y^1\tilde\partial_3 ,\qquad
    \nabla^\f_EE'= %\nabla_EE'+i\nu\nabla_EH-2\nu^2\nabla_EE
    -2y^1\tilde\partial_1-2y^2\tilde\partial_2
    +4i\nu\tilde\partial_3
    +4\nu^2y^1\tilde\partial_3 ,\\[6pt] 
\nabla^\f_EH= %\nabla_EH+2i\nu\nabla_EE=
    4y^2\tilde\partial_3    -4i\nu y^1\tilde\partial_3    ,\qquad\qquad\qquad
    \nabla^\f_{E'}E%=\nabla_{E'} E
    =-2y^3\tilde\partial_3-2y^2\tilde\partial_2,\\[6pt]
    \nabla^\f_{E'}E'
    %\nabla_{E'}E'    -i\nu\nabla_{E'}H
    =-2y^3\tilde\partial_1
    +4i\nu y^2\tilde\partial_1,\qquad
    \nabla^\f_{E'}H=% \nabla_{E'}H-2i\nu\nabla_{E'}E=
    -4y^2\tilde\partial_1
    +4i\nu(y^2\tilde\partial_2+y^3\tilde\partial_3),\\[6pt]
    \nabla^\f_HE=
    2y^1\tilde\partial_2,\qquad         
    \nabla^\f_HE'=
    -2y^3\tilde\partial_2,\qquad 
    \nabla^\f_HH
    =4y^1\tilde\partial_1+4y^3\tilde\partial_3.
\ea
\eea

We recall that a sheet of the hyperboloid $M_c$, $c<0$, is equivalent to a hyperbolic plane. Other deformation quantizations of the latter have been done,
in particular that of \cite{BieDetSpi2009} in the framework
\cite{Rieffel,BieBonMae2007} (cf. the introduction). However, while the  $\star$-product \cite{BieDetSpi2009} is
$U\mathfrak{k}$-equivariant, i.e. relation (\ref{Leibniz}) (which is the 'infinitesimal' version of the invariance  property (10) in  \cite{BieDetSpi2009} or (1) of \cite{BieBonMae2007}) holds, our $\star$-product
is $U\mathfrak{k}^{\scriptscriptstyle {\cal F}}$-equivariant
i.e. relation (\ref{TwistedLeibniz})  holds.

\subsubsection{Additional twist deformation of the cone (h) 
%\texorpdfstring{: $a_2,-a_3>0$, $a_{03}=a_{00}=0$}{}
}
\label{Cone}

The equation of the cone $M_0$ in canonical form  is  (\ref{eq19}) with $c=0$.
In addition to the tangent vector fields  $L_{ij}$ 
or $H,E,E'$ fulfilling (\ref{so(2,1)}) also 
the generator ${\sf D}=%\sum_{i=1}^3
x^i\partial_i=y^i\tilde{\partial}_i$ of dilatations
is  tangent to  $M_0$ (only), ${\sf D}\in\Xi_{\Ms_0}$,
since ${\sf D}(f)=2f$; furthermore it commutes with all $L_{ij}$.
Hence the anti-Hermitian elements $H,E,E',{\sf D}$ span a Lie algebra
$ \mathfrak{g}\simeq\mathfrak{so}(2,1)\times\RR$. The actions of
$H,E,E'$ on ${\cal Q}_{\scriptscriptstyle M}$ are as in cases (e-f),
while that of ${\sf D}$ is determined by
\be
{\sf D}\rhd y^i:=[{\sf D},y^i]=y^i,\qquad {\sf D}\rhd \eta^i:=d({\sf D}\rhd y^i)=\eta^i
,\qquad {\sf D}\rhd \tilde{\partial}_i:=[{\sf D},\tilde{\partial}_i]=-\tilde{\partial}_i.
\ee
Therefore, we can build also abelian twist deformations
of $M_0$ of the form $\mathcal{F}=\exp(i\nu {\sf D}\otimes g)$, $g\in\g$.  
Here we choose $g=\frac{L_{13}}{\sqrt{b}}=\frac H2$,
i.e. $\mathcal{F}=\exp(i\nu {\sf D}\otimes\frac H2)$.
The cases with $L_{23},L_{12}$ are similar.
Setting $\mu_1=1=-\mu_3$ and $\mu_2=0$, for $u^i,v^i\in\{y^i,\eta^i\}$ we find
\begin{align*}
    \bF(\rhd\ot \rhd) (u^i\otimes v^j)
    =&e^{-i\nu\mu_j}u^i\otimes v^j,\qquad
    \bF(\rhd\ot \rhd)(u^i\otimes\tilde{\partial}_j)
    =e^{i\nu\mu_j}u^i\otimes\tilde{\partial}_j,\\[4pt]
    \bF(\rhd\ot \rhd)(\tilde{\partial}_i\otimes u^j)
    =&e^{i\nu\mu_j}\tilde{\partial}_i\otimes u^j,\qquad
    \bF(\rhd\ot \rhd)(\tilde{\partial}_i\otimes\tilde{\partial}_j)
    =e^{-i\nu\mu_j}\tilde{\partial}_i\otimes\tilde{\partial}_j.
\end{align*}
Having this in
mind, in the appendix we easily determine the twist deformed structures.
\begin{prop}\label{prop06}
$\F=\exp(i\nu {\sf D}\otimes H/2)$  is a unitary abelian twist
 inducing  the following twisted deformation of $U\g$,
 of $\Q^\bullet$ on $\RR^3$
and of $\QM^\bullet$  on the  cone $M_0$. 
%$\beta=S(\beta)=\exp(-i\nu L_{13}L_{23})$.
The $U\g^\f$ counit, coproduct, antipode on $\{{\sf D},H,E,E'\}$
coincide with the undeformed ones, except
\begin{equation} \label{AbTwistDeltaSCone}
\ba{llllll}
    \Delta_\f(E)
    &=&E\otimes \1+\exp(i\nu {\sf D})\otimes E,\qquad
  &  S_\f(E)
    &=&-E\exp(-i\nu {\sf D}),\\[6pt]
  \Delta_\f(E')
    &=&E'\otimes \1+\exp(-i\nu {\sf D})\otimes E',\qquad
  &     S_\f(E')
    &=&-E'\exp(i\nu {\sf D}).
\ea
\end{equation}
The twisted star products %$L_{ij}\star L_{hk}=L_{ij}L_{hk}$ 
 among ${\sf D},L_{ij} $ coincide with the untwisted ones.
The twisted star products of ${\sf D},L_{ij} $ with  %coordinate functions 
$u^i\in\{y^i,\eta^i\}$, $\tilde{\partial}_i$ coincide with the untwisted ones, except 
\be      \label{AbTwistStarProdCone1}
\ba{ll}
    u^i\star E
    =e^{-i\nu}u^iE,\qquad\qquad &
    u^i\star E'
    =e^{i\nu}u^iE',\\[6pt]
        \tilde{\partial}_i\star E
    =e^{i\nu}\tilde{\partial}_iE,\qquad\qquad &
    \tilde{\partial}_i\star E'
    =e^{-i\nu}\tilde{\partial}_iE'.
\ea\ee
The twisted star products among $y^i,\eta^i,\tilde{\partial}_i$ read
\be   \label{AbTwistStarProdCone2}
\ba{ll}
    u^i\star v^j
    = e^{-i\nu\mu_j}u^iv^j,\qquad\qquad &
   \tilde{\partial}_i\star\tilde{\partial}_j
    = e^{-i\nu\mu_j}\tilde{\partial}_i\tilde{\partial}_j,\\[6pt]
      u^i\star\tilde{\partial}_j
    = e^{i\nu\mu_j}u^i\tilde{\partial}_j,\qquad\qquad &
    \tilde{\partial}_i\star u^j
    = e^{i\nu\mu_j}\tilde{\partial}_iu^j,
\ea\ee
with $u^i\!,\!v^i\!\in\!\{y^i,\!\eta^i\}$.
Hence the $\star $-commutation relations of the $U\g^\f\!$-equivariant 
algebra $\Q_\star$ are
\begin{equation}   \label{AbTwistStarComRelCone1}
\begin{split}
    y^i\star y^j
    =&e^{i\nu(\mu_i-\mu_j)}y^j\star y^i,\\
    y^i\star\eta^j
    =&e^{i\nu(\mu_i-\mu_j)}\eta^j\star y^i,\\
  \tilde{\partial}_j\star y^i  
    =&e^{i\nu\mu_i}\delta^i_j\1+e^{i\nu(\mu_i-\mu_j)}y^i\star\tilde{\partial}_j,
\end{split}
\hspace{1cm}
\begin{split}
    \eta^i\star\eta^j
    =&-e^{i\nu(\mu_i-\mu_j)}\eta^j\star\eta^i,\\
    \eta^i\star\tilde{\partial}_j
    =&e^{-i\nu(\mu_i-\mu_j)}\tilde{\partial}_j\star y^i,\\
    \tilde{\partial}_i\star\tilde{\partial}_j
    =&e^{i\nu(\mu_i-\mu_j)}\tilde{\partial}_j\star\tilde{\partial}_i.
\end{split}
\end{equation}
%
%except
%\bea\ba{l}  x^3\star \xi^3=\xi^3\star x^3+i\nu a\,  (\xi^1\star  x^2
%-\xi^2 \star x^1),\qquad  \xi^3\star \xi^3=-i\nu a\,  \xi^1\star  \xi^2, \\[6pt]
%\partial_1\star \partial_2=\partial_2\star \partial_1-i\nu a\,
%\partial_3\star\partial_3,\\[6pt]
%\partial_j\star x^i=\delta^i_j\1+x^i\star  \partial_j+
%\delta^i_3 i\nu a\,(\delta_{1j} x^2-\delta_{2j} x^1)\star\partial_3,\\[6pt]
%\partial_i\star \xi^j=\xi^j\star \partial_i+
%\delta^i_3 i\nu a\,(\delta_{1j} \xi^2-\delta_{2j} \xi^1)\star\partial_3.
%\ea\eea
%
The $*$-structures on $U\g^\f$, $\Q^\bullet_\star, {{\cal Q}_{{\scriptscriptstyle M}\star}}$ are
undeformed, except
$$
 (\tilde{\partial}_i)^{*_\star}
=-e^{-i\nu\mu_i}\tilde{\partial}_i,\qquad\qquad
(u^i)^{*_\star}
=e^{-i\nu\mu_i}u^i, \quad u^i=y^i,\eta^i,
$$
which are nontrivial for $i=1$ and $i=3$.
In terms of star products \ ${\sf D}=\sum_{i=1}^3e^{-i\nu\mu_i} y^i\star\tilde{\partial}_i$,
$H=2(e^{-i\nu} y^1\star\tilde{\partial}_1-e^{i\nu} y^3\star\tilde{\partial}_3)$, 
$E=\frac{e^{-i\nu}}{\sqrt{a}}y^1\star\tilde{\partial}_2-2\sqrt{a}y^2\star\tilde{\partial}_3$, 
$E'=\frac{e^{i\nu}}{\sqrt{a}}y^3\star\tilde{\partial}_2 -2\sqrt{a}y^2\star\tilde{\partial}_1$, \
and the relations  characterizing the $U\g^\f$-equivariant 
$*$-algebra  $\QMst^\bullet$, i.e. equation  (\ref{eq19})$_{c=0}$, %of the parabolic cylinder 
 its differential   and the linear dependence relations 
 %(\ref{DepRel})$_2$,
  become
\be      \label{AbTwistCharCone1}
\ba{l}
 \displaystyle   f(y)
    \equiv\frac{1}{2}e^{-i\nu}y^1\star y^3+\frac{a}{2}y^2\star y^2=0,\\[6pt]
 \displaystyle    \mathrm{d}f \equiv\frac{1}{2}(e^{i\nu}y^3\star\eta^1
    +e^{-i\nu}y^1\star\eta^3)
    +ay^2\star\eta^2=0,\\[6pt]
    \epsilon^{ijk}L_{jk}\star f_i    =0.
\ea\ee
\end{prop}

\noindent
{\bf Acknowledgments.} 
We are indebted with F. D'Andrea for stimulating discussions, critical advice and useful suggestions at all stages of the work.  The third author would like
to thank P.~Aschieri   for his kind hospitality at the
University of Eastern Piedmont "Amedeo Avogadro".

\appendix

\section{Real Nullstellensatz}
\label{RealNullstellensatz}

First of all, we recall some basic notions and notation in algebraic geometry
 that we are using in this subsection. 
In what follows we fix a ground field $\KK$ of any characteristic (even though we work only over real and complex fields, 
all the notions and definitions we are going to review hold true in a much wider generality).
\begin{enumerate}
\item (Algebraic Sets \cite[p. 2]{Hartshorne})	A subset of $\KK^n$ is an \textit{algebraic set} if it is defined as the set of common solutions of a  system of polynomial equations. By Hilbert basis theorem \cite[Theorem 3.3]{Matsumura}, algebraic sets can be also defined as
$$Z(I):=\{x\in \KK^n \mid \,\, P(x)=0, \,\,\forall P\in I \},$$
where $I$ denotes an ideal of the polynomial ring $\KK[x^1, \dots, x^n]$.
\item (Zariski topology \cite[p. 2]{Hartshorne}) The affine space $\KK^n$ can be endowed with a topology, the so called Zariski topology, where closed sets coincide with algebraic sets.  In this section we will equip algebraic sets with the induced topology.
\item (Algebraic, or affine, varieties) An algebraic variety is an \textit{irreducible} algebraic set, i.e. an algebraic set which is not the union of two proper (i.e. strictly contained)
closed subsets. It turns out \cite[ Exercise I.1.6]{Hartshorne} that a non-empty open set of an affine variety is irreducible. 
\item (Decomposition of algebraic sets) An algebraic set  $M$ can be expressed uniquely as a union of varieties, no one containing another \cite[Corollary 1.6]{Hartshorne}. Such varieties are called \textit{irreducible components}
of $M$.
\item (Radicals) For any ideal $I\leq \KK[x^1, \dots, x^n]$ the \textit{radical} of $I$ \cite[p. 3]{Matsumura},  \cite[\S 1.3]{Fulton} is the ideal defined as
$$\Rad (I):=\{P\in \KK[x^1, \dots, x^n]:\,\, \exists k>0 \,\, \mid \,\, P^k\in I \}.$$
A \textit{radical ideal} is an ideal $I$ s.t. $I=\Rad(I)$. By the very definition of \textit{prime ideal} \cite[p. 2]{Matsumura}, any such an ideal is radical.
\item (Correspondence among varieties and prime ideals) An affine algebraic set is a 
variety if and only if its ideal is a prime ideal \cite[Corollary 1.4]{Hartshorne}.
\item (Associated primes) Consider an algebraic set $M=Z(I)$. An \textit{associated prime} is a prime ideal of  $\KK[x^1, \dots, x^n]$
which is the annihilator $ann(x)$ of some element  $x\in \frac{\KK[x^1, \dots, x^n]}{I}$. It turns out \cite[\S 6]{Matsumura} that there are two kinds of associated primes: \textit{minimal associated primes} (they are  in one to one
correspondence with the irreducible components of $M$), and  \textit{embedded associated primes} (they have NOT a simple geometric interpretation).
\item (Hilbert's Nullstellensatz \cite[\S 5]{Matsumura}) Assume now $\KK$  algebraically closed and define 
$$ 
\ic(S):=\{P\in \KK[x^1, \dots, x^n]\mid \,\, P\mid_S \equiv 0 \, \},$$
 for any subset $S\subseteq \KK^n$.
  Then we have
  $$\ic(Z(I))= \Rad (I), \,\,\, \forall I\leq \KK[x^1, \dots, x^n].$$
  A weak form of this result says that $Z(I)\not= \emptyset $, for any proper ideal $I\leq \KK[x^1, \dots, x^n]$ 
  \cite[\S 1.7]{Fulton}.
\item (Regular sequences) A set of polynomials $P_1, \dots , P_k$ form a \textit{regular sequence}
 in  $\frac{\KK[x^1, \dots, x^n]}{I}$ \cite[\S 16]{Matsumura}, if every $P_i$ is not a zero divisor in
 $\frac{\KK[x^1, \dots, x^n]}{(I, P_1, \dots , P_{i-1})}$.
 \item (Cohen-Macaulay property \cite[\S 17]{Matsumura}) An affine variety $M=Z(I)$, such that $\dim M=m$, is said \textit{Cohen-Macaulay} at $x\in M$ if there is a 
 regular sequence $P_1, \dots , P_m$ in  $\frac{\KK[x^1, \dots, x^n]}{I}$, such that $P_i(x)=0$, $\forall i$. 
 An affine variety $M=Z(I)$  is said Cohen-Macaulay if it is Cohen-Macaulay at any point.
 \end{enumerate}

\medskip
Now consider an algebraic submanifold, i.e. a smooth algebraic variety,   $M\subseteq\RR^n$, defined by a system of polynomial equations (\ref{DefIdeal}), with
$f^1, \dots , f^k\in \RR[x^1, \dots, x^n]$. Assume that $\dim M=n-k$. Then, the hypersurfaces defined by each of the equations in  (\ref{DefIdeal}) meet transversally 
et each point of $M$; in other words, the Jacobian matrix is of rank $k$ at each point of $M$. Consider a further polynomial $Q\in \RR[x^1, \dots, x^n]$ and assume that 
$$
Q\mid_M\equiv 0.
$$ 
One may wonder whether the irreducibility in $\RR[x^1, \dots, x^n]$ of each  polynomial $f^1, \dots , f^k$  is  a sufficient condition in order that $Q$ lies in $(f^1, \dots , f^k)$, the ideal generated by $f^1, \dots , f^k$. The following example answers in the negative.

\medskip

\begin{example}
\label{counterexample} Consider in $\RR^3$ the variety defined by the system
\begin{equation}\label{ce}
   \begin{cases}
   2x^3-y^3=1,\\ 
      y=1,
   \end{cases}
\end{equation}
where the first equation represents a cubic cylinder ${\sf C}$.
Since the curve defined by $$ 2x^3-y^3-z^3=0$$
 is smooth in $\Pd$, the cylinder ${\sf C}$ is smooth and the polynomial $ 2x^3-y^3-1$ is irreducible in
$\RR [x,y,z]$ (the same conclusion is obvious for $y-1$).
The real variety defined by (\ref{ce}) is the line
$$l:= \{(1,1,t): \,\, t\in \RR \},$$
which is obviously smooth.  Furthermore, the equation of the tangent plane to the cylinder ${\sf C}$ at the point $(1,1,t)\in l$ is $2(x-1)-(y-1)=0$, hence the intersection is transversal at each point of $l$. On the other hand, the plane $\pi$ defined by $x+y-2=0$ contains $l$ but 
$$x+y-2\not \in (2x^3-y^3-1, y-1),$$
 since both $2x^3-y^3-1$ and $y-1$ do vanish at the points $(\exp{\frac{2}{3}\pi i},1,t)$, $\forall t\in \RR$, and conversely $x+y-2$ does not.
\end{example}

In view of the previous example, it is interesting to ask for some sufficient condition in order that $Q\in (f^1, \dots , f^k)$. An answer is provided by Theorem~\ref{DecoCpol}, which we now prove.

\begin{comment}
\begin{prop}
Consider as above a smooth algebraic variety  $M\subset \RR^n$, obtained by intersecting $k$ hypersurfaces meeting transversally at each point of $M$.
Assume additionally that the variety defined by (\ref{DefIdeal}) is irreducible in $\CC^n$. Then we have
$$
Q\in (f^1, \dots , f^k)\leq \RR[x^1, \dots, x^n],
$$
  where $(f^1, \dots , f^k)\leq \RR[x^1, \dots, x^n]$ denotes the ideal generated by $f^1, \dots , f^k$ in the ring $ \RR[x^1, \dots, x^n]$.
 For instance, this happens as soon as there exists an affine subspace $\pi \subseteq \RR^n$,
such that $\dim \pi = k$,  meeting $M$ in $s:=\Pi _{i=1}^k \deg f^i$ points. 
\end{prop}
\end{comment}

\subsubsection*{Proof of Theorem \ref{DecoCpol}} 

Denote by
$$
I:=(f^1, \dots , f^k)\cdot  \CC[x^1, \dots, x^n]\leq \CC[x^1, \dots, x^n]
$$  
the ideal of $\CC[x^1, \dots, x^n]$ generated by $f^1, \dots , f^k$.
Since we are assuming that the zero locus of $I$
is irreducible, there is only one minimal  prime associated to the ideal $I$.
The hypothesis that the hypersurfaces corresponding to the generators of $I$ meet transversally at $M$ imply that $f^1, \dots , f^k$
form a \textit{regular sequence} in $\CC[x^1, \dots, x^n]$ \cite[\S 16]{Matsumura}, hence the zero locus $Z(I)$ of $I$ is a complete intersection, i.e. an affine variety defined as the intersection of as many hypersurfaces as its codimension.
This implies that $Z(I)$ is Cohen-Macaulay, hence there is no embedded associated prime 
\cite[Theorem 17.3]{Matsumura} and 
the ideal $I$ is primary, i.e. there is only one associated prime.
Again, the hypothesis that the hypersurfaces defined by the equations in (\ref{DefIdeal}) meet transversally et each point of $M$ imply that $I$ is a prime ideal in 
$\CC[x^1, \dots, x^n]$.

On the other hand, by Hilbert's Nullstellensatz \cite[Exercise 7.14]{AM}, \cite[\S 1.7]{Fulton},
\cite[Theorem 5.4]{Matsumura}, the hypothesis $Q\mid_M\equiv 0$ amounts to 
$$Q\in \Rad(I):=\{P\in \CC[x^1, \dots, x^n]:\,\, \exists n>0 \,\, \mid \,\, P^n\in I \}=I,$$
where $\Rad (I)$ denotes the \textit{radical} of $I$ \cite[Exercise 1.12]{AM},  \cite[\S 1.3]{Fulton} and where the last equality follows because $I$ is prime.
This shows that $Q\in I\cap \RR[x^1, \dots, x^n]= (f^1, \dots , f^k)$.
Finally, for a complex-valued $h=Q_1+iQ_2$ vanishing on $M$ both $Q_1,Q_2$ do,
and therefore $h$ belongs to the complexification of $(f^1, \dots , f^k)$.

As for the last statement, the projective closure $X\subseteq \Pn$ of the zero locus of (\ref{DefIdeal}) in $\Pn$ has degree at least $s$. On the other hand, $s$ is the maximum degree so $X$ is a complete intersection in $\Pn$. Then, there cannot be other components and the variety defined by (\ref{DefIdeal}) is irreducible in $\Pn$. The statement follows at once, since a non-empty open set of an irreducible variety is irreducible
\cite[Exercise I.1.6]{Hartshorne}. \hfill $\Box$

\begin{remark}
Consider an algebraic smooth hypersurface $M$, defined by a single equation $f(x)=0$, with $d:= \deg f$. By the result above, in order that any polynomial $h$, such that $h\mid _M\equiv 0$, is a multiple of $f$ it suffices that there exists a line 
meeting $M$ in $ d$ points.
\end{remark}

\section{Proofs of sections~ \ref{Preli}, \ref{TwistDiffGeomAlgSubman}, \ref{quadricsR^3}}
\label{OtherProofs}

\begin{comment}
\subsection{Proof of Proposition \ref{IsomorHopf}}
The proposition is a straightforward consequence of the relation
\be
D\circ *_\star=*\circ D, \qquad \mbox{or equivalently }\qquad
*_\star=D^{-1}\circ *\circ D.
\ee
This is almost the same as eq. (31) in \cite{Fiore2010}. We prove it again using 
the relation
\be 
\Delta(\beta)=\bF(\beta \ot \beta )[(S\ot S)\bF_{21}]
=\bF_{21}(\beta \ot \beta )[(S\ot S)\bF],   \label{deltabeta}
\ee
which is proved e.g.  in Lemma 2.2. in \cite{GurMajid1994}
[or see eq. (126) in \cite{Fiore2010}]. The claim can be derived as follows:
\bea
D(\xi^{*_\star}) \!&\!=\!&\! D\left[S(\beta)\trc \xi^*\right]= 
D\left[S\!\left(\beta_{(1)}\right)  \xi^* \beta_{(2)} \right]=
\F_1 S\left(\beta_{(1)}\right)  \xi^* \beta_{(2)}  S(\F_2) \beta^{-1}\stackrel{(\ref{deltabeta})}{=}\! S(\beta)S(\bF_2)\, \xi^*\,\bF_1\nn
\!&\!=\!&\! S(\beta)\left[\F_1\, \xi \,S(\F_2)\right]^*
=S(\beta)\left[D( \xi)\beta\right]^*=S(\beta)S\big(\beta^{-1}\big)\left[D( \xi)\right]^*
=\left[D( \xi)\right]^* 
\eea
As particular consequences,
$\Delta_\star\circ *_\star=(*_\star\otimes *_\star)\circ\Delta_\star $, 
$S_\star\circ *_\star\circ S_\star\circ *_\star=\id$ follow from
$\Delta_\f\circ *=(*\ot *)\circ\Delta_\f$, 
$S_\f\circ *\circ S_\f\circ *=\id$.
\end{comment}

\subsection{Proof of  Proposition \ref{propQR^nstar}}

Using the definition \ $\beta:=\F_1\cdot S(\F_2)$, \ $a'\star a  =  (\bR_1\trc a)\star(\bR_2\trc a')$ % Make again formula {braid1}?
and the relation
\be
(S_\f \otimes \mbox{id})(\R)=\R_{21}= (\mbox{id}\otimes S_\f^{-1})(\R).
\qquad\Rightarrow   \quad (S_\f \otimes S_\f)(\R)=\R,    \label{SR}
\ee
valid for all triangular Hopf algebras,
one can prove relations (\ref{DCcomrelstar1}-\ref{DCcomrelstar2}) as follows:
\bea
x^i\!\star\! x^j & = & %\stackrel{(\ref{braid})}{=}
\left(\R_2\trc x^j \right) \star  \left(\R_1\trc x^i\right) =x^\mu\star x^\nu\,\check\tau^{\mu j}\left(\R_2\right)  \check\tau^{\nu i}\left(\R_1\right)=R^{\nu\mu}_{ij}x^h\!\star\! x^k,\nn[4pt]
x^i\!\star\! \xi^j & = & %\stackrel{(\ref{braid})}{=}
\left(\R_2\trc \xi^j \right) \star  \left(\R_1\trc x^i\right) =\xi^h\star x^k\,\tau^{hj}\left(\R_2\right)  \tau^{ki}\left(\R_1\right)=R^{kh}_{ij}\xi^h\!\star\! x^k,\nn[4pt]
\xi^i\!\star\! \xi^j & = & %\stackrel{(\ref{braid})}{=}
-R^{kh}_{ij}\xi^h\!\star\! \xi^k \qquad \mbox{obtained from the previous one applying }\mathrm{d}\nn[4pt]
\partial_i'\star  \partial_j' & = & %\stackrel{(\ref{braid})}{=}
\left[\R_2\trc \partial_j'\right]\star  \left[\R_1\trc \partial_i'\right]=
\tau^{jh}\!\left[S_\f\!\left(\R_2\right)\!\right]  \tau^{ik}\!
\left[S_\f\!\left(\R_1\right)\!\right]\partial_h'\star \partial_k'\nn 
& \stackrel{(\ref{SR})}{=} & 
\tau^{jh}\left[\R_2\right]  \tau^{ik} \left[\R_1\right]\partial_h'\star \partial_k'
=R_{kh}^{ij}\partial_h'\star \partial_k',\nn[4pt]
  \partial_i'\star x^j &\stackrel{(\ref{starprod})}{=} &
\left[\bF_1\trc\partial_i'\right] 
 \left[\bF_2\trc x^j\right]=\tau^{ik}\!\left[\beta S\!\left(\bF_1\right)\!\right]
\check\tau^{\mu j}\!\left[\bF_2\right]\partial_k x^\mu\nn[4pt] 
\quad  & = & \tau^{ik}\!\left[\beta S\!\left(\bF_1\right)\!\right]\left[\check\tau^{0 j}\!\left(\bF_2\right)\1\partial_k+\check\tau^{h j}\!\left(\bF_2\right)(\delta^h_k\1+x^h\partial_k)\!\right]\nn 
\quad  & = & \tau^{ik}\!\left[\beta S\!\left(\bF_1\right)\!\right]\left[\1\tau^{k j}\!\left(\bF_2\right)+
\check\tau^{\mu j}\!\left(\bF_2\right)x^\mu\partial_k\!\right]\nn 
\quad & \stackrel{(\ref{starprod})}{=} & \tau^{ij}\!\left[\beta S\!\left(\bF_1\right)\bF_2\!\right]\1+
\tau^{ik}\!\left[\beta S\!\left(\bF_1\right)\!\right]\check\tau^{\mu j}\!\left(\bF_2\right)
\left(\F_1\trc x^\mu\right)\star\left(\F_2\trc\partial_k\!\right)\nn 
\quad  & = & 
\tau^{ij}\!\left[\beta S\!\left(\bF_1\right)\bF_2\!\right]\1
+\check\tau^{\mu j}\!\left[\F_{1'}\bF_2\right]\tau^{ik}\!\left[\beta S\!\left(\F_{2'}\bF_1\right)\!\right]x^\mu\star\partial_k\nn 
\quad  & = & \delta^i_j\1+\check\tau^{\mu j}\!\left[\R_2\right]\tau^{ik}\!\left[\beta S\!\left(\R_1\right)\!\right]x^\mu\star\partial_k =
\delta^i_j\1+\check\tau^{\mu j}\!\left[\R_2\right]\tau^{ik}\!\left[ S_\f\!\left(\R_1\right)\!\beta\right]x^\mu\star\partial_k\nn 
\quad  & = & \delta^i_j\1+\check\tau^{\mu j}\!\left[\R_2\right]\tau^{ik}\!
\left[S_\f\!\left(\R_1\right)\!\right] x^\mu\!\star\!\partial_k'
 \stackrel{(\ref{SR})}{=}\delta^i_j\1+\check\tau^{\mu j}\!\left[\R_1\right]\tau^{ik}\!
\left[\R_2\right] x^\mu\!\star\!\partial_k'\nn 
 & = & \delta^i_j\1+R_{jk}^{\mu i} x^\mu\star\partial_k'.\nonumber
\eea
%CONTROLLA: DAVVERO C'E' $\mu$ AL POSTO DI $h$ NELL'ULTIMA? SI'!
%in JPA2010 avevo scritto where $R^{ab}_{hk}=(\tau^{a}_{h}\ot \tau^{b}_{k})(\R)$,  and
%$R^{ab}_{k(m\!+\!1)}=\delta^a_k\delta^b_{m\!+\!1}=R^{ba}_{(m\!+\!1)k}$.
%% è lo stesso risultato: là gli indici potevano assumere anche il valore (m+1), corrispondente
%a 0 nel lavoro presente visto, che $x^{m+1}=\1$ là  e $x^0=\1$ qua

 By (\ref{lintransf1})  the action of either leg $\F_1,\F_2$ of the twist, or  $\bF_1,\bF_2$  of its inverse, as well as of any tensor factor in the (iterated) coproducts of $\F_1,\F_2,\bF_1,\bF_2$, maps every homogeneous polynomial in $\xi^i$ or $\partial_i$ into another one of the same degree, and  every polynomial in $x^i$  into another one of the same degree: hence 
 (\ref{Qpqr=Qpqrstar}), (\ref{Qproduct+decostar}) follow. Finally, 
the relations $g\trc \partial_i'=\tau^{ij}[S_\f(g)]\partial_j'$,
(\ref{*_FR^n}) are straightforward consequences of 
(\ref{inter-2}), (\ref{eq03}), (\ref{lintransf1}):
\bea
&& g\trc \partial_i'=gS(\beta)\trc \partial_i=\tau^{ij}[\beta S(g)]  \partial_j
=\tau^{ij}\left[S_\f(g)\beta\right]  \partial_j =\tau^{ij}\left[S_\f(g)\right]  \partial_j' \nn[4pt]
&&x^i\,{}^{*_\star} =S(\beta)\trc x^i{}^{*}=
S(\beta)\trc x^i=x^\mu\check\tau^{\mu i} \left[ S(\beta)\right],\nn[4pt]
&&\xi^i\,{}^{*_\star} =S(\beta)\trc \xi^i{}^{*}=
S(\beta)\trc \xi^i=\xi^k\tau^{ki} \left[ S(\beta)\right],\nn[4pt]
&&\partial_i'{}^{*_\star} = S(\beta)\trc\left[S(\beta)\trc\partial_i \right]^*=S(\beta)\beta^*\trc\partial_i^* =-
S(\beta)S(\beta^{-1})\trc \partial_i=-\partial_i=-\tau^{ik}\left(\beta^{-1}\right)\partial_k'. 
\nonumber \eea

\subsection{Proof of  Proposition \ref{propQM_cstar}}

All  statements up to (\ref{QMpqr=QMpqrstar}) and the statement that the
$\star$-polynomials $\hat f_c^J(a\star)$ have the same degrees in 
$x^i,\xi^i,e_\alpha$ as the polynomials $f_c^J(a)$ are straightforward 
consequences of  (\ref{lintransf1})  and of what precedes the proposition.
Under  $U\g$ the $f_i$ transform as the $\partial_i$; in fact,
since  $g\trc f  =\varepsilon(g)f$, we find
$g\trc f_i=g\trc(\partial_i f-f\partial_i )=(g\trc\partial_i) f-f(g\trc\partial_i )=
\tau^{ih}(Sg)(\partial_h f-f\partial_h)=\tau^{ih}(Sg)f_h$.
Hence 
\bea
 g\trc L_{ij} &=& g\trc(f_i\partial_j-f_j\partial_i)=\tau^{ih}(Sg_{(1)})\tau^{jk}(Sg_{(2)})[f_h\partial_k-f_k\partial_h]=\tau^{ih}(Sg_{(1)})\tau^{jk}(Sg_{(2)})L_{hk},\nn[4pt]
L_{ij}^{*_\star}  &=& -S(\beta)\trc L_{ij}=- \tau^{ih}(\beta_{(1)})\tau^{jk}(\beta_{(2)})L_{hk};
\nonumber
\eea
this can be computed more explicitly using the relation (see e.g. eq. (126) in \cite{Fiore2010})
\be 
\Delta(\beta)=\F^{-1}(\beta \ot \beta )[(S\ot S)\F^{-1}_{21}]
=\F^{-1}_{21}(\beta \ot \beta )[(S\ot S)\F^{-1}].   \label{deltabeta}
\ee

\begin{comment}
By (\ref{lintransf1}),
the  polynomials $f_c^J(a)$  can be expressed as  $\star$-polynomials  $\hat f_c^J(a\star)$ of the same degrees in 
$x^i,\xi^i,e_\alpha$ ($\alpha\le B$), and the set of characterizing relations $f_c^J(a)=0$  
is equivalent to the set of relations $\hat f_c^J(a\star)=0$.
In particular, the set of (\ref{DCcomrel1}) is equivalent to the set of (\ref{DCcomrelstar1}).
\end{comment}

\subsection{(a)   Family of parabolic cylinders}

\subsubsection*{Proof of Proposition~\ref{prop01'}}

Since $L_{13}$ and $L_{23}$ are commuting anti-Hermitian vector fields
it follows that $\mathcal{F}$ is a unitary abelian twist on $U\g$.
We find $S(\beta)=\exp(-i\nu L_{12}L_{23})$, and
$$
    L_{13}^nL_{12}    =L_{12}L_{13}^n    +nL_{23}L_{13}^{n-1}
$$
for all $n>0$, since $[L_{13},L_{12}]=L_{23}$, $[L_{13},L_{23}]=0$.
This implies
\begin{align*}
    \mathcal{F}(L_{12}\otimes \1)
    =&\sum_{n=0}^\infty\frac{(i\nu)^n}{n!}L_{13}^nL_{12}\otimes L_{23}^n\\
      =&L_{12}\otimes \1
    +\sum_{n=1}^\infty\frac{(i\nu)^n}{n!}L_{12}L_{13}^n\otimes L_{23}^n
    +\sum_{n=1}^\infty\frac{(i\nu)^n}{n!}nL_{23}L_{13}^{n-1}\otimes L_{23}^n\\
    =&(L_{12}\otimes \1)\mathcal{F}
    +i\nu(L_{23}\otimes L_{23})
    \sum_{n=1}^\infty\frac{(i\nu)^{n-1}}{(n-1)!}L_{13}^{n-1}
    \otimes L_{23}^{n-1}\\
    =&(L_{12}\otimes \1)\mathcal{F}
    +i\nu(L_{23}\otimes L_{23})\mathcal{F},
\end{align*}
$$    \Delta_\f(L_{12})
    =\mathcal{F}(L_{12}\otimes \1+\1\otimes L_{12})\mathcal{F}^{-1}\\
      =L_{12}\otimes \1+\1\otimes L_{12}
    +i\nu L_{23}\otimes L_{23},
$$
where in the last equation we use $\mathcal{F}(\1\otimes L_{12})
=(\1\otimes L_{12})\mathcal{F}$ since the second leg of the twist is central.
Moreover
$\mathcal{F}(L_{13}\otimes \1)
=\sum_{n=0}^\infty\frac{(i\nu)^n}{n!}L_{13}^{n+1}\otimes L_{23}^n
=(L_{13}\otimes \1)\mathcal{F}$ and $\mathcal{F}(L_{23}\otimes \1)
=(L_{23}\otimes \1)\mathcal{F}$ show that
$\Delta_\f(L_{13})=\Delta(L_{13})$ and
$\Delta_\f(L_{23})=\Delta(L_{23})$. We have thus proved  
  the claimed coproducts $\Delta_\f(L_{ij})$. Next, the latter
and the antipode property
$ \mu[(S_\f\otimes\mathrm{id})\Delta_\f(g)]
   =\epsilon(L_{ij})\1=0$ easily determine
the twisted antipode $S_\f(L_{12})$ as in (\ref{DeltaSParCyl}),  and other ones
$S_\f(L_{ij})=S(L_{ij})=-L_{ij}$. 
Furthermore, since the $L_{23}$ contained in the second leg of the twist commutes with $L_{k\ell}$ we conclude that
$L_{ij}\star L_{k\ell}=L_{ij} L_{k\ell}$ and
$[L_{ij},L_{k\ell}]_\star=[L_{ij},L_{k\ell}]$ for all $1\leq i<j\leq 3$
and $1\leq k<\ell\leq 3$. For the same reason one gets  for all $1\leq i,j,k\leq 3$
$$
x^i\star L_{jk}=x^iL_{jk},\quad\xi^i\star L_{jk}=\xi^iL_{jk},\quad\partial_i\star L_{jk}=\partial_iL_{jk}.
$$
On the other hand by (\ref{ParCyl-gaction}) we obtain
\begin{align*}
L_{ij}\star x^k =& L_{ij}x^k-i\nu b\epsilon_{ij3}\delta^k_2L_{23},\quad\:
L_{ij}\star\xi^k =L_{ij}\xi^k,\quad\: L_{ij}\star\partial_k =L_{ij}\partial_k,\\[6pt]
x^i\star x^j    =&x^ix^j-i\nu b\delta^j_2(\delta^i_1b+\delta^i_3x^1),\quad
   \partial_i\star\partial_j    =\partial_i\partial_j,\quad    \xi^i\star\xi^j
    =\xi^i\xi^j,\\[6pt]
    \partial_i\star x^j    =&x^j\partial_i+i\nu b\delta_{i1}\delta^j_2\partial_3,\qquad
 x^i\star\partial_j    =x^i\partial_j, \qquad   \xi^i\star\partial_j    =\xi^i\partial_j,\\[6pt]
       \xi^i\star x^j    =&x^j\xi^i-i\nu b\delta^i_3\delta^j_2\xi^1,\qquad
    x^i\star\xi^j    =x^i\xi^j,\qquad
    \partial_i\star\xi^j    =\partial_i\xi^j,
\end{align*}
for all $1\leq i,j,k\leq 3$. The commutation relations respectively follow.
Furthermore this means we can express the generators of the Lie algebra
in terms of the twisted module action, namely 
$L_{12}=x^1\partial_2=x^1\star\partial_2$, 
$L_{13}=x^1\partial_3+b\partial_1=x^1\star\partial_3+b\partial_1$ and
$L_{23}=b\partial_2$, while
$f_c(x)=\frac{1}{2}(x^1)^2-bx^3-c
=\frac{1}{2}x^1\star x^1-bx^3-c$,
$\mathrm{d}f_c=x^1\xi^1-b\xi^3
=x^1\star\xi^1-b\xi^3$ and
$$
\epsilon^{ijk}f_iL_{jk}
=2(x^1L_{23}-bL_{12})
=2(x^1\star L_{23}-bL_{12})
$$
hold. Again by (\ref{ParCyl-gaction})
\ $L_{12}L_{23}\trc x^i= 0$,  $L_{12}L_{23}\trc \xi^i= 0$, 
$L_{12}L_{23}\trc \partial_i= 0$ for all $i=1,2,3$,  
which implies that the $*$-structure on $\Q^\bullet_\star$
remains the same as on $\QM^\bullet$:
$*_\star=*$.

\subsection{(b)   Family of elliptic paraboloids}

\subsubsection*{Proof of Proposition~\ref{prop07}}

Since $L_{13},L_{23}$ are commuting anti-Hermitian vector fields
$\mathcal{F}$ is a unitary abelian twist on $U\g$.
By a direct calculation one finds
$ \beta
    =S(\beta)
    =\mathcal{F}_2S(\mathcal{F}_1)
%    =\sum_{n=0}^\infty\frac{(-i\nu)^n}{n!}L_{23}^nL_{13}^n
    =\exp(-i\nu L_{13}L_{23}).
$
The commutation relations (\ref{eq04'}) also imply \ 
$\mathcal{F}\Delta(L_{13})=\Delta(L_{13})\mathcal{F}$,
$\mathcal{F}\Delta(L_{23})=\Delta(L_{23})\mathcal{F}$, resulting in
$\Delta_\f(L_{13})=\Delta(L_{13})$ and
$\Delta_\f(L_{13})=\Delta(L_{13})$, respectively. Moreover,
$$    
L_{13}^nL_{12}    =L_{12}L_{13}^n+nL_{23}L_{13}^{n-1}, \quad
    \text{ and }\quad
    L_{23}^nL_{12}
    =L_{12}L_{23}^n-naL_{13}L_{23}^{n-1}
$$%
for $n>0$, which follow by iteratively applying eq.(\ref{eq04'}). Then
\begin{align*}
    \mathcal{F}(L_{12}\otimes \1)
    =&\sum_{n=0}^\infty\frac{(i\nu)^n}{n!}L_{13}^nL_{12}
    \otimes L_{23}^n\\
    =&L_{12}\otimes \1+\sum_{n=1}^\infty\frac{(i\nu)^n}{n!}L_{12}L_{13}^n
    \otimes L_{23}^n
    +\sum_{n=1}^\infty\frac{(i\nu)^n}{n!}nL_{23}L_{13}^{n-1}
    \otimes L_{23}^n\\
    =&(L_{12}\otimes \1)\mathcal{F}
    +i\nu(L_{23}\otimes L_{23})\mathcal{F},
\end{align*}
\begin{align*}
    \mathcal{F}(\1\otimes L_{12})
    =&\1\otimes L_{12}
    +\sum_{n=1}^\infty\frac{(i\nu)^n}{n!}L_{13}^n
    \otimes L_{23}^nL_{12}\\
    =&\1\otimes L_{12}
    +\sum_{n=1}^\infty\frac{(i\nu)^n}{n!}L_{13}^n
    \otimes L_{12}L_{23}^n
    -a\sum_{n=1}^\infty\frac{(i\nu)^n}{n!}nL_{13}^n
    \otimes L_{13}L_{23}^{n-1}\\
    =&(\1\otimes L_{12})\mathcal{F}
    -i\nu a(L_{13}\otimes L_{13})\mathcal{F}
\end{align*}
imply (\ref{DSEllypticParab})$_1$. The
twisted antipodes   (\ref{DSEllypticParab})$_2$ follow
using the properties 
$\mu\circ(S_\f\otimes\mathrm{id})\circ\Delta_\f
=\eta\circ\epsilon
=\mu\circ(\mathrm{id}\otimes S_\f)\circ\Delta_\f$.
The twisted tensor  and star products
%of the generators of $\mathfrak{g}$ 
coincide with the untwisted ones
as soon as one of the factors is $L_{13}$ or $L_{23}$. This
is because the latter commute with both legs of the twist. Among all star products
of generators of $\g$ only the one
\begin{align*}
    L_{12}\star L_{12}
    =\sum_{n=0}^\infty\frac{(-i\nu)^n}{n!}
    (L_{13}^n\rhd L_{12})(L_{23}^n\rhd L_{12})
    =L_{12} L_{12}
    +i\nu a L_{23} L_{13}
\end{align*}
is different. By a similar direct calculation one can prove
(\ref{starprodEllypticParab}). The latter imply
Eq. (\ref{starcomEllypticParab1}-\ref{starcomEllypticParab2}) 
and that 
the submanifold constraints coincide with their twisted analogues, namely
(\ref{characterizingEllPar}) holds.
The twisted star involutions coincide with the untwisted ones, since
$L_{13}L_{23}\rhd x^i=L_{13}L_{23}\rhd\xi^i=L_{13}L_{23}\rhd\partial_i=0$.
This concludes the proof of the proposition.

\subsection{(c)   Family of elliptic cylinders}

\subsubsection*{Proof of Proposition 16 in \cite{FioreWeber2020}, see section \ref{prop03}
}

Since $[\partial_3,L_{12}]=0$ and $\partial_3,L_{12}$ are anti-Hermitian,
$\mathcal{F}=\exp(i\nu\partial_3\otimes
L_{12})$ is a unitary abelian twist on $U\g$. \  As
$\partial_3,L_{12}$ commute with both legs of the twist, 
$\Delta_\f(\partial_3)=\Delta(\partial_3)$,
$\Delta_\f(L_{12})=\Delta(L_{12})$,
$S_\f(\partial_3)=S(\partial_3)$,
$S_\f(L_{12})=S(L_{12})$ and all twisted tensor and star products as well as
Lie brackets where one of the factors is $\partial_3,L_{12}$ coincide with the untwisted
ones. Furthermore the star products of $x^i,\xi^j,\partial^k,L_{12}$
coincide with the classical ones unless the first leg of the twist acts on $x^3$.
Consequently, eq. (\ref{EllParL12})$_{b=0}$ implies (\ref{eq07}) and
the equations (\ref{eq08}) coincide with their classical analogues.
The twisted star involutions are trivial since
$\partial_3L_{12}\rhd x^i=\partial_3L_{12}\rhd\xi^i
=\partial_3L_{12}\rhd\partial_i=0$. This concludes the proof.

\subsection{(d)   Family of hyperbolic  paraboloids}

\subsubsection*{Proof of Proposition~\ref{prop08}}

The anti-Hermitian vector fields $H$ and $E$ satisfy eq.(\ref{eq11}),
which implies that $\mathcal{F}=\exp(H/2\otimes\log(\1+i\nu E))$ is a
unitary Jordanian twist on $U\mathfrak{g}$. We note that
$$
    E^m H= H E^m-2m E^m
$$
for all $m>0$, which follows by iteratively applying (\ref{eq11}).
In particular this implies
\begin{align*}
   & \log(\1+i\nu E) H
    =-\sum_{m=1}^\infty\frac{(-i\nu)^m}{m}E^m H=
- H\sum_{m=1}^\infty\frac{(-i\nu)^m}{m}E^m
    +2\sum_{m=1}^\infty\frac{(-i\nu)^m}{m}mE^m\\
    & =H\log(\1+i\nu E)
    +2(-i\nu)E\sum_{m=1}^\infty(-i\nu E)^{m-1}= H\log(\1+i\nu E)
    -2i\nu E\frac{\1}{\1+i\nu E},
\end{align*}
where we have made use of the expansions
$$    
\log(\1+i\nu E)    =-\sum_{m=1}^\infty\frac{(-i\nu E)^m}{m},\qquad\qquad
    \frac{\1}{\1+i\nu E}    =\sum_{m=0}^\infty(-i\nu E)^m.
$$
 Both are well-defined in
the $\nu$-adic topology. Applying this result iteratively we obtain
$$
   [ \log(\1+i\nu E)]^n H    = H[\log(\1+i\nu E)]^n
    -2n \frac{i\nu E}{\1+i\nu E}[\log(\1+i\nu E)]^{n-1}
$$
for all $n>0$, whence
\begin{align*}
    \mathcal{F}(\1\otimes  H)
    =&\sum_{n=0}^\infty\frac{1}{n!}\bigg(\frac{H}{2}\bigg)^n
    \otimes[\log(\1+i\nu E)]^n H\\
     =&\sum_{n=0}^\infty\frac{1}{n!}\bigg(\frac{H}{2}\bigg)^n
    \otimes  H[\log(\1+i\nu E)]^n
-2\sum_{n=1}^\infty\frac{n}{n!}\bigg(\frac{H}{2}\bigg)^n
    \otimes\frac{i\nu E}{\1+i\nu E}[\log(\1+i\nu E)]^{n-1}\\
    =&(\1\otimes  H)\mathcal{F}-
    \bigg(H\otimes\frac{i\nu E}{\1+i\nu E}\bigg)\mathcal{F}.
\end{align*}
This and $\F(H\ot \1)=(H\ot \1)\F$ determine
$\Delta_\f(H)$ as in (\ref{DeltaSHypPar}). 
For the twisted coproduct of $E$ we first remark that
$\left(\frac{H}{2}\right)^nE
    =E\left(\frac{H}{2}+\1\right)^n$
for all $n\geq 0$, which is proven   by induction. Then
\begin{align*}
    \mathcal{F}(E\otimes \1)
    =&\sum_{n=0}^\infty\frac{1}{n!}\bigg(\frac{H}{2}\bigg)^nE
    \otimes\log(\1+i\nu E)^n\\
    =&(E\otimes \1)\sum_{n=0}^\infty\frac{1}{n!}\bigg(\frac{H}{2}+\1\bigg)^n
    \otimes\log(\1+i\nu E)^n\\
    =&(E\otimes \1)\exp\bigg(\bigg(\frac{H}{2}+\1\bigg)
    \otimes\log(\1+i\nu E)\bigg)\\
    =&(E\otimes \1)\exp\bigg(\1\otimes\log(\1+i\nu E)\bigg)\exp\bigg(\frac{H}{2}
    \otimes\log(\1+i\nu E)\bigg)\\
    =&[E\otimes (\1+i\nu E)]\F.
\end{align*}
This and $\F(\1\ot E)=(\1\ot E)\F$ determine
$\Delta_\f(E)$ as in (\ref{DeltaSHypPar}). 
Similarly one proves 
\bea
\mathcal{F}(E'\otimes \1)
        &=&(E'\otimes \1)\exp\bigg(-\1\otimes\log(\1+i\nu E)\bigg)\exp\bigg(\frac{H}{2}
    \otimes\log(\1+i\nu E)\bigg)\nn
    &=&\left(E'\otimes \frac \1{\1+i\nu E}\right)\F             \label{FE'ot1}
\eea
and $\F(\1\ot E')=(\1\ot E')\F$, which determine
$\Delta_\f(E')$ as in (\ref{DeltaSHypPar}). 
Next, it is straightforward to check that   the coproducts $\Delta_\f(g)$,
with $g=H,E,E'$, and the antipode property
$ \mu[(S_\f\otimes\mathrm{id})\Delta_\f(g)]
   =\epsilon(g)\1=0$ determine
the twisted antipodes $S_\f(g)$ as in (\ref{DeltaSHypPar}). 

To compute the twisted tensor and star products we first make only the first leg $\bF_1$
of $\bF$ to act on its eigenvectors \ $H,E,E',u^i,\partial_i$ (generators of $U\g$ and of $\Q$), and find 
\bea
\bF_1\trc H\otimes \bF_2 &=&  H\otimes \1,\nn
\bF_1\trc E\otimes \bF_2
&=& E\otimes\exp\bigg(-\log(\1+i\nu E)\bigg)= E\otimes(\1+i\nu E)^{-1},\nn
  \bF_1\trc E'\otimes \bF_2
&=& E'\otimes\exp\bigg(\log(\1+i\nu E)\bigg)= E'\otimes(\1+i\nu E),\nn
\bF_1\trc u^i\otimes \bF_2
&=& u^i\otimes\exp\bigg(-\frac{\lambda_i}{2}\log(\1+i\nu E)\bigg)=
 u^i\otimes(\1+i\nu E)^{-\frac{\lambda_i}{2}},\nn
%    &=& \begin{cases}
%    u^1\otimes(1+i\nu E)^{-1} & \text{ if }i=1\\
%    u^2\otimes 1 & \text{ if }i=2\\
%    u^3\otimes(1+i\nu E) & \text{ if }i=3
%    \end{cases}\nn
\bF_1\trc  \partial_i\otimes \bF_2
&=& \partial_i\otimes\exp\bigg(\frac{\lambda_i}{2}\log(\1+i\nu E)\bigg)=
  \partial_i\otimes(\1+i\nu E)^{\frac{\lambda_i}{2}},
% \nn      &=& \begin{cases}
%    \partial_1\otimes(1+i\nu E) & \text{ if }i=1\\
%    \partial_2\otimes 1 & \text{ if }i=2\\
%    \partial_3\otimes(1+i\nu E)^{-1} & \text{ if }i=3
%    \end{cases} 
\nonumber
\eea
for all $u^i\in\{y^i,\eta^i\}$,  $1\leq i \leq 3$; note that the exponents $\pm \lambda_i/2$ take the values
$\pm 1,0$. This simplifies the computation
of the action of the second leg $\bF_2$ on the second factor; using (\ref{eq11}) and (\ref{gQaction}) and noting that 
only the terms of degree lower than two in the power
expansion of $\1/(\1+i\nu E)$ contribute to its action on the $H,E,E',u^i,\partial_i$, by a direct computation one thus finds
the star products (\ref{starprodgHypPar}-\ref{starprodgQHypPar}). 
In particular the twisted tensor and star products are trivial if $H,u^3$ or $\tilde \partial_3$ appears in the first factor. 
The  twisted commutation relations (\ref{starcomrelQHypPar})-(\ref{starcomrelQgHypPar}) 
and the twisted submanifold constraints (\ref{starcomrelQMHypPar}) follow. 
For the twisted star involution note that
\bea
    S(\beta)\rhd y^i    =\sum_{n=0}^\infty\frac{1}{n!}[\log(\1+i\nu E)]^n
    \bigg(-\frac{H}{2}\bigg)^n\rhd y^i=(\1+i\nu E)^{-\frac{\lambda_i}{2}}\rhd y^i=y^i+i\nu 2b\delta^i_2\nn
    S(\beta)\rhd \tilde{\partial}_i    =\sum_{n=0}^\infty\frac{1}{n!}[\log(\1+i\nu E)]^n
    \bigg(-\frac{H}{2}\bigg)^n\rhd \tilde{\partial}_i=(\1+i\nu E)^{\frac{\lambda_i}{2}}\rhd \tilde{\partial}_i=\tilde{\partial}_i-i\nu\delta_{i1}\tilde{\partial}_3, \nonumber
\eea
since $E\rhd y^i=\delta^i_3y^1+2b\delta^i_2$ and $\lambda_3=0$, $\lambda_2=-1$, while
$E\rhd\tilde{\partial}_i=-\delta_{i1}\tilde{\partial}_3$, and $\lambda_1=1$. 

\subsection{(d-e-f)   Elliptic cone and hyperboloids}

\subsubsection{Proof of Proposition 17 in \cite{FioreWeber2020}, see section~\ref{prop05}}

From the anti-Hermiticity of the vector fields $H,E$ and from $[H,E]=2E$ it 
follows that $\mathcal{F}=\exp(H/2\otimes\log(\1+i\nu E))$ is a
unitary Jordanian twist on $U\mathfrak{g}$ and that the coproducts and antipodes
of $H,E$ are exactly as in case (d). Similarly, (\ref{FE'ot1}) holds, because it is only based
on the relation $[H,E']=-2E'$. 

To compute $\Delta_\f(E')$ we first  determine 
 $\F(\1\ot E')$.  We use \ $[H,E]=2E$, $EE'=E'E-H$, \ and   find  by induction first  that \ $ E^nH    =(H-2n)E^n$, \ then 
\begin{align*}
    E^nE'
    =&E^{n-1}E'E-E^{n-1}H=E^{n-2}E'E^2-E^{n-2}HE-E^{n-1}H\\
    =&\ldots\\
     =&E'E^n-HE^{n-1}-EHE^{n-2}-\ldots -E^{n-2}HE-E^{n-1}H\\
     =&E'E^n+[-nH+2(1+2+\ldots+n-1)]E^{n-1}\\
       =&E'E^n-nHE^{n-1}
    +n(n-1)E^{n-1}
\end{align*}
for all $n\geq 0$, by the
``little Gauss" $\sum_{h=1}^{n-1}h=\frac{n^2-n}{2}$. Consequently,
using the series expansions
$$
\log(\1\!+\!i\nu E)
=-\sum_{n=1}^\infty\frac{(-i\nu E)^n}{n},\quad
\frac{\1}{\1\!+\!i\nu E}
=\sum_{n=0}^\infty(-i\nu E)^n,\quad
\frac{\1}{(\1\!+\!i\nu E)^2}
=\sum_{n=1}^\infty n(-i\nu E)^{n-1},
$$
we obtain
\bea
    \log(\1+i\nu E)H
    &=&-\sum_{n=1}^\infty\frac{(-i\nu E)^n}{n}H=-\sum_{n=1}^\infty(H-2n)\frac{(-i\nu E)^n}{n}=H\log(\1+i\nu E) -\frac{2i\nu E}{1+i\nu E},\nn
    \log(\1+i\nu E)E'
    &=&-\sum_{n=1}^\infty\frac{(-i\nu)^n}{n}E^n E'\nn
    &=&-E'\sum_{n=1}^\infty\frac{(-i\nu)^n}{n}E^n
    -i\nu H\sum_{n=1}^\infty(-i\nu E)^{n-1}+i\nu\sum_{n=2}^\infty (n\!-\!1)(-i\nu E)^{n-1}\nn
    &=&E'\log(\1+i\nu E)+\frac{\nu^2E}{(\1+i\nu E)^2} -H\frac{i\nu}{\1+i\nu E} ,\nonumber
\eea
and in turn
\begin{align*}
    [\log(\1+i\nu E)]^nH
    =&[\log(\1+i\nu E)]^{n-1}H\log(\1+i\nu E)
    -\frac{2i\nu E}{\1+i\nu E}[\log(\1+i\nu E)]^{n-1}\\
    =& \ldots = H[\log(\1+i\nu E)]^{n}
    -2n\frac{i\nu E}{\1+i\nu E}[\log(\1+i\nu E)]^{n-1},
\end{align*}
\begin{align*}
   [ \log(\1+i\nu E)]^nE'
    =&\: [\log(\1+i\nu E)]^{n-1}\left\{E'\log(\1+i\nu E)+\frac{\nu^2E}{(\1+i\nu E)^2} -H\frac{i\nu}{\1+i\nu E}\right\}\\
    =&\:   [\log(\1+i\nu E)]^{n-1}E'\log(\1+i\nu E)+\frac{\nu^2E[\log(\1+i\nu E)]^{n-1}}{(\1+i\nu E)^2} \\
    &-i\nu H\frac{[\log(\1+i\nu E)]^{n-1}}{\1+i\nu E}
      -2(n-1)\frac{\nu^2 E}{(\1+i\nu E)^2}[\log(\1+i\nu E)]^{n-2}\\
      =&\:  \ldots=   E'\log(\1\!+\!i\nu E)]^n + n\frac{\nu^2E[\log(\1\!+\!i\nu E)]^{n-1}}{(\1\!+\!i\nu E)^2}
- n i\nu H\frac{[\log(\1\!+\!i\nu E)]^{n-1}}{\1\!+\!i\nu E}\\
    &     -2[(n-1)+(n-2)+\cdots+1]\frac{\nu^2 E[\log(\1+i\nu E)]^{n-2}}{(\1+i\nu E)^2}
\end{align*}
$$ =\: E'\log(\1\!+\!i\nu E)]^n
    +n\left[\frac{\nu^2E}{\1\!+\!i\nu E}\!-\!i\nu H\right]\frac{[\log(\1\!+\!i\nu E)]^{n-1}}{\1\!+\!i\nu E }
    -n(n-1)\frac{\nu^2 E[\log(\1\!+\!i\nu E)]^{n-2}}{(\1\!+\!i\nu E)^2}.
$$
Hence
\bea
    \mathcal{F}(\1\otimes E')
    & = &    \sum_{n=0}^\infty\frac{1}{n!}\bigg(\frac{H}{2}\bigg)^n
    \otimes[\log(\1+i\nu E)]^nE'\nn
   & = &    \sum_{n=0}^\infty\frac{1}{n!}\bigg(\frac{H}{2}\bigg)^n
    \otimes\left\{E'\log(\1\!+\!i\nu E)]^n
 +n\left[\frac{\nu^2E}{\1\!+\!i\nu E}\!-\!i\nu H\right]\frac{[\log(\1\!+\!i\nu E)]^{n-1}}{\1\!+\!i\nu E }\right.\nn
&&\left.   \qquad\qquad \qquad \qquad -n(n-1)\frac{\nu^2 E\log(\1\!+\!i\nu E)]^{n-2}}{(\1\!+\!i\nu E)^2}\right\}\nn
    & = &\left\{\1\ot E'+\frac{H}{2}\ot
\left[\frac{\nu^2E}{\1\!+\!i\nu E}\!-\!i\nu H\right]\frac{\1}{\1\!+\!i\nu E }
-\bigg(\frac{H}{2}\bigg)^2\ot \frac{\nu^2 E}{(\1\!+\!i\nu E)^2}\right\}\mathcal{F}.\nonumber
\eea
On the other hand,  using $\big(\frac{H}{2}\big)^nE'=E'\big(\frac{H}{2}-\1\big)^n$ we obtain
\bea
    \mathcal{F}(E'\otimes \1)
    &=&\sum_{n=0}^\infty\frac{1}{n!}\bigg(\frac{H}{2}\bigg)^n E'
    \otimes[\log(\1+i\nu E)]^n=\sum_{n=0}^\infty\frac{1}{n!}E'\bigg(\frac{H}{2}-\1\bigg)^n 
    \otimes[\log(\1+i\nu E)]^n\nn
    &=&(E'\otimes \1)
    \exp\bigg[\bigg(\frac{H}{2}-\1\bigg)\otimes\log(\1+i\nu E)\bigg]
=\left(E'\otimes \frac{\1}{\1\!+\!i\nu E }\right)\F. \nonumber
\eea
Summing the last two equations we find that  $\Delta_\f(E')$ is as in (\ref{copr}).
The antipode $S_\f(E')$ follows from  the antipode property
$ \mu[(S_\f\otimes\mathrm{id})\Delta_\f(E')]=0$. 

To compute the twisted tensor and star products we first make only the first leg $\bF_1$
of $\bF$ to act on its eigenvectors \ $H,E,E',u^i$ (generators of $U\g$ and of $\Q$), and find 
\bea
\ba{rcl}
\bF_1\trc H\otimes \bF_2 &=&  H\otimes \1,\\[6pt]
\bF_1\trc E\otimes \bF_2
&=& E\otimes\exp\left[-\log(\1+i\nu E)\right]= E\otimes(\1+i\nu E)^{-1},\\[6pt]
  \bF_1\trc E'\otimes \bF_2
&=& E'\otimes\exp\left[\log(\1+i\nu E)\right]= E'\otimes(\1+i\nu E),\\[6pt]
\bF_1\trc u^i\otimes \bF_2
&=& u^i\otimes\exp\left[-\frac{\lambda_i}{2}\log(\1+i\nu E)\right]=
 u^i\otimes(\1+i\nu E)^{-\frac{\lambda_i}{2}},\\[6pt]
     &=& \begin{cases}
    u^1\otimes(\1+i\nu E)^{-1} & \text{ if }i=1,\\
    u^2\otimes \1 & \text{ if }i=2,\\
    u^3\otimes(\1+i\nu E) & \text{ if }i=3,
    \end{cases}
\ea \label{utile}
\eea
where $u^i\in\{y^i,\eta^i,\tilde{\partial}^i\}$,  $1\leq i \leq 3$; 
note that the exponents $\pm \lambda_i/2$ take the values $\pm 1,0$. 
By the first relation the twisted tensor or star products are trivial if
$H$ or some $u^2$ is the first factor. The following two  imply
(\ref{gstargEllHyp}). Moreover, 
for all $u^i,v^i\in\{y^i,\eta^i,\tilde{\partial}^i\}$, we find
\begin{comment}
\bea
u^i\otimes_\star v^j &=&\begin{cases}
    u^1\otimes\left(1-i\nu E
    -\nu^2 E^2\right) \trc v^j& \text{ if }i=1\\
    u^2\otimes v^j & \text{ if }i=2\\
    u^3\otimes(1+i\nu E)\rhd v^j & \text{ if }i=3
    \end{cases}\nn
 &=&\begin{cases}
    u^1\otimes\left[v^j-i\nu\left(\delta^j_2\frac{1}{\sqrt{a}}v^1
    -2\delta^j_3\sqrt{a}v^2\right)
    +2\nu^2 \delta^j_3 v^1\right] & \text{ if }i=1\\
    u^2\otimes v^j & \text{ if }i=2\\
    u^3\otimes\left[v^j+i\nu  \left(\delta^j_2\frac{1}{\sqrt{a}}v^1
    -2\delta^j_3\sqrt{a}v^2\right)\right]  & \text{ if }i=3
    \end{cases}
\eea
\end{comment}
\bea
u^i\star v^j =\left\{\!\!\ba{l}
    u^1 \left(\1\!-\!i\nu E
    \!-\!\nu^2 E^2\right) \rhd v^j\\
    u^2 v^j \\
    u^3(\1+i\nu E)\rhd v^j 
          \ea     \right.\!\!=\begin{cases}
    u^1\left[v^j\!-\!i\nu\left(\frac{\delta^j_2}{\sqrt{a}}v^1
    \!-\!2\delta^j_3\sqrt{a}v^2\right)
    \!+\!2\nu^2 \delta^j_3 v^1\right] & \:\:\text{ if }i=1,\\
    u^2 v^j & \:\:\text{ if }i=2,\\
    u^3\left[v^j\!+\!i\nu  \left(\delta^j_2\frac{1}{\sqrt{a}}v^1
    \!-\!2\delta^j_3\sqrt{a}v^2\right)\right]  & \:\:\text{ if }i=3.
    \end{cases}\nonumber
\eea
By explicit calculations these imply relations (\ref{starEllHyp}-\ref{gstarEllHyp}), as
well as (\ref{QMstarEllHyp}), once one notes that
$$
\frac2{\sqrt{ab}}\varepsilon^{ijk}f_iL_{jk}=y^3E-y^1E'-\sqrt{a}y^2H
=y^3\star E-y^1\star E'-\sqrt{a}\,y^2\star H+i\nu y^1\star H-2i\nu (1+i\nu) y^1\star E.
$$

To determine $u^i{}^{*_\star}= S(\beta)\rhd u^i{}^*$ recall that 
$\beta=\mathcal{F}_1S(\mathcal{F}_2)$. Then
\begin{align*}
    S(\beta)\rhd u^i
    =&(\mathcal{F}_2S(\mathcal{F}_1))\rhd u^i=\bigg(\sum_{n=0}^\infty\frac{1}{n!}
    \log(\1+i\nu E)^n\bigg(-\frac{H}{2}\bigg)^n\bigg)\rhd u^i\\
    =&\bigg(\sum_{n=0}^\infty\frac{1}{n!}
    \log(\1+i\nu E)^n\bigg(-\frac{\lambda_i}{2}\bigg)^n\bigg)\rhd u^i=
(\1+i\nu E)^{-\frac{\lambda_i}{2}}\rhd u^i\\
    =&\left\{\!\ba{l}
    \sum_{n=0}^\infty(-i\nu E)^n\rhd u^1 \\
    u^2 \\
    (\1+i\nu E)\rhd u^3 
          \ea     \right.=\begin{cases}
     u^1 & \text{ if }i=1,\\
    u^2 & \text{ if }i=2,\\
     u^3 -2i\nu\sqrt{a} u^2& \text{ if }i=3.
    \end{cases}
\end{align*}

\subsubsection{Metric and principal curvatures on the circular hyperboloids and cone}
\label{ClassicalCircularHyperboloids}

One can easily check the statements of the first paragraph of section \ref{CircHyperb} using e.g. the basis 
$S:=\{v_1,v_2,v_3\}$ of $\Xi$, % (alternative to $\{\partial_1,\partial_2,\partial_3\}$),
where
\bea
v_1:=L_{12}%=(x^1\partial_2-x^2\partial_1)
,\qquad
v_2:=\rho^2\partial_3+x^3(x^1\partial_1\!+\!x^2\partial_2),\qquad
v_3:=\Vp=f_j\eta^{ji}\partial_i=x^i\partial_i;
\eea
we have abbreviated $\rho^2\equiv (x^1)^2\!+\!(x^2)^2$.
$S$ is orthogonal with respect to $\gm$, while $S_t:=\{v_1,v_2\}$  is 
an orthogonal basis of $\Xi_t$ with respect to  $\gm_t$, since, by an easy computation,
\bea
\gm(v_i,v_j)=\left\{\ba{ll}
\rho^2 \quad &\mbox{ if }i=j=1,\\
-\E \rho^2\quad &\mbox{ if }i=j=2,\\
\E \quad &\mbox{ if }i=j=3,\\
0  \quad &\mbox{ otherwise},
 \ea\right. \label{metricv_i}
\eea
where $\E(x)\equiv f^i(x)f_i(x)=x^ix_i$. Since  $\E=2c$ on $M_c$, $M_c$ is indeed Riemannian if $c<0$, Lorentzian if $c>0$,  whereas
the metric induced on the cone $M_0$  is degenerate. 
One can easily check (\ref{IIFF}), (\ref{curvatures})  on such a $S_t$ by explicit computations.
The dual basis consists of
\be
%\vartheta^1=\frac{x^1\xi^2-x^2\xi^1}{\rho^2},\qquad\vartheta^2=
%\frac{x^3(x^1\xi^1+x^2\xi^2)-\rho^2\xi^3}{\E\rho^2},\qquad\vartheta^3=\frac{df}{\E}.
\vartheta^1=\frac 1{\rho^2}(x^1\xi^2-x^2\xi^1),\qquad\vartheta^2=\frac 1{\E\rho^2}
[x^3(x^1\xi^1+x^2\xi^2)-\rho^2\xi^3],\qquad\vartheta^3=\frac 1 \E df.
\ee
The principal curvatures on $M_c$ 
are indeed \ $-\mbox{sign}(c)/\sqrt{|2c|}$, $1/\sqrt{|2c|}$, because in the `orthonormal' 
basis  $S:=\{e_1,e_2,e_3\}$, with
$e_1:=\frac 1{\rho} v_1$, $e_2:=\frac 1{\rho\sqrt{|\E|}} v_2$, $e_3:=\Np=\frac 1{\sqrt{|\E|}} \Vp$, one finds
\bea
\gm(e_i,e_j)=\left\{\!\ba{ll}
1 \:\: &\mbox{ if }i=j=1,\\
-\mbox{sign}(\E) \:\: &\mbox{ if }i=j=2,\\
\mbox{sign}(\E) \:\: &\mbox{ if }i=j=3,\\
0  \:\:&\mbox{ otherwise},
 \ea\right.\qquad\quad 
II(e_\alpha,e_\beta)=-\frac {\mbox{sign}(\E)}{\sqrt{|\E|}}\gm(e_\alpha,e_\beta)e_3.
%,\quad  \alpha,\beta\in\{1,2\}.
\eea 
On $H,E,E'$ the metric gives
\bea
\ba{lll}
\gm(E,E)=(y^1)^2,\qquad &\gm(E',E')=(y^3)^2,\qquad &\gm(H,H)=-8y^1y^3,\\[8pt]
\gm(E,E')%=\gm(E',E)
=\E+(y^2)^2\qquad &\gm(E,H)=-2y^1y^2,\qquad &\gm(E',H)=-2y^2y^3,
\ea\label{metricHEE'}
\eea
and the same results in the last line if we flip the arguments.
To prove (\ref{gmtstar}) we use (\ref{II^F}), %(\ref{metricHEE'}),
 (\ref{twistedmetric2}), (\ref{utile}), the $\X$-linearity of $\gm$. 
To prove (\ref{IIfstar})  we use (\ref{II^F}), (\ref{utile}).
The undeformed version of (\ref{curvaturesstar}) follows from  (\ref{curvatures})
by $\X$-linearity. We prove (\ref{curvaturesstar}) using  (\ref{II^F}),  (\ref{cocycle}),
the definition of $\bR$:
\bea
\rR^\f_{t\star}(X,Y,Z) = \rR_{t}\!\left(\bF_1\trc X,\bF_2\trc Y,\bF_3\trc Z\right)
= (A-B)/2c,\qquad\mbox{where}\nonumber
\eea
\vskip-.8cm
\bea
A &:=&\left(\bF_2\trc Y\right) \gm\!\left(\bF_1\trc X,\bF_3\trc Z\right)
=\left(\bF_{1(2)}\bF_2'\trc Y\right) \gm\!\left(\bF_{1(1)}\bF_1'\trc  X,\bF_2\trc Z\right)\nn
&=& \left(\bF_{1(1)}\bF_1''\F_1'''\bF_2'\trc Y\right) \gm\!\left(\bF_{1(2)}\bF_2''\F_2'''\bF_1'\trc  X,\bF_2\trc Z\right)\nn
&=& \left(\bF_{1}\F_1'''\bF_2'\trc Y\right) \gm\!\left(\bF_{2(1)}\bF_1''\F_2'''\bF_1'\trc  X,\bF_{2(2)}\bF_2''\trc Z\right)\nn
&=& \left(\bF_{1}\bR_1\trc Y\right) \gm\!\left(\bF_{2(1)}\bF_1\bR_2\trc  X,\bF_{2(2)}\bF_2''\trc Z\right)
=\left(\bF_{1}\bR_1\trc Y\right) \bF_{2}\trc\gm\!\left(\bF_1\bR_2\trc  X,\bF_2''\trc Z\right)\nn
&=& \left(\bR_1\trc Y\right) \star\gm_{t\star}\!\left(\bR_2\trc  X, Z\right),\nn
B&:=&\left(\bF_1\trc X\right)\gm\!\left(\bF_2\trc Y,\bF_3\trc Z\right)
=\left(\bF_1\trc X\right)  \gm\!\left(\bF_{2(1)}\bF_1'\trc Y,\bF_{2(2)}\bF_2'\trc Z\right)\nn
&=& \left(\bF_1\trc X\right)\bF_2\trc \gm\!\left(\bF_1'\trc Y,\bF_2'\trc Z\right)
=X\star\gm_{t\star}(Y,Z).\nonumber
\eea
To prove (\ref{TwistedConnection}) we note that classically $\nabla$ is
the Levi-Civita covariant derivative $\nabla_{X^i\partial_i}(Y^i\partial_i)
=X^i\partial_i(Y^j)\partial_j$ and $\bF(\trc\ot\trc)(H\ot X)=H\ot X$,
$\bF(\trc\ot\trc)(X\ot E)=X\ot E$ for all $X\in\Xi$, while by (\ref{utile}),  (\ref{so(2,1)})
$\bF(\trc\ot\trc)(E\ot H)
=E\ot H+2i\nu E\ot E$,
$\bF(\trc\ot\trc)(E\ot E')
=E\ot E'+i\nu E\ot H-2\nu^2E\ot E$,
$\bF(\trc\ot\trc)(E'\ot H)
=E'\ot H-2i\nu E'\ot E$ and
$\bF(\trc\ot\trc)(E'\ot E')
=E'\ot E'-i\nu E'\ot H$.

\subsubsection{Proof of Proposition~\ref{prop06}}

\ $[D,\g]=0$ implies $\F(g\ot \1)=(g\ot \1)\F$ for all $g\in\g$.
Moreover, since $\F(\1\ot H)=(\1\ot H)\F$, $\F(\1\ot D)=(\1\ot D)\F$,   it follows that $\Delta_\f(H)=\Delta(H)$, $S_\f(H)=S(H)=-H$, $\Delta_\f(D)=\Delta(D)$, $S_\f(D)=S(D)=-D$. On the other hand, $HE=E(H+2)$,
 $HE'=E'(H-2)$ imply
\bea
&&\F(\1\ot E)=\exp(i\nu D\otimes H/2)(\1\ot E)=(\1\ot E)\exp[i\nu D\otimes (H/2+\1)]=(e^{i\nu D}\ot E)\F,\nn[4pt]
&& \F(\1\ot E')=\exp(i\nu D\otimes H/2)(\1\ot E')=(\1\ot E')\exp[i\nu D\otimes (H/2-\1)]=(e^{-i\nu D}\ot E')\F,
\nonumber
\eea
which together with $\F(g\ot \1)=(g\ot \1)\F$ and 
$ \mu[(S_\f\otimes\mathrm{id})\Delta_\f(g)]=0$ for $g=E,E'$
imply (\ref{AbTwistDeltaSCone}).

 $[D,\g]=0$ also implies $\bF_1\rhd g\otimes \bF_2=g\otimes \1$ for all $g\in\g$,
whence \ $g\star \alpha=g\alpha$ for all $\alpha\in U\g,\Q$,
in particular for the $\alpha$ appearing in the formulas of the proposition.
$D\rhd\tilde{\partial}_i=-\tilde{\partial}_i$, $D\rhd u^i=u^i$ for $u^i=y^i,\eta^i$
 imply 
$\bF_1\rhd \tilde{\partial}_i\otimes \bF_2=\tilde{\partial}_i\otimes e^{i\nu H/2}$,
$\bF_1\rhd u^i\ot \bF_2=u^i\ot e^{-i\nu H/2}$,
whence \ $\tilde{\partial}_i\star\alpha=\tilde{\partial}_i (e^{i\nu H/2}\rhd\alpha)$,
$u^i\star\alpha=u^i (e^{-i\nu H/2}\rhd\alpha)$. 
Since $D,H,E,E'$ and $y^i,\eta^i,\tilde{\partial}_i$ (generators of $\Q$) are all 
eigenvectors of $H\rhd$, choosing $\alpha$ as
each of them, we immediately find the remaining formulae in
(\ref{AbTwistStarProdCone1}-\ref{AbTwistStarProdCone2}). 
One  finds the involution $*_\star$ using the following results:
\bea
\ba{l} S(\beta)=\F_2S(\F_1)=\sum_{n=0}^\infty  \frac {(-i\nu H)^n}{2}\frac {D^n}{n!}=e^{-\frac i2\nu HD}\qquad
\qquad\Rightarrow\qquad \\[4pt]
 S(\beta)\rhd u^i=e^{-\frac i2\nu H}\rhd u^i=e^{-i\nu\mu_i}u^i, \qquad
S(\beta)\rhd \tilde{\partial}_i=e^{\frac i2\nu H}\rhd \tilde{\partial}_i=e^{-i\nu\mu_i}\tilde{\partial}_i. 
\ea\nonumber
\eea
The commutation relations (\ref{AbTwistStarComRelCone1}),
the realization of $D,H,E,E'$ as combinations of $y^i\star\tilde{\partial}_i$, and the relations
(\ref{AbTwistCharCone1}) characterizing $\QMst$  follow from (\ref{AbTwistStarProdCone1}-\ref{AbTwistStarProdCone2}) by direct computations.

\end{document}